\documentclass[a4paper,10pt]{article}
\usepackage{amsmath,amssymb,subfigure,graphicx,setspace,mathptmx,chicago}

\usepackage[geometry]{ifsym}
\usepackage[text={6in,9in},right=1in]{geometry}

\newcommand{\bfmu}{\hbox{\boldmath$\mu$}}
\newcommand{\likelihood}[1]{\ensuremath{\bfL\left(#1\right)}}
\newcommand{\bfE}{\hbox{\boldmath$E$}}
\newcommand{\bfR}{\hbox{\boldmath$R$}}
\newcommand{\bfX}{\hbox{\boldmath$X$}}
\newcommand{\bfu}{\hbox{\boldmath$u$}}
\newcommand{\bftheta}{\hbox{\boldmath$\theta$}}
\newcommand{\bfL}{\hbox{\boldmath$L$}}
\newcommand{\dnorm}[2]{\ensuremath{\mathcal{N}\left(#1, #2\right)}}
\newcommand{\bfzeta}{\hbox{\boldmath$\zeta$}}
\newcommand{\bfeta}{\hbox{\boldmath$\eta$}}
\newcommand{\bfalpha}{\hbox{\boldmath$\alpha$}}
\newcommand{\bfbeta}{\hbox{\boldmath$\beta$}}
\newcommand{\bfzero}{\hbox{\boldmath$0$}}
\newcommand{\bfI}{\hbox{\boldmath$I$}}
\newcommand{\dgamma}[2]{\ensuremath{\text{Gamma}\left(#1, #2\right)}}
\newcommand{\bfm}{\hbox{\boldmath$m$}}
\newcommand{\bfv}{\hbox{\boldmath$v$}}

\title{Discrimination for Two Way Models with Insurance Application}
\author{\centering G. O. Brown \footnote{Corresponding Author:
    Statistical Laboratory, Centre for Mathematical Sciences,
    Cambridge CB3 0WB, UK. Email: \texttt{gob20@statslab.cam.ac.uk}},
   W. S. Buckley \footnote{College of Business Administration, Florida International University,
        Miami, Florida 33199, USA. Email: \texttt{wbuck001@fiu.edu}}}

\begin{document}
\maketitle

\begin{abstract}
  In this paper, we review and apply several approaches to model selection for analysis 
  of variance models which are used in a credibility and insurance context.
   The reversible jump algorithm is employed for model selection, where 
   posterior model probabilities are computed.
    We then apply this method to insurance data from workers' compensation insurance schemes. The
  reversible jump results are compared with the Deviance Information
  Criterion, and are shown to be consistent.
\end{abstract}


\section{Introduction}

In this paper, we address a problem posed by ~\shortciteN{klugman1987}.
 We consider an example using the efficient proposals reversible jump method.
 In this example, we consider a complex two way analysis of variance model
 using loss ratio. We introduce alternative models of describing the process 
 and perform model discrimination using the reversible jump algorithm.
 
Throughout our discussion we consider data $\bfR$ which are insurance loss
ratios. The motivation for working with loss ratios are given by
\citeN{hogg1984} and \citeN{klugman1987}. The higher levels will reflect the
group to group variations in the departure from the expected losses. This
will be more stable than the group to group variations in the absolute level
of losses. Also we use normal models since we want to compare classical
credibility models.
By assuming a linear least squares
approach, as in classical approach, there is a tacit assumption of normality
underlying the modelling process.

Suppose that $\bfR_{obs}$ are the observed loss ratios, and we seek to
minimise the predicted future loss ratios $\bfR_{new}$.
The minimum expected loss is the conditional variance of $\bfR_{new}$ given
$\bfR_{obs}$ and this minimum variance occurs when the predictor is the
regression of $\bfR_{new}$ on $\bfR_{obs}$ i.e. the conditional expectation
$\mathbb{E}(\bfR_{new}|\bfR_{obs} )$. Using this decision theoretic approach
we could specify a collection of candidate models, $\mathcal{M}=\{M_i\}$ say, then
construct a decision principle based on some collection of utility functions
and select the model which minimises the expected loss. In some cases,
however, the specification of a utility function is not always possible and
we must seek alternative approaches.  In this paper, we show how an approach
based on the deviance function can be used for model selection.  It is assumed that 
 a collection of plausible models exist, and we begin by asking the questions:
\begin{enumerate}
  \item Which model explains the data we have observed?
  \item Which model best predicts future observations?
  \item Which model best describes the underlying process which generated the
    data?
\end{enumerate}
We briefly review several
perspectives on model selection and the connection between them before presenting our models and results.

\section{General Perspective}
We consider joint modelling of the parameter vector $\bftheta_k$ and the model $M_k$.
As noted by \citeN{rubin1995}, the Bayes factor is based on the assumption
that one of the models being compared is the true model.  However, we
cannot assume this to be  generally true, and we do no make this
assumption. \citeN{carlin1996} discusses several methods using Markov chain
methods for model assessment and selection.
 We analyse
credibility models using some of these methods.   We consider model
selection using posterior model probabilities based on joint modelling over
the model space and parameter space.  Prediction is often
the ultimate goal in credibility theory. 
 We consider model selection using
predictive ability and the overall complexity of the model. We intend to use
a decision theoretic approach to prediction using utility theory.
  We begin by motivating a decision theoretic approach and then show
how this approach can be implemented using Markov Chain Monte Carlo (MCMC) methods.

\citeN{bernardo1994:smith} discusses several alternative views of model
comparison. They are separated into three principal classes. The first is
called the $\mathcal{M}$--closed system; it assumes that one of the models is
the true model generating the observed data; however, it does not specifying which
model is the true model. In this case, the marginal likelihood of the data is
averaged over the specified model. Thus
\begin{equation*}
  p(\bfR) = \sum_{M_i\in \mathcal{M}} p(M_i) p(\bfR | M_i).
\end{equation*}
In addition \citeN{madigan1994} show that in posterior predictive terms if
$\gamma$ is a quantity of interest, averaging over the candidate models
produces better results than relying on any single model.
\begin{equation}\label{eq:posteriorpredictive}
  \pi(\gamma | \bfR) = \sum_{i=1}^K p(\gamma| M_i, \bfR) \pi(M_i | \bfR)
\end{equation}
where $\pi(M_k | \bfR)$ is the posterior probability of model $M_k$ given the
observed data and
\begin{equation}\label{eq:postgammacond}
  p(\gamma | M_k, \bfR) = \int p(\gamma | \bfR, \bftheta_k, M_k)
  \pi(\bftheta_k | \bfR, M_k) d\bftheta_k .
\end{equation}
For a general review of Bayesian modelling averaging, see \citeN{clyde1999}, and
\shortciteN{hoeting1999}.   However, when the set of candidate
models $\mathcal{M}$ is not exhaustive,  we might not be able to average
over all possible models.  In that context,  placing a prior distribution
on $\mathcal{M}$ does not apply, and since we are interested only in
predicting future unknown values, this might be more appropriate than
selecting a single model.

The second alternative is the so called $\mathcal{M}$-completed view,
which simply seeks to compare a set of models which are available at that
time. In this case $\mathcal{M} = \{ M_i \}$ simply constitute a range of
specified models to be compared. From this perspective, assigning the
probabilities $\{ P(M_i),\, M_i \in \mathcal{M} \}$ does not make sense and
the actual overall model specifies beliefs for $\bfR$ of the form $p(\bfR) =
p(\bfR | M_t)$. Typically, $\{ M_i \}$ will have been proposed largely
because they are attractive from the point of view of tractability of
analysis or communication of results, compared with the actual belief model
$M_t$.

The third alternative is the $\mathcal{M}$-open view. In an
$\mathcal{M}$-open system it is assumed that none of the models being
considered is the true model which generated the observations. In this case, our goal 
is to select some model or subset of models which best describe the data.
For the $\mathcal{M}$-completed and $\mathcal{M}$-open views, assigning prior
probabilities on the model space $\mathcal{M}$ is inappropriate since
statements like $p(M_k)= c$ do not make sense. However, in the
$\mathcal{M}$-open case, there is not separate overall belief specification.

\section{Decision Theoretic Approach}
\shortciteN{key1999} argue that any criteria for model comparison should
depend on the decision context in which the comparison is taking place, as
well as the perspective from which the models are viewed. In particular, an
appropriate utility structure is required, making explicit those aspects of
the performance of the model that is most important.  Using a decision theoretic
approach, we can assign utilities to the choice of model $M_i$, $u(M_i,
\gamma)$, where $\gamma$ is some unknown of interest.  The general decision
problem is then to choose the optimal model, $M^*$, by maximising expected
utilities
\begin{equation*}
  \bar{u} (M^* | \bfR) = \sup_{\substack{M_i}} \bar{u}(M_i | \bfR),
\end{equation*}
where 
\begin{equation*}
  \bar{u}(M_i |\bfR) = \int u(M_i, \gamma) \pi(\gamma | \bfR) d\gamma
\end{equation*}
with $\pi(\gamma | \bfR)$ representing actual beliefs about $\gamma$ after
observing $\bfR$ in Equation~\eqref{eq:posteriorpredictive}.

\shortciteN{spiegelhalter2002} propose their deviance information criterion,
$DIC$, as an alternative to Bayes' factors. In \shortciteN{spiegelhalter2002},
the $DIC$ is developed to address how well the posterior might predict future
data generated by the same mechanism that gave rise to the observed data. Our
motivation is that likelihood ratio tests cannot be used when there are
unobservables, and that they apply only to nested models. Also likelihood
ratio based tests are inconsistent, since as the sample size tends to infinity, the
probability that the full model is selected does not approach zero
~\cite{gelfand1996:mcmcip}. 

The likelihood ratio gives too much weight to the
higher dimensional model, which motivates the discussion on penalised
likelihoods using penalty functions. A good penalty function should depend on
both the sample size and the dimension of the parameter vector.  The decision
theoretic approach is general enough to include traditional model selection
strategies, such as choosing the model with the highest posterior probability.
For example in the $\mathcal{M}$--closed system, where we assume that
$\mathcal{M}$ contains the true model, if we assume a utility function of the
form
\begin{equation*}
  u(M_i, \gamma) = \begin{cases}1 & \text{if $\gamma = M_i$}\\
    0 & \text{if $\gamma\neq M_i$}, \end{cases}
\end{equation*}
then from \eqref{eq:postgammacond}
\begin{equation*}
  p(\gamma | \bfR, M_k) = \begin{cases} 1 & \text{if $\gamma = M_i$}\\
    0 & \text{if $\gamma\neq M_i$}\end{cases}
\end{equation*}
and
\begin{equation*}
  \pi(\gamma | \bfR )=\begin{cases} \pi(M_i|\bfR) & \text{if $\gamma = M_i$}\\
    0 & \text{if $\gamma\neq M_i$}.\end{cases}
\end{equation*}
The expected utility is then
\begin{align*}
  \bar{u}(M_i | \bfR) &= \int u(M_i, \gamma) \pi(\gamma | \bfR) \, d\gamma\\
  &= \pi(M_i | \bfR).
\end{align*}
Therefore, the optimal decision is to choose the model with the highest
posterior probability. For the $\mathcal{M}$--completed case,
\citeN{bernardo1994:smith} shows that the cross validation predictive density
yields similar results. The connection between $DIC$ and the utility approach
using cross validation predictive densities, has been studied by
\citeN{vehtari2002}, and \citeN{vehtari2002:spiegelhalter} who use cross
validation to estimate expected utility directly, and also the effective
number of parameters. The main differences are, that cross validation can be
less numerically stable than the $DIC$ and can also require more computation.
However, $DIC$ can underestimate the expected deviance. For a list of specific
utilities used when choosing models, see \shortciteN{key1999}.

\section{Computing Posterior Model Probabilities}\label{sec:transmcmc}

\subsection{Reversible Jump Algorithm}\label{sec:reversible}
We assume there is a countable collection of candidate models, indexed by
$M\in\mathcal{M}=\{M_1$, $M_2$,\ldots , $M_k\}$. We further assume that for
each model $M_i$, there exists an unknown parameter vector $\bftheta_i \in
\mathbb{R}^{n_i}$ where $n_i$, the dimension of the parameter vector, can
vary with $M_i$.

Typically, we are interested in finding which models have the greatest
posterior probabilities, in addition to estimates of their parameters. Thus the
unknowns in this modelling scenario will include the model index $M_i$, as
well as the parameter vector $\bftheta_i $. We assume that the models and
corresponding parameter vectors have a joint density $\pi(M_i,\bftheta_i )$.
The reversible jump algorithm constructs a reversible Markov chain on the
state space $\mathcal{M} \times \bigcup_{M_i\in\mathcal{M}} \mathbb{R}^{n_i}$
which has $\pi$ as its stationary distribution \cite{green1995}.  In many
instances, and in particular for Bayesian problems, this joint distribution
is
\begin{equation*}
  \pi(M_i, \bftheta_i) = \pi(M_i, \bftheta_i\vert
  \bfR)\propto
  \bfL(\bfR\vert M_i, \bftheta_i) \; 
  p(M_i, \bftheta_i ) ,
\end{equation*}
where the prior on $(M_i,\bftheta_i)$ is often of the form
\begin{equation*}
  p(M_i, \bftheta_i) = p(\bftheta_i\vert M_i) \;p(M_i)
\end{equation*}
with $p(M_i)$ being the density of some counting distribution.

Suppose we are at model $M_i$, and a move to model $M_j$ is proposed
with probability $r_{ij}$. The corresponding move from
$\bftheta_i$ to $\bftheta_j$ is achieved by using a
deterministic transformation $h_{ij}$, such that
\begin{equation}\label{eq:revjumpmapping}
  (\bftheta_j , \bfv) = h_{ij}(\bftheta_i, \bfu ),
\end{equation}
where $\bfu$ and $\bfv$ are random variables introduced to ensure dimension
matching necessary for reversibility. To ensure dimension matching, we must
have
\begin{equation*}
  \dim(\bftheta_j)+\dim(\bfv)=\dim(\bftheta_i)+\dim(\bfu). 
\end{equation*}
For discussions about possible choices for the function $h_{ij}$, we refer the
reader to \citeN{green1995}, and \shortciteN{brooks2003}.  If we denote the
ratio
\begin{equation}\label{eq:acceptratio}
  \frac{\pi(M_j, \bftheta_j)}{\pi(M_i, \bftheta_i )}  
  \frac{q(\bfv)}{q(\bfu)} 
  \frac{r_{ji}}{r_{ij}}\hskip .15cm
  \biggl\lvert\frac{\partial h_{ij}(\bftheta_i, \bfu)}
  {\partial (\bftheta_i,\bfu)} \biggr\rvert
\end{equation}
by $A(\bftheta_i, \bftheta_j )$, the acceptance
probability for a proposed move from model $(M_i, \bftheta_i)$ to
model $(M_j, \bftheta_j)$ is:
\begin{equation*}
  \min\left\{1, A(\bftheta_i, \bftheta_j )  \right\}
\end{equation*} 
where $q(\bfu)$ and $q(\bfv)$ are the respective proposal densities for $\bfu$
and $\bfv$, and $\lvert \partial h_{ij}(\bftheta_i,\bfu) / \partial
(\bftheta_i,\bfu) \rvert$ is the Jacobian of the transformation induced by
$h_{ij}$. It can be shown that the algorithm constructed above is reversible
\cite{green1995} which, again, follows from the detailed balance equation
\begin{equation*}
  \pi(M_i, \bftheta_i) q(\bfu) r_{ij} = 
  \pi(M_j, \bftheta_j) q(\bfv) r_{ji} \hskip .1cm \biggl\lvert
  \frac{\partial h_{ij}(\bftheta_i,\bfu )} 
  {\partial (\bftheta_i, \bfu)}
  \biggr\rvert .
\end{equation*}
Detailed balance is necessary to ensure reversibility and is a sufficient
condition for the existence of a unique stationary distribution.
For the reverse move from model $M_j$ to model $M_i$ it is easy to see that
the transformation used is $(\bftheta_i, \bfu) =
h_{ij}^{-1}(\bftheta_j, \bfv)$, and the acceptance probability for
such a move is
\begin{equation*}
  \min\left\{1, 
    \frac{\pi(M_i, \bftheta_i)}{\pi(M_j, \bftheta_j)}  
    \frac{q(\bfu)}{q(\bfv)} 
    \frac{r_{ij}}{r_{ji}}\hskip .15cm
    \biggl\lvert\frac{\partial h_{ij}(\bftheta_i,\bfu)}
    {\partial (\bftheta_i,\bfu)} \biggr\rvert^{-1} \right\} 
  =\min\left\{1, A(\bftheta_i, \bftheta_j)^{-1} \right\}.
\end{equation*} 
For inference regarding which model has the greater posterior probability, we
can base our analysis on a realisation of the Markov chain constructed above.
The marginal posterior probability of model $M_i$ 
\begin{equation*}
  \pi(M_i\vert \bfR) = \frac{p(M_i) f(\bfR\vert M_i)}
  {\sum_{M_j\in\mathcal{M}} p(M_j) f(\bfR\vert M_j) },
\end{equation*}
where 
\begin{equation*}
  f(\bfR\vert M_i) =\int \bfL(\bfR\vert 
  M_i,\bftheta_i)
  p(\bftheta_i|M_i)\, d\,\bftheta_i
\end{equation*} 
is the marginal density of the data after integrating over the unknown
parameters $\bftheta$. In practice, we estimate $\pi(M_i| \bfR)$ by counting
the number of times the Markov chain visits model $M_i$ in a single long run
after becoming stationary.

\subsection{Efficient Proposals for TD MCMC} \label{sec:efficientprop}
In practice, the between--model moves can be small resulting in poor mixing of
the resulting Markov chain. In this section, we discuss recent attempts at
improving between--model moves by increasing the acceptance probabilities for
such moves. Several authors have addressed this problem, including
\citeN{troughton1997}, \citeN{giudici1998}, \citeN{godsill2001},
\citeN{rotondi2002}, and \shortciteN{alawadhi2004}. \citeN{green2001:mira}
proposes an algorithm so that when between--model moves are first rejected, a
second attempt is made. This algorithm allows for a different proposal to be
generated from a new distribution, that depends on the
previously rejected proposal. Methods to improve mixing of reversible jump
chains have also been proposed by \citeN{green2002} and
\shortciteN{brooks2003}; these are extended by \citeN{ehlers2002}.

One  strategy proposed by \shortciteN{brooks2003}, and extended to more
general cases by \citeN{ehlers2002}, is based on making the term
$A_{ij}(\bftheta_i,\bftheta_j)$ in the acceptance probability for
 between--model moves given in Equation~\eqref{eq:acceptratio}, as close as possible to
1. The motivation is that if we make this term as close as
possible to 1, then the reverse move acceptance governed by $1 /
A_{ij}(\bftheta_i,\bftheta_j)$ will also be maximised resulting in easier
between--model moves. In general, if the move from $(M_i,
\bftheta_i)\Rightarrow(M_j,\bftheta_j)$ involves a change in dimension, the
best values of the parameters for the densities $q(\bfu)$ and $q(\bfv)$ in
Equation~\eqref{eq:acceptratio}, will generally be unknown, even if their
structural forms are known. 

Using some known point $(\widetilde \bfu,
\widetilde \bfv)$, which we call the centering point, we can solve
$A_{ij}(\bftheta_i,\bftheta_j)=1$ to get the parameter values for these
densities. Setting $A_{ij}=1$ at some chosen centering point is called the
zeroth-order method. Where more degrees of freedom are required, we can expand
$A_{ij}$ as a Taylor series about $(\widetilde\bfu, \widetilde \bfv)$ and
solve for the proposal parameters. 
New parameters are proposed so that the mapping function in
Equation~\eqref{eq:revjumpmapping} is the identity function, i.e.,
\begin{equation*}
  (\bftheta_j, \bfv) = h_{ij}(\bftheta_i, \bfu) = (\bfu, \bftheta_i)
\end{equation*}
and the acceptance ratio term $A_{ij}(\bftheta_i, \bftheta_j)$ probability in
Equation~\eqref{eq:acceptratio} becomes
\begin{align*}
  A_{ij}(\bftheta_i, \bftheta_j) &=
  \frac{\pi(M_j, \bftheta_j)}{\pi(M_i, \bftheta_i)} 
  \frac{r_{ji}}{r_{ij}} 
  \frac{q (\bfv)}{ q(\bfu)} \\
  &=
  \frac{\pi(M_j, \bftheta_j)}{\pi(M_i, \bftheta_i)} 
  \frac{r_{ji}}{r_{ij}} 
  \frac{q (\bftheta_i)}{ q(\bftheta_j)} .
\end{align*}

 Several authors have
proposed simulation methods to construct Markov chains which can explore such
state spaces. These include the product space formulation given in
\citeN{carlin1995}, the reversible jump (RJMCMC) algorithm of
\citeN{green1995}, the jump diffusion method of ~\citeN{grenander1994}, and
\citeN{phillips1996:mcmcip} and the continuous time birth-death method of
\citeN{stephens2000}. Also for particular problems involving the size of the
regression vector in regression analysis, there is the stochastic search
variable selection method of \citeN{george1993}.
practice trans--dimensional algorithms work by updating model parameters for
the current model, then proposing to change models with some specified
probability.

\subsection{Deviance Information Criterion}
The $DIC$ is based on using the residual information in $X$ conditional on
$\bftheta$, defined up to a multiplicative constant as $-2\log L(\bfX |
\bftheta)$. If we have some estimate $\widetilde\bftheta =
\widetilde\bftheta(\bfX)$ of the true parameter, $\bftheta^t$, then the
excess residual information is
\begin{equation*}
  d(\bfX, \bftheta^t, \widetilde\bftheta) = -2\log \bfL(\bfX | \bftheta^t) +
  2\log\bfL(\bfX | \widetilde\bftheta) 
\end{equation*}
This can be thought of as the reduction in uncertainty due to estimation or
the degree of overfitting due to $\widetilde\bftheta$ adapting to the data
$\bfX$. From a Bayesian perspective $\bftheta^t$ may be replaced by some
random variable $\bftheta\in\Theta$. Then $d(\bfX, \bftheta^t,
\widetilde\bftheta)$ can be estimated by its posterior expectation with
respect to $\pi(\bftheta|\bfX)$ denoted 
\begin{align*}
  p_D(\bfX, \Theta, \widetilde\bftheta) &=
  \mathbb{E}_{\bftheta|\bfX} d(\bfX, \bftheta, \widetilde\bftheta)\\
  &=  \mathbb{E}( -2\log\bfL(\bfX|\bftheta) ) + 
  2\log\bfL(\bfX|\widetilde\bftheta) .
\end{align*}
$p_D$ is then proposed as the effective number of parameters with respect to
a model with focus $\Theta$. Thus, if we take $h(\bfX)$ as some fully
specified standardising term that is a function of the data alone, then $p_D$ may
be written as
\begin{align*}
  p_D &= \overline{D(\bftheta)} - D(\bar\bftheta) \\
  &= \mathbb{E}_{\bftheta|\bfX}(D(\bftheta)) - 
  D(\mathbb{E}_{\bftheta|\bfX}(\bftheta))
\end{align*}
where
\begin{equation}\label{eq:saturateddeviance}
  D(\bftheta) = -2\log\bfL(\bfX|\bftheta) + 2\log h(\bfX) .
\end{equation}
Using Bayes' theorem we have
\begin{equation*}
  p_D = \mathbb{E}_{\bftheta|\bfX} -2 \log\left( 
    \frac{\pi(\bftheta|\bfX)}{p(\bftheta)}
  \right) +
  2\log\left(
    \frac{\pi(\widetilde\bftheta|\bfX)}{p(\widetilde\bftheta}
  \right)
\end{equation*}
which can be viewed as the posterior estimate of the gain in information
provided by the data about $\bftheta$, minus the plug--in estimate of the
gain in information. Having an estimate for the effective number of
parameters, $p_D$, the quantity
\begin{align*}
  DIC &= D(\bar\bftheta) + 2 p_D \\
  &= \overline{D(\bftheta)} + p_D
\end{align*}
can then be used as a Bayesian measure of fit, which when used in models with
negligible prior information will be approximately equivalent to the $DIC$
criterion.

If $D(\cdot)$ in Equation~\eqref{eq:saturateddeviance} is available in
closed form, $p_D$ may easily be computed using samples from an MCMC
run. This is what we propose to do to measure each models complexity
and then rank the models in terms of their complexity. Even though we
have defined $p_D$ in terms of the expectation with respect to some
density,
other measures such as the mode or median can be used instead.

\section{Discrimination for ANOVA Type Models}
Quite often, the hierarchical credibility model of \citeN{jewell1975} can be
formulated as an analysis of variance type model.  In this paper, we use
reversible jump techniques to compute posterior model probabilities and
compare various analysis of variance models. The reversible jump results are
also compared with the results obtained by using the $DIC$.

Hierarchical models in credibility theory have been considered by
\citeN{jewell1975}, \citeN{taylor1979}, \citeN{zehnwirth1982}, and
\citeN{norberg1986}. Recent reviews of linear estimation for such models has
been presented by \citeN{goovaerts1987} and \shortciteN{dannenburg1996}.
The results in this paper also have implications for other problems such as
the claims reserving run-off triangle method, which we have not considered. 
 This formulation has already been exploited by
\citeN{kremer1982:scandactj} and \citeN{ntzoufras2002}, who use MCMC to
estimate claim lags.

In this paper, we address a problem posed by ~\shortciteN{klugman1987}
and we consider an example using the efficient proposals reversible
jump method. This example is a complex two--way analysis of variance
model involving loss ratios . We introduce alternative models for
describing the process which generated the data, and perform model
discrimination using the reversible jump algorithm.

This paper contributes to the literature on model discrimination based
on reversible jumps for reparameterised B{\"u}hlmann--Straub model, a 
two--way model, and the hierarchical model of \citeN{jewell1975}.  The general
question is whether there is any advantage gained by using a two--way model
rather than a simple random effects model in analysing the data. 
Even though the
one--way model is a nested sub-model of the two--way model, the resulting
parameter estimates can be different under both models since they have
different interpretations.  In this example, we see that the the
two--way model is vastly superior.  In the context of the Bayesian paradigm, we
are able to derive posterior model probabilities and use these to discriminate
between competing models.  For each algorithm, the between--model moves are
augmented with within--model moves which can be used to estimated model
parameters for each model. 

In Section~\ref{sec:hierarchicalcentering}, we
therefore discuss how the choice of parameterisation affects the convergence
of the Markov chain algorithm for within--model simulations.  The between--model 
moves are done using the Taylor series expansion of the between--model
acceptance probabilities near to some point called the centering point.
In some cases using weak
non--identifiable centering does not work well. Another approach, which we
employ in this example, is the conditional maximisation approach, where
the centering point is selected to maximise the posterior density.

\subsection{The Basic Two--Way Model}
\begin{figure}[h]
  \centering
    \setlength{\unitlength}{1cm}
  \begin{picture}(6,2.5)(0,1)
    \put (1,2){\circle{1}}
    \put (3,2){\circle{1}}
    \put (1.5,2) {\vector(2,0){1}}
    \put (3.5,2) {\vector(2,0){1}}
    \put (0.9,1.85) {$\boldsymbol\theta$}
    \put (2.9,1.85) {$\boldsymbol X$}
    \put (4.5,1.5) {\framebox(1,1){$\boldsymbol Y$}}
  \end{picture}
  \caption{Centred parameterisation}\label{fig:centred}
\end{figure}
\begin{figure}[h]
  \centering
  \setlength{\unitlength}{1cm}
  \begin{picture}(6,3)(-1, -0.5)
    \put (0,0) {\circle {1}}
    \put (0,2) {\circle {1}}
    \put (2,1) {\circle{1}}
    \put (3.5,0.5) {\framebox(1,1){$\boldsymbol Y$}}
    \put (2.5,1){\vector(2,0){1}}
    \put (-0.1,1.85) {$\widetilde{\boldsymbol X}$}
    \put (-0.1,-0.15) {$\boldsymbol\theta$}
    \put (1.8,0.85) {$\boldsymbol X$}
    \put (0.5,0){\vector(1,1){1}}
    \put (0.5,2){\vector(1,-1){1}}
  \end{picture}
  \caption{Non-centred parameterisation}\label{fig:noncentred}
\end{figure}
The generic hierarchical model can be described as a connected graph
as shown in Figure~\ref{fig:centred}.  Let $\theta$ denote the
collection of parameters, $Y$ represent the observed data, and $X$ can
take the role of missing data or other possibilities. The algorithm
for sampling from the joint distribution of $\theta$, $X$, given the
observed data might proceed by alternating
\begin{enumerate}
\item Update $\theta$ from a Markov chain with stationary distribution
  $\theta | X$
\item Update X from a Markov chain with stationary distribution $X | \theta,
  Y$
\end{enumerate} 
The rate of convergence of the Gibbs sampler is directly related to the
choice of parameterisation for such problems. On the other hand, we might be
able to find an alternative parameterisation, $(X, \theta)\rightarrow (\tilde
X, \theta)$, of the model in Figure~\ref{fig:centred} where the new missing
data is some function of the previous missing data $X$ and the parameters
$\theta$, such that $\tilde X$ is a priori independent of $\theta$. The type
of parameterisation shown in Figure~\ref{fig:noncentred} is called
non--centred parameterisation. The corresponding algorithm for simulating
from the posterior distribution of $(\tilde X, \theta)$, is then
\begin{enumerate}
\item Update $\theta$ from a Markov chain with stationary distribution
  $\theta | \tilde X, Y$
\item Update $\tilde X$ from a Markov chain with stationary distribution
  $\tilde X | \theta, Y$ .
\end{enumerate}
For more general discussions, see \citeN{gelfand1999} and
\shortciteN{papaspilioppoulos2003}.

The general form of the two--way model considered herein is the non--centred
parameterisation:
\begin{equation}\label{eq:noncentred}
  y_{ijt} = \mu + \alpha_i + \beta_j + \gamma_{ij} + \epsilon_{ijt}
  \quad i=1,\ldots,m;\, j=1,\ldots,n;\,t=1,\ldots, s,
\end{equation}
in which there are $s$ replications for factors $i$ and $j$. The error terms
in the observations are assumed to be normally distributed and can depend on
other known values. Quite often we assume that $s=1$.  The interpretation of
this model is that there is some overall level common to all observations,
$\mu$, and then there are treatment effects that depend on the factors $i$
and $j$, denoted $\alpha_i$ and $\beta_j$, respectively.  The $\gamma_{ij}$
are the interactions between the factors and they are assumed
identically equal to zero. 

 Bayesian analysis of one-way and two-way models
and general mixed linear models are studied by \citeN{scheffe1959},
\citeN{box1973}, and \citeN{smith1973}. The analysis of
\citeN{smith1973} is based on the more general normal linear model of
~\citeN{lindley1972}. The error term $\epsilon_{ijt}$, is assumed to be
normally distributed with $\epsilon_{ijt}\sim\dnorm{0}{( \sigma E_{ijt})^{-1}
}$, where $E_{ijt}$ is some scale factor associated with observation
$y_{ijt}$. The effects $\alpha_i$ and $\beta_j$ are assumed to have prior
variances $1 / \tau_{\alpha}$ and $1/\tau_{\beta}$, respectively.  Similar
models have been analysed by \citeN{nobile2000},   who modelled the
factor terms as mixtures of normal distributions using reversible jump
methods to select the number of components in the mixture.
\shortciteN{ahn1997}  uses classical methods to compare their models. For
the within--model parameter updates, we use the Gibbs sampler algorithm.
We briefly discuss the choice of parameterisation and how
different updating schemes can affect the within model convergence
properties.

Before discussing how the choice of parameterisation affects the Gibbs
sampler for linear mixed models, we note that the centering discussed in this
section is related to the parameterisation of the models discussed, and not to
the choice of centering point discussed in relation to the efficient
proposals methods.  For example, let 
\begin{align*}
  \eta_i & = \mu + \alpha_i \\
  \zeta_{ij}& = \eta_i + \beta_j. 
\end{align*}
The above stated model could be reparameterised so that
\begin{align*}
  y_{ijt} & = \zeta_{ij} + \epsilon_{ijt} \\
   \zeta_{ij} & \sim \dnorm{\eta_i}{\tau_{1}^{-1}} \\
   \eta_i &\sim \dnorm{\mu}{\tau_{2}^{-1}} .
\end{align*}
This new $(\mu, \bfeta, \bfzeta)$ parameterisation is then called the centred
parameterisation, since the $\zeta_{ij}$ are centred about the $\eta_{i}$ and
the $\eta_{i}$ are also centred about $\mu$.  The original $(\mu, \bfalpha,
\bfbeta)$ parameterisation in \eqref{eq:noncentred} is called the non--centred
parameterisation.  Partial centerings are also possible, see
\citeN{gilks1996:mcmcip} for further discussion.

\subsection{Hierarchical Centering and Gibbs Updating Schemes}
\label{sec:hierarchicalcentering}
\shortciteN{gelfand1995} consider general parameterisations and a
hierarchically centred parameterisation by increasing the number of levels in
a Bayesian analysis.
They show that, if $\tau_{\beta}\rightarrow 0$ with
$\tau_{\alpha}$ and $\sigma$ fixed, then the centred parameterisation will be
better. If, however, $\sigma\rightarrow 0$ with $\tau_{\alpha}$ and
$\tau_{\beta}$ fixed, then the non-centred parameterisation will be better.
They make no optimality claims for such centerings, and generally
recommend  centering the random effects with the largest posterior
variance to improve convergence.  Thus, in the two--way model, we would centre
either the $\alpha_i$s or the $\beta_j$s, provided that their variability
dominated at the data level. In problems where the variance components
are unknown this would necessitate a preliminary run of the algorithm to
determine the variance components.

\citeN{roberts1997} show that when the target density is Gaussian a
deterministic scheme is most optimal for fast convergence of the Gibbs
sampling algorithm for a class of structured hierarchical models.  This
updating scheme is also optimal for Gaussian target densities when the
components can be arranged in blocks and where there is negative partial
correlation between the blocks. The model parameters in the hierarchically
centred parameterisation have different interpretations than those in the
non--centred implementation, so direct comparison is not possible. We,
however, compare both implementations using the methods of
~\citeN{roberts1997}, whose results extend the results of
~\shortciteN{gelfand1995}. Note that with the blocked parameterisation, the
$\alpha_i$'s are conditionally independent given $\mu$, $\beta_j$ and
$\sigma$. Therefore blocking them together does not alter the performance of the
Gibbs algorithm. Blocking does not completely overcome the problems.

Block updating of the parameters should result in smaller posterior
correlations \shortcite{amit1991,liu1994}. \citeN{roberts1997} and
\citeN{whittaker1990} show that for the parameterisation given in
Equation~\eqref{eq:noncentred},  the partial correlation between any component
of one block and any component of another block, is negative. In this case a
random scan Gibbs algorithm or a random permutation Gibbs sampling algorithm
would be expected to perform better than the deterministic scan algorithm
that we use.  Where the target densities are Gaussian, \citeN{amit1991}
recommend the use random updating strategies. However, for unknown variance
components, this is not necessarily true.  

When the variance components are
unknown, the posterior distribution will cease to be Gaussian. The variance
components will be included in the model with their respective prior
specifications. The Gibbs sampler needs to sample from the joint posterior
distribution of the $\mu$, $\bfalpha$, and $\bfbeta$ and the variance
components. However, the conditional distribution of $\mu$, $\bfalpha$, and
$\bfbeta$ given the variance component will still be Gaussian. Consequently,
the behaviour of the Gibbs sampler should still be guided by the above
considerations. 

 Another reason for choosing this parameterisation is that it
allows for easy implementation in reversible jump schemes. It allows us to
easily construct algorithms to move between models with no parameters in
common as we now show, since for the one--way model, the more efficient
parameterisation depends on the ratio of variances. Thus, we choose this model
only because it allows for easier moves in the reversible jump scheme. For
general discussions about parameterisation and MCMC implementation in linear
models, see \citeN{hills1992}, \citeN{gilks1996:mcmcip},
\shortciteN{gelfand1995}, \citeN{gelfand1996}, and \shortciteN{gelfand2001}.
The case of generalized linear models is considered by \citeN{gelfand1999}.

We adopt the non--centred parameterisation for the models analysed in this
paper partly because the variance components are unknown. Also the non--centred
 parameterisation seems more readily implemented for reversible jump algorithm, 
 since there are usually fewer model parameters. In addition, for non--centred
models, the proposal distribution can easily be computed using the
efficient proposals methods.

\section{Example : Workers' Compensation Insurance}
In this section
we analyse a set of insurance data from a Workers' compensation scheme, using
a hierarchical random effects model. A typical workers' compensation scheme
exists to provide workers who are injured in the workplace with a guaranteed source of
income, until they recover and re-enter the work-force.

\subsection{The Data and Model Specification}
Our model is fully parametric and can be used to describe data representing
workers compensation for 25 classes of occupations across 10 U.S. States,
over a period of 7 years. The losses represent frequency counts on workers'
compensation insurance on permanent partial disability and the exposures are
scaled payroll totals that have been inflated to represent constant dollars.
We  use the first 6 years of data for parameter estimation of the model; 
the 7th year of data is used to test the accuracy of the predictive distribution obtained. We
need to estimate the class and occupation parameters, so that we have a basis
for estimating future observations.  The dataset has previously been analysed
by \citeN{klugman1992} using numerical approximations. Our approach will be
hierarchical Bayesian using Markov chain Monte Carlo integration to estimate
the model parameters.

The results of \citeN{klugman1992} are based on matrix analytic arguments and
numerical approximations of the posterior estimates of the parameters. In
particular \citeN{klugman1992} uses the method of Gaussian quadrature to
approximate the posterior distributions of the model parameters.
We present a MCMC based analysis based on the loss ratios, defined as
\emph{loss} / \emph{exposure}. We let
\begin{align*}
  L_{ijt} &=\text{ losses for State $i$, Occupation $j$ for year $t$ }\\
  E_{ijt} &=\text{ exposure for State $i$, Occupation $j$ for year $t$ }\\
  i&=1\ldots, 10,\quad j=1,\ldots, 25,\quad t= 1\ldots, 7.
\end{align*}
and the corresponding loss-ratios by $R_{ijt}$, where $R_{ijt}=
L_{ijt}/E_{ijt}$.

There is one occupation class with $E_{ijt}=0$ for all $i$ and $t$ ;
we removed this value of $j$ from our analysis so that data for $24$
occupation classes are left.  We begin by showing how MCMC can be used
to implement the original model in ~\citeN{klugman1992}, which is a
hierarchically centred model.  In our analysis, we reparameterise the
model and employ a non--centred model so that each level can then be
compared with the first level. Other parameterisations are possible
(See, for example, ~\citeN{venables1999}).  The choice of
parameterisation does not affect the result, since it is the sum,
$\alpha_i+\beta_j$, that really matters.

\subsection{Short Review of the Klugman Model}\label{sec:klugmanreview}
The model described by \citeN{klugman1992}, which is a special case of
\citeN{jewell1975}, has first level given by
\begin{multline}\label{eq:dset4first}
  R_{ijt}\vert\bfalpha, \bfbeta, \sigma\sim
  \dnorm{\alpha_i+\beta_j}{(\sigma E_{ijt})^{-1}}, \,\,\,
  i= 1, \ldots, 10 ; 
  j=1,\ldots, 24 ;
  t=1,\ldots,6 ,
\end{multline}
and prior structure
\begin{equation}\label{eq:ds4sec}
  \alpha_i\vert\mu, \tau_{\alpha}\sim
  \dnorm{\tfrac1{2}\mu}{\tau_{\alpha}^{-1}}
  \quad i=1\ldots,10 ,
\end{equation}
\begin{equation}\label{eq:ds4sec1}
  \beta_j\vert\mu,\tau_{\beta} \sim
  \dnorm{\tfrac1{2}\mu}{\tau_{\beta}^{-1}}
  \quad j= 1,\ldots, 24.
\end{equation}
For the hyper-parameters $\sigma$, $\tau_{\alpha}$, and $\tau_{\beta}$, we use
conjugate and diffuse \dgamma{a}{b} priors. For $\mu$, we use a diffuse
Gaussian prior with mean $0$ and variance $c^{-1}$.  The model is a two--way
model where we have made the assumption that there is no interaction between
classes and occupation. The first level \eqref{eq:dset4first} reflects what
we think about the data; that the observations are normally distributed about
some mean, which does not change with time, but depends only on the class ($i$)
and occupation ($j$).  

We also assume the variance of any particular
observation about its mean is proportional to some measure of the exposure.
This assumption is popular among insurance practitioners such as
\shortciteN{ledolter1991}, \shortciteN{klugman1992}, and
\shortciteN{ramlau1982}.  The second level comprising
Equations~\eqref{eq:ds4sec} and \eqref{eq:ds4sec1}, allows for any interaction
between the class and occupation parameters $\bfalpha
=(\alpha_1,\ldots,\alpha_{10})'$ and $\bfbeta=(\beta_1,\ldots,\beta_{24})'$
respectively. We assume they are independent, and normally
distributed with mean equal to one-half the overall mean. There is 
apparently no special reason for choosing such a prior for the $\alpha_i$ or
$\beta_j$ parameters, other than their sum should equal the overall mean
$\mu$. For our implementation we choose $a=b=c=0.001$.

\subsection{The Posterior Conditional Distributions}
The joint posterior distribution of the parameters, given the data, up to a
constant of proportionality, takes the form:
\begin{multline}\label{eq:dset4model}
  \pi(\mu, \sigma, \tau_{\alpha},\tau_{\beta}, \bfalpha, \bfbeta\vert
  \mathbf{E, R})
  \propto \\
  p(\mu)p(\tau_{\alpha})p(\tau_{\beta})p(\sigma)
  \prod_{i=1}^{10}p(\alpha_i\vert\mu, \tau_{\alpha})
  \prod_{j=1}^{24}p(\beta_j\vert\mu, \tau_{\beta})
  \prod_{i, j, t} f(R_{ijt}\vert\alpha_i, \beta_j, \sigma) .
\end{multline}

\noindent From Equation~\eqref{eq:dset4model}, we can determine the following posterior
conditional distributions for implementing a Gibbs updating scheme:
The posterior conditional for $\alpha_i$ is
\begin{align*}
  \pi(\alpha_i\vert\mu, \bfbeta,\tau_{\alpha},\sigma,\mathbf{E,R})
  &\propto p(\alpha_i\vert\mu, \tau_{\alpha})
  \prod_{jt}f(R_{ijt}\vert\alpha_i,\beta_j,\sigma)
  \nonumber \\
  \alpha_i\vert\mu, \bfbeta,\tau_{\alpha},\sigma,\mathbf{E,R}
  &\sim \dnorm{
    \frac{\tau_{\alpha}\mu /2 + \sigma\sum_{jt}E_{ijt}(R_{ijt}- \beta_j) }
    {\tau_{\alpha} + \sigma\sum_{jt}E_{ijt}}}
  {\frac1{\tau_{\alpha} + \sigma\sum_{jt}E_{ijt}}} .\\
\intertext{The posterior conditional distribution for $\beta_j$ is}
  \pi(\beta_j\vert\mu, \bfalpha, \tau_{\beta}, \sigma,\mathbf{E,R})
  &\propto p(\beta_j\vert\mu,\tau_{\beta})
  \prod_{it} f(R_{it}\vert\alpha_i,\beta_j,\sigma)
  \nonumber \\
  \beta_j\vert\mu, \bfalpha, \tau_{\beta}, \sigma,\mathbf{E,R}
  &\sim \dnorm{ 
    \frac{\tau_{\beta}\mu /2 + \sigma\sum_{it}E_{ijt}(R_{ijt}-\alpha_i) }
    {\tau_{\beta} + \sigma\sum_{it}E_{ijt}}}
  {\frac1{\tau_{\beta} + \sigma\sum_{it}E_{ijt} }} .\\
\intertext{The posterior conditional for $\mu$ is}
  \pi(\mu\vert
  \bfalpha,\bfbeta,\tau_{\alpha},\tau_{\beta},\mathbf{E,R})
  &\propto p(\mu)\prod_{ij}p(\alpha_i\vert\mu, \tau_{\alpha})
  p(\beta_j\vert\mu,\tau_{\beta}) \\
  \mu\vert \bfalpha,\bfbeta,\tau_{\alpha},\tau_{\beta},\mathbf{E,R}
  &\sim \dnorm{\frac{\tau_{\alpha}/2\sum_i\alpha_i+\tau_{\beta}/2\sum_j\beta_j}
    {c+0.25m\tau_{\alpha} + 0.25n\tau_{\beta}}}
  {\frac1{c+0.25m\tau_{\alpha} + 0.25n\tau_{\beta}}} .\\
\intertext{The posterior conditional for $\sigma$ is}
  \pi(\sigma \vert\bfalpha, \bfbeta,\mathbf{E,R})
  &\propto p(\sigma)\prod_{ijt}f(R_{ijt}\vert\alpha_i,\beta_j,\sigma)
  \nonumber\\
  \sigma \vert\bfalpha, \bfbeta,\mathbf{E,R}
  &\sim \dgamma{a+\frac{mns}{2}}
  {b + \tfrac1{2}\sum_{ijt}E_{ijt}(R_{ijt}-\alpha_i-\beta_j)^2} .\\
\intertext{The posterior conditional distribution for $\tau_{\alpha}$ is}
  \pi(\tau_{\alpha}\vert\mu, \tau_{\alpha},\mathbf{E,R})
  &\propto p(\tau_{\alpha})
  \prod_{i}p(\alpha_i\vert\mu, \tau_{\alpha})
  \nonumber \\
  \tau_{\alpha}\vert\mu, \tau_{\alpha},\mathbf{E,R}
  &\sim \dgamma{a+\frac{m}{2}}
  {b + \tfrac1{2}\sum_i(\alpha_i-\tfrac1{2}\mu)^2 } .\\
\intertext{The posterior conditional distribution for $\tau_{\beta}$ is}
  \pi(\tau_{\beta} \vert\bfbeta, \mu,\mathbf{E,R})
  &\propto p(\tau_{\beta})\prod_j p(\beta_j,\vert\mu,\tau_{\beta})
  \nonumber \\
  \tau_{\beta} \vert\bfbeta, \mu,\mathbf{E,R}
  &\sim \dgamma{a+\frac{n}{2}}
  {b + \tfrac{1}{2}\sum_j(\beta_j-\tfrac1{2}\mu)^2} .
\end{align*}

\subsection{Results}
\begin{figure}
  \subfigure[$\alpha_{1}$.]{
    \label{fig:acf-klalpha:01}
    \begin{minipage}[b]{0.33\textwidth}
      \centering \includegraphics[width=\textwidth]{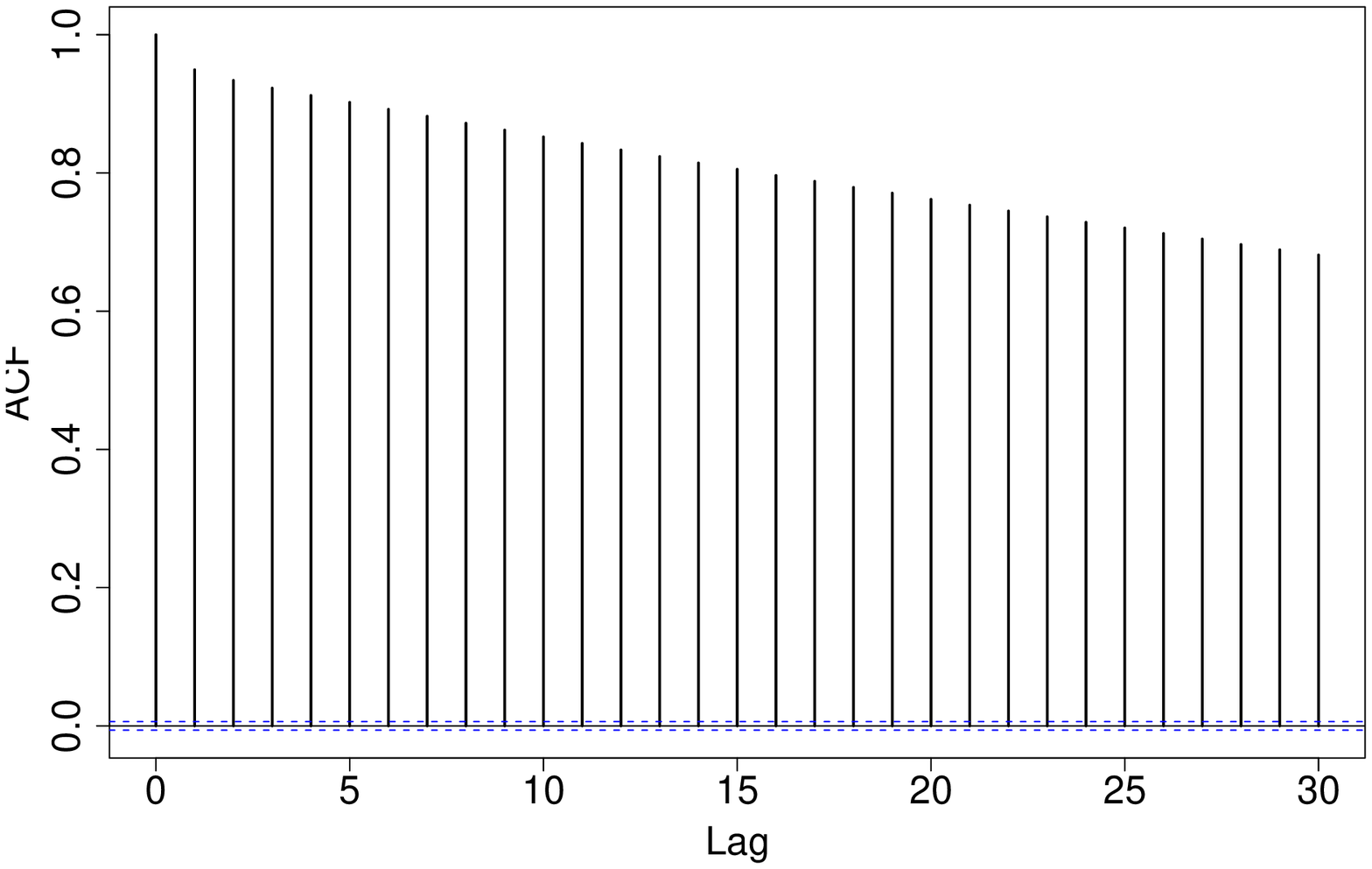}
    \end{minipage}}%
  \subfigure[$\alpha_{2}$.]{
    \label{fig:acf-klalpha:02}
    \begin{minipage}[b]{0.33\textwidth}
      \centering \includegraphics[width=\textwidth]{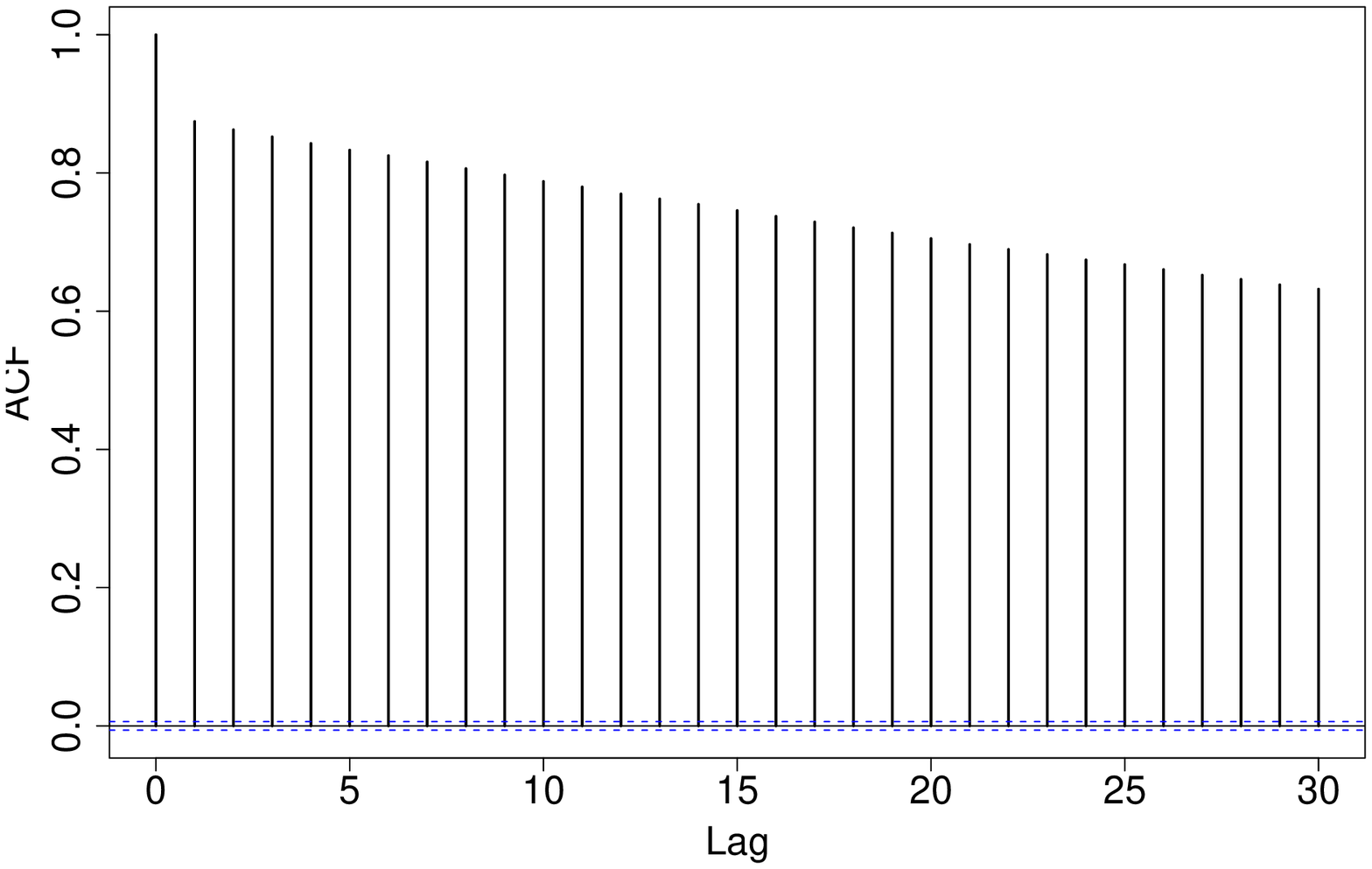}
    \end{minipage}}%
  \subfigure[$\alpha_{3}$.]{
    \label{fig:acf-klalpha:03}
    \begin{minipage}[b]{0.33\textwidth}
      \centering \includegraphics[width=\textwidth]{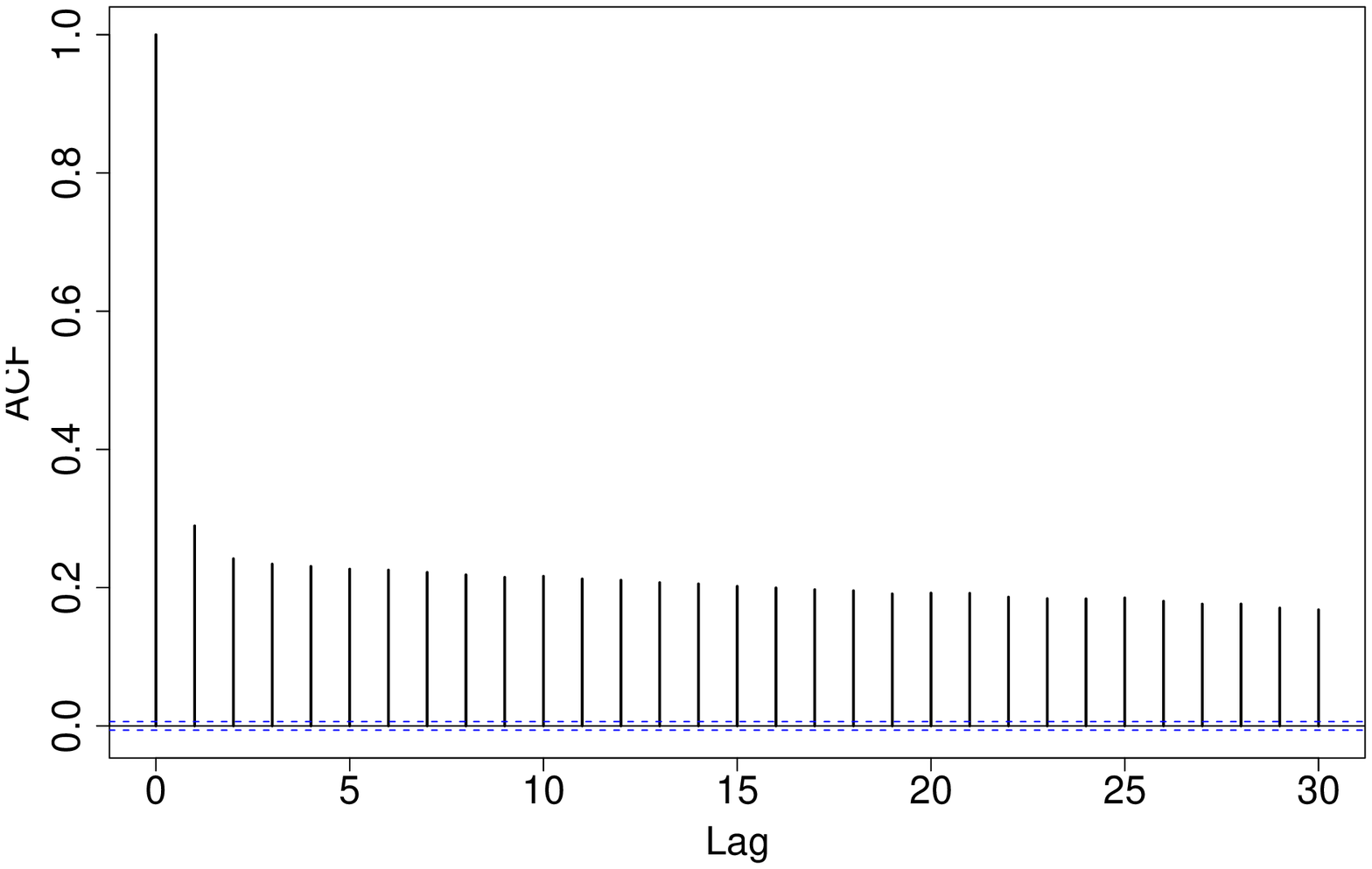}
    \end{minipage}}\\
  \subfigure[$\alpha_{4}$.]{
    \label{fig:acf-klalpha:04}
    \begin{minipage}[b]{0.33\textwidth}
      \centering \includegraphics[width=\textwidth]{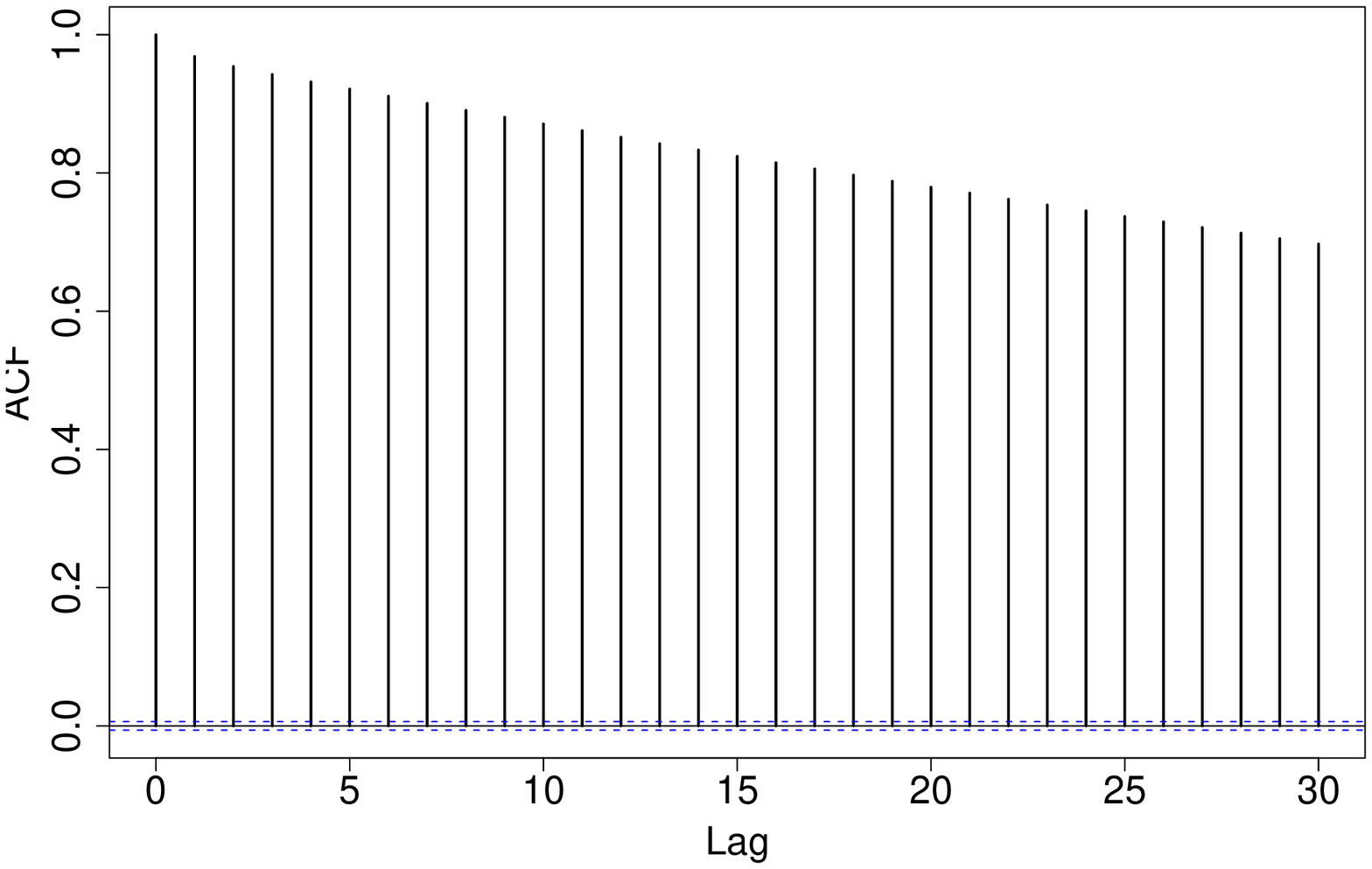}
    \end{minipage}}%
  \subfigure[$\alpha_{5}$.]{
    \label{fig:acf-klalpha:05}
    \begin{minipage}[b]{0.33\textwidth}
      \centering \includegraphics[width=\textwidth]{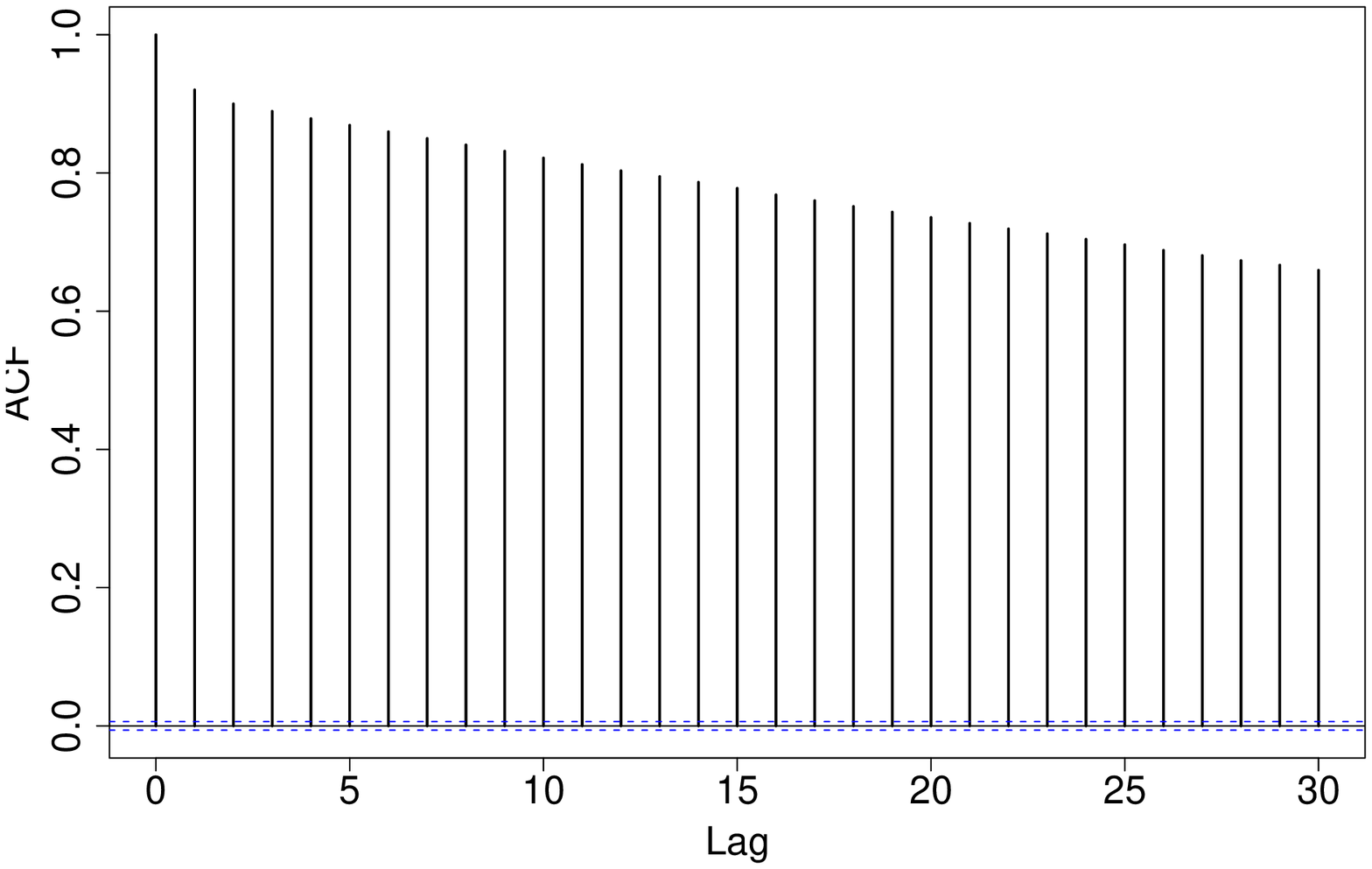}
    \end{minipage}}%
  \subfigure[$\alpha_{6}$.]{
    \label{fig:acf-klalpha:06}
    \begin{minipage}[b]{0.33\textwidth}
      \centering \includegraphics[width=\textwidth]{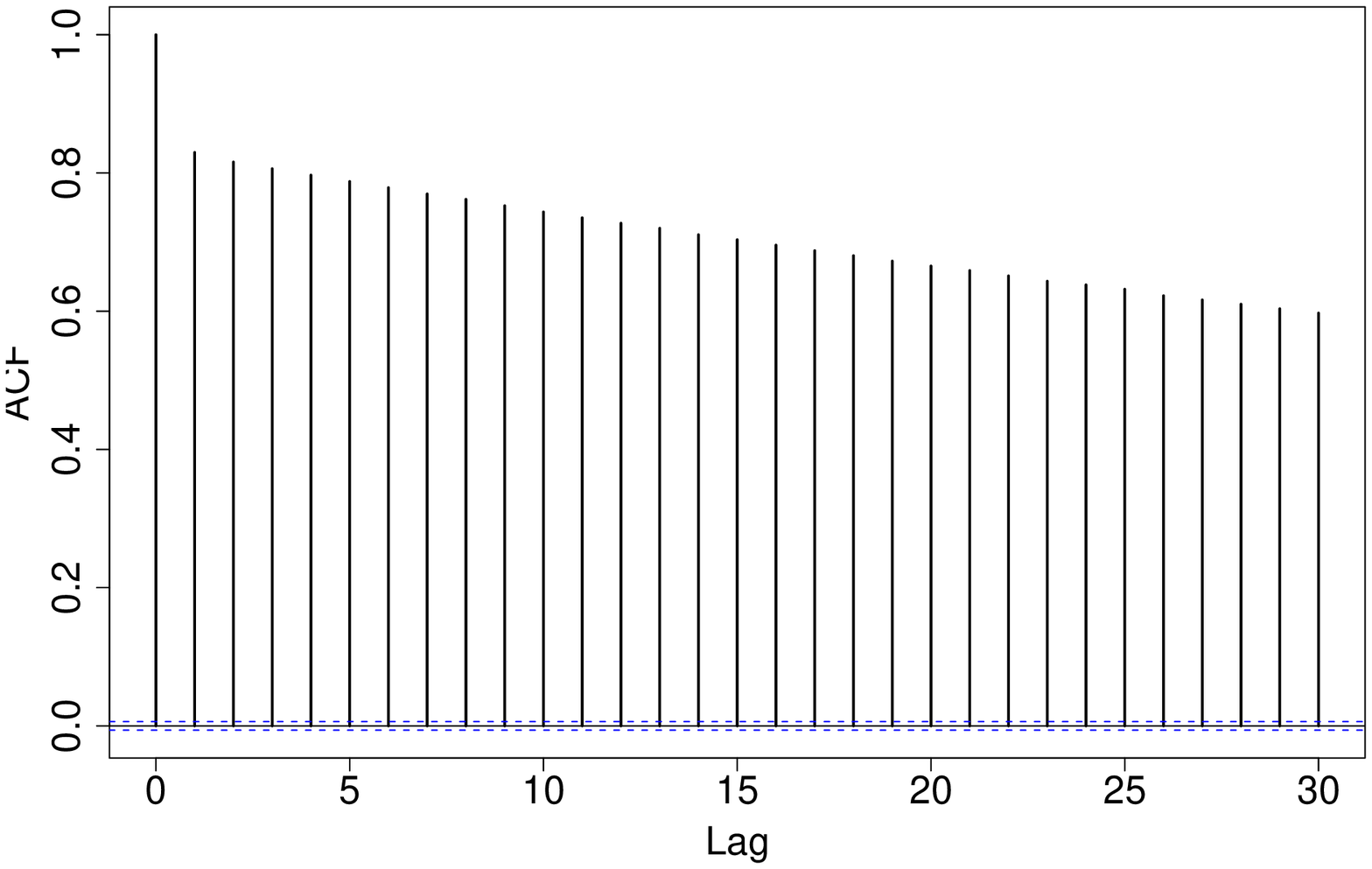}
    \end{minipage}}\\
  \subfigure[$\alpha_{7}$.]{
    \label{fig:acf-klalpha:07}
    \begin{minipage}[b]{0.33\textwidth}
      \centering \includegraphics[width=\textwidth]{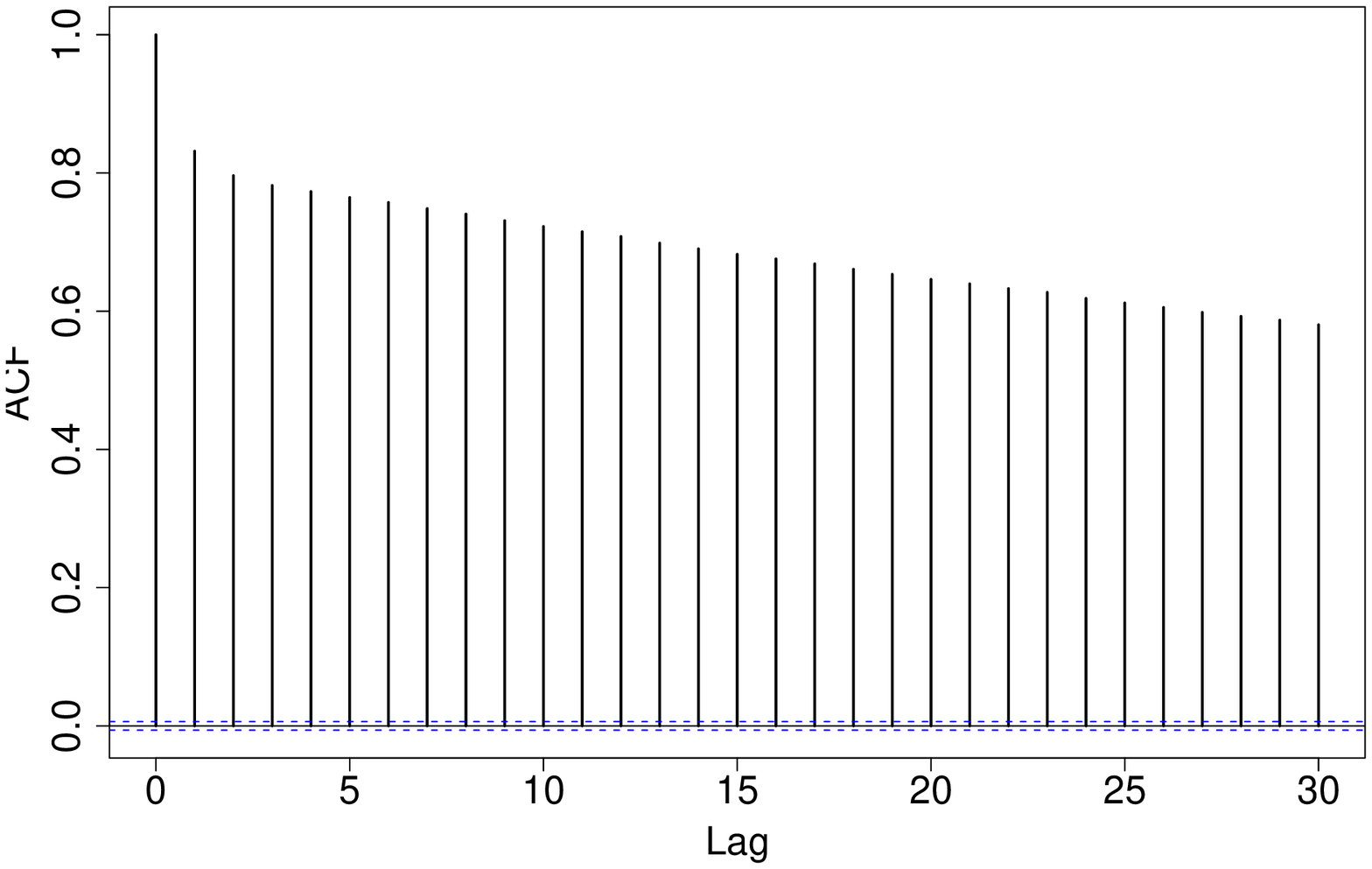}
    \end{minipage}}%
  \subfigure[$\alpha_{8}$.]{
    \label{fig:acf-klalpha:08}
    \begin{minipage}[b]{0.33\textwidth}
      \centering \includegraphics[width=\textwidth]{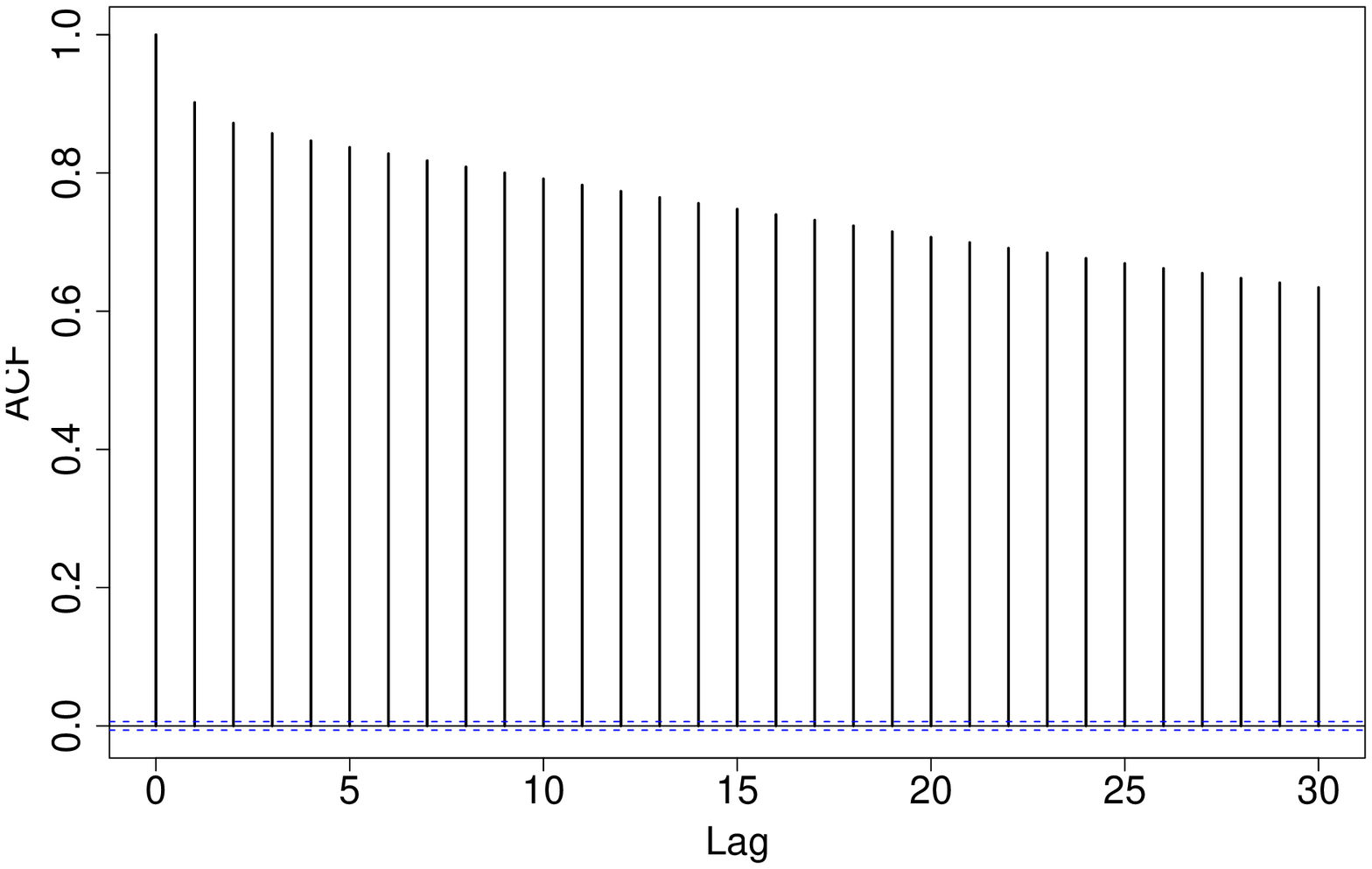}
    \end{minipage}}%
  \subfigure[$\alpha_{9}$.]{
    \label{fig:acf-klalpha:09}
    \begin{minipage}[b]{0.33\textwidth}
      \centering \includegraphics[width=\textwidth]{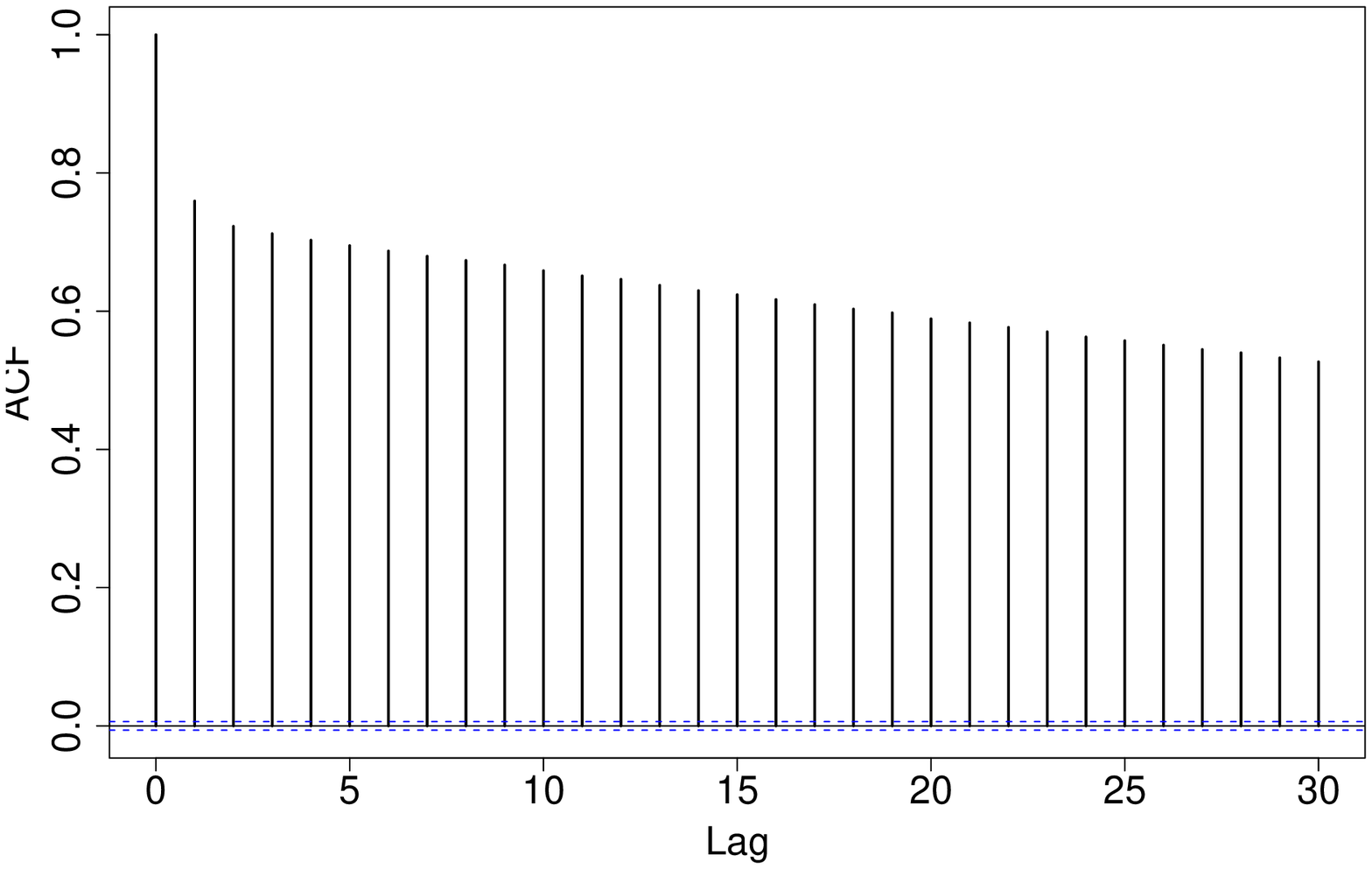}
    \end{minipage}}\\
  \subfigure[$\alpha_{10}$.]{
    \label{fig:acf-klalpha:10}
    \begin{minipage}[b]{0.33\textwidth}
      \centering \includegraphics[width=\textwidth]{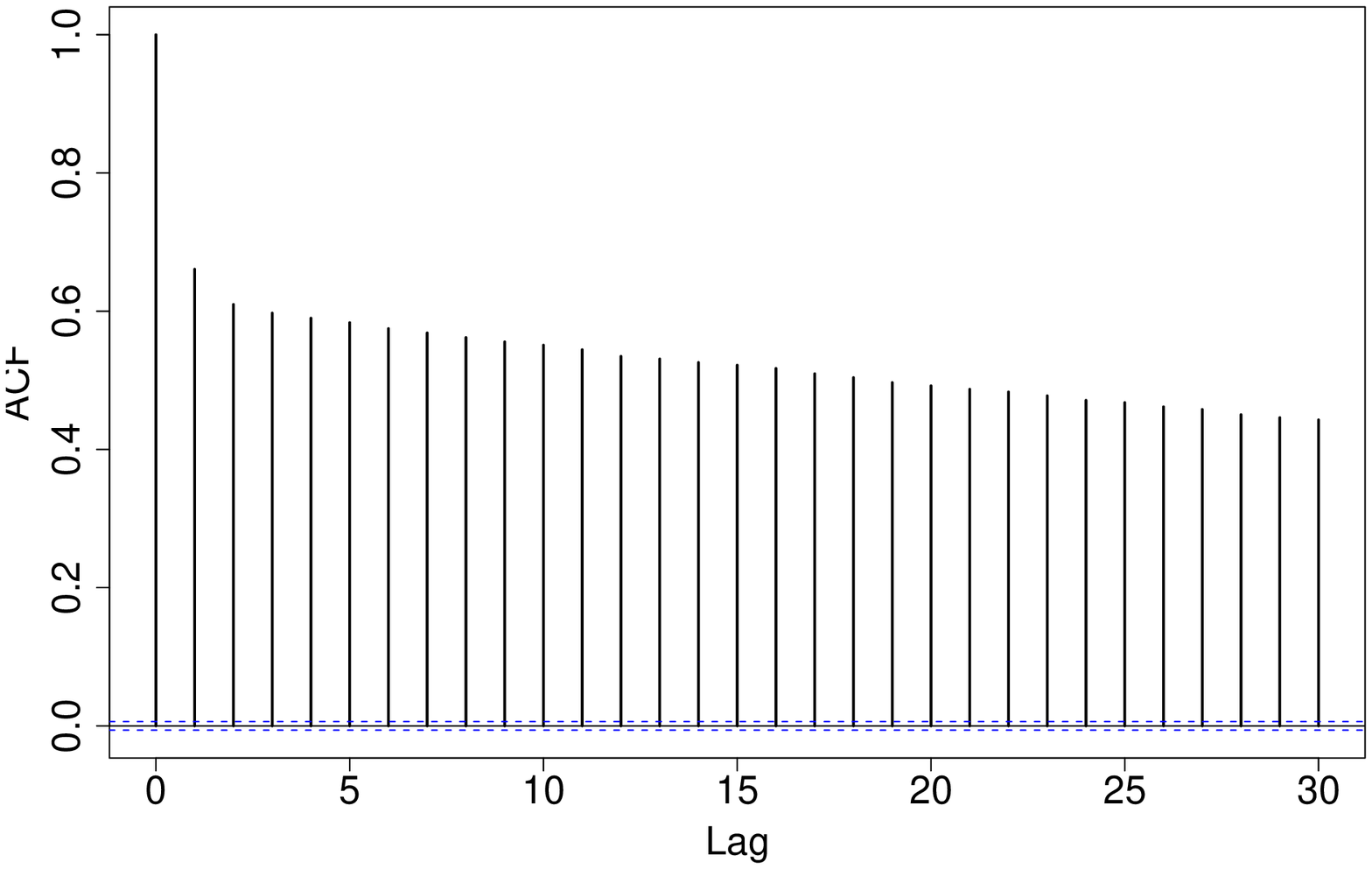}
    \end{minipage}}%
  \caption{Autocorrelation plots for $\alpha$.}
  \label{fig:acf-klalpha} 
\end{figure}

\begin{figure}
  \subfigure[$\beta_{1}$.]{
    \label{fig:acf-klbeta:01}
    \begin{minipage}[b]{0.23\textwidth}
      \centering \includegraphics[width=\textwidth]{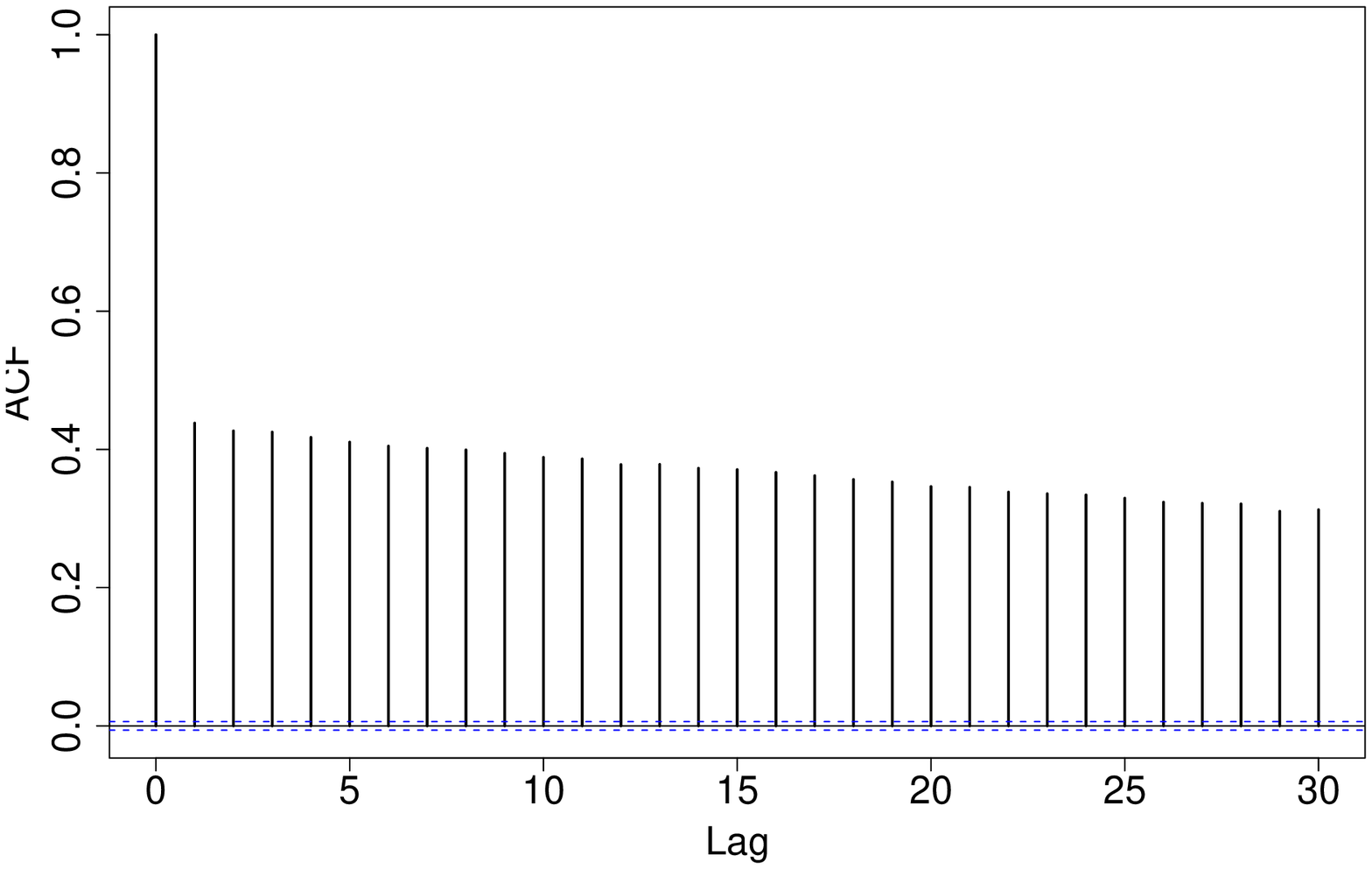}
    \end{minipage}}%
  \subfigure[$\beta_{2}$.]{
    \label{fig:acf-klbeta:02}
    \begin{minipage}[b]{0.23\textwidth}
      \centering \includegraphics[width=\textwidth]{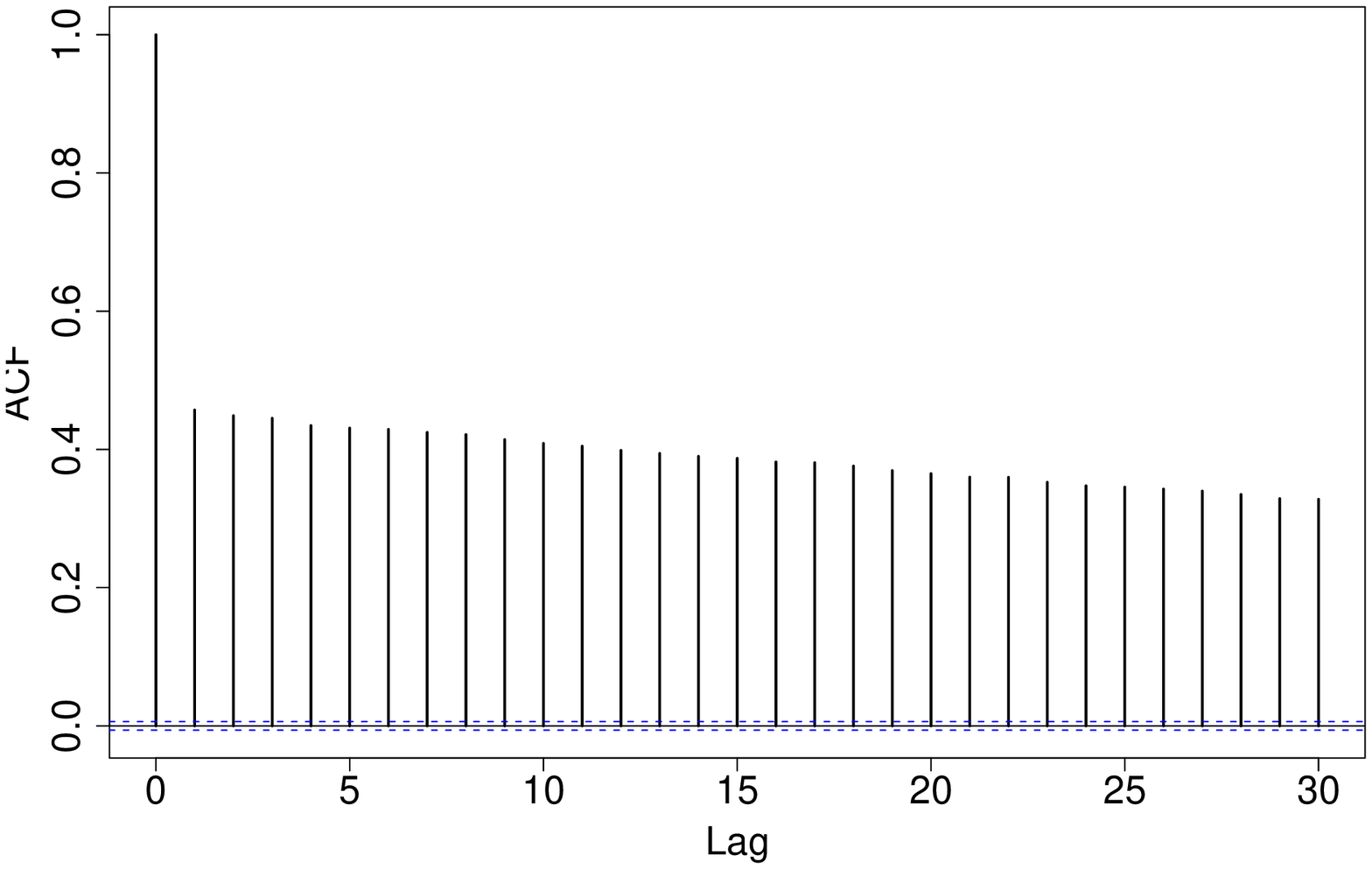}
    \end{minipage}}%
  \subfigure[$\beta_{3}$.]{
    \label{fig:acf-klbeta:03}
    \begin{minipage}[b]{0.23\textwidth}
      \centering \includegraphics[width=\textwidth]{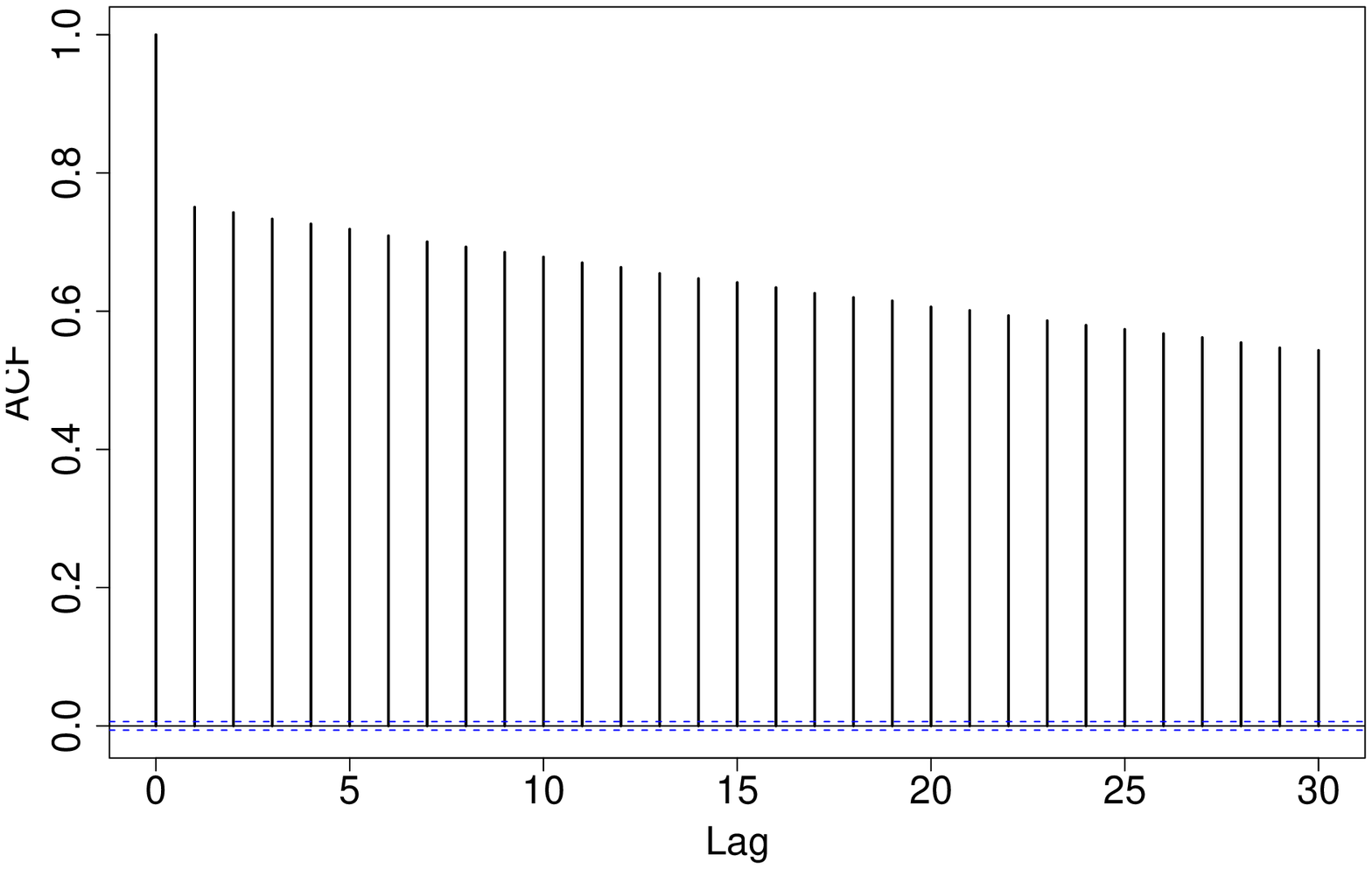}
    \end{minipage}}%
  \subfigure[$\beta_{4}$.]{
    \label{fig:acf-klbeta:04}
    \begin{minipage}[b]{0.23\textwidth}
      \centering \includegraphics[width=\textwidth]{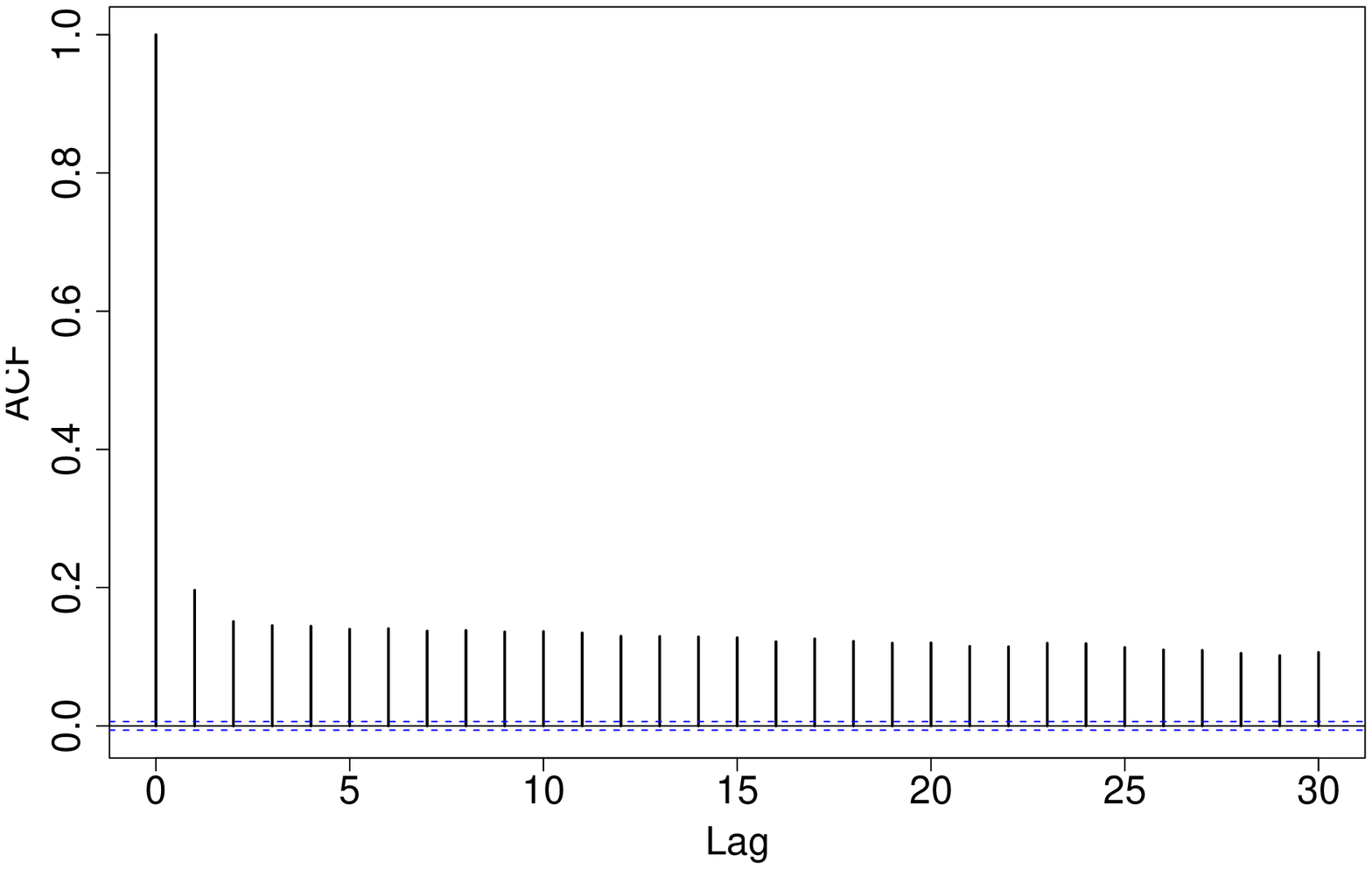}
    \end{minipage}}\\
  \subfigure[$\beta_{5}$.]{
    \label{fig:acf-klbeta:05}
    \begin{minipage}[b]{0.23\textwidth}
      \centering \includegraphics[width=\textwidth]{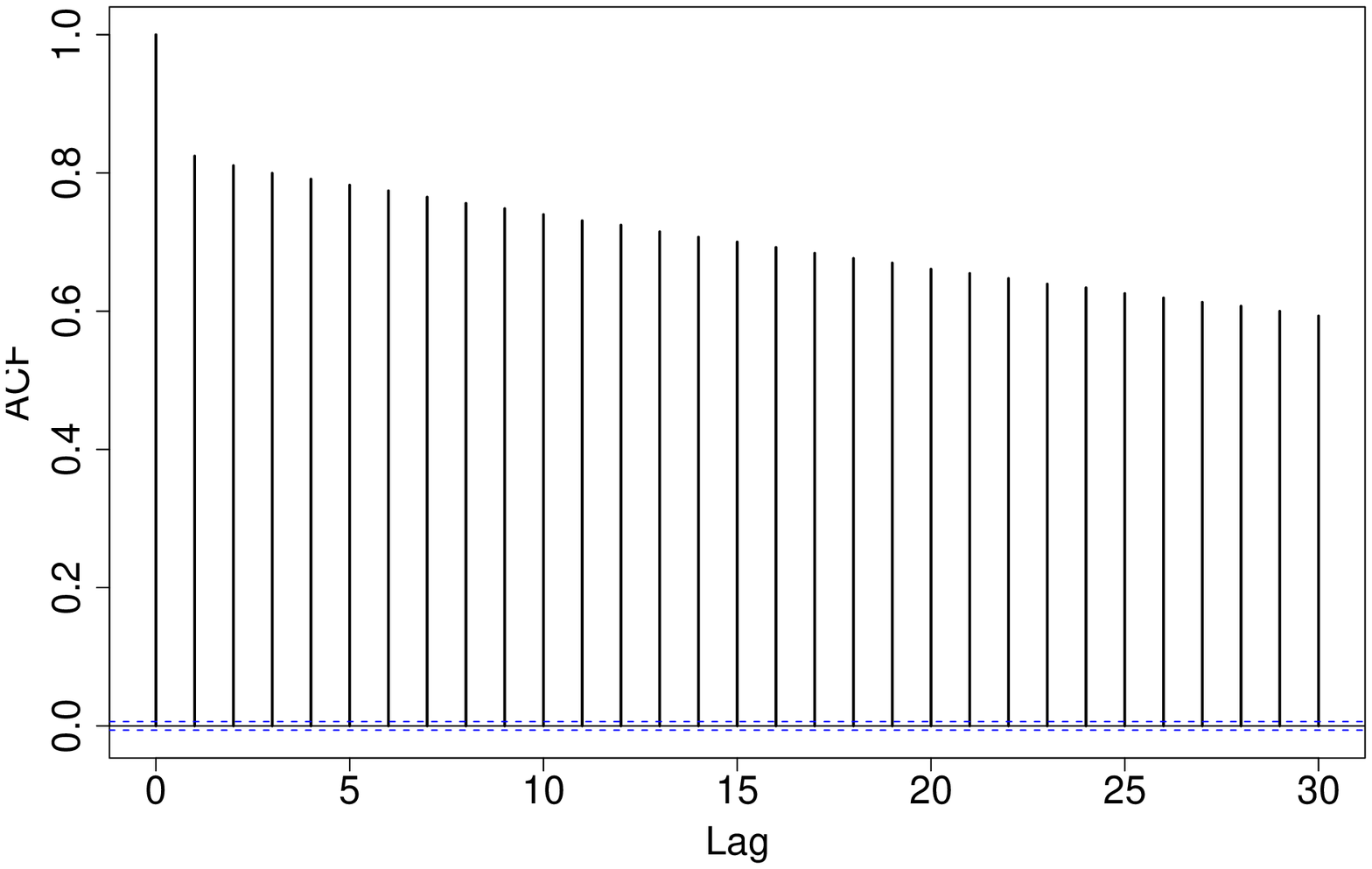}
    \end{minipage}}%
  \subfigure[$\beta_{6}$.]{
    \label{fig:acf-klbeta:06}
    \begin{minipage}[b]{0.23\textwidth}
      \centering \includegraphics[width=\textwidth]{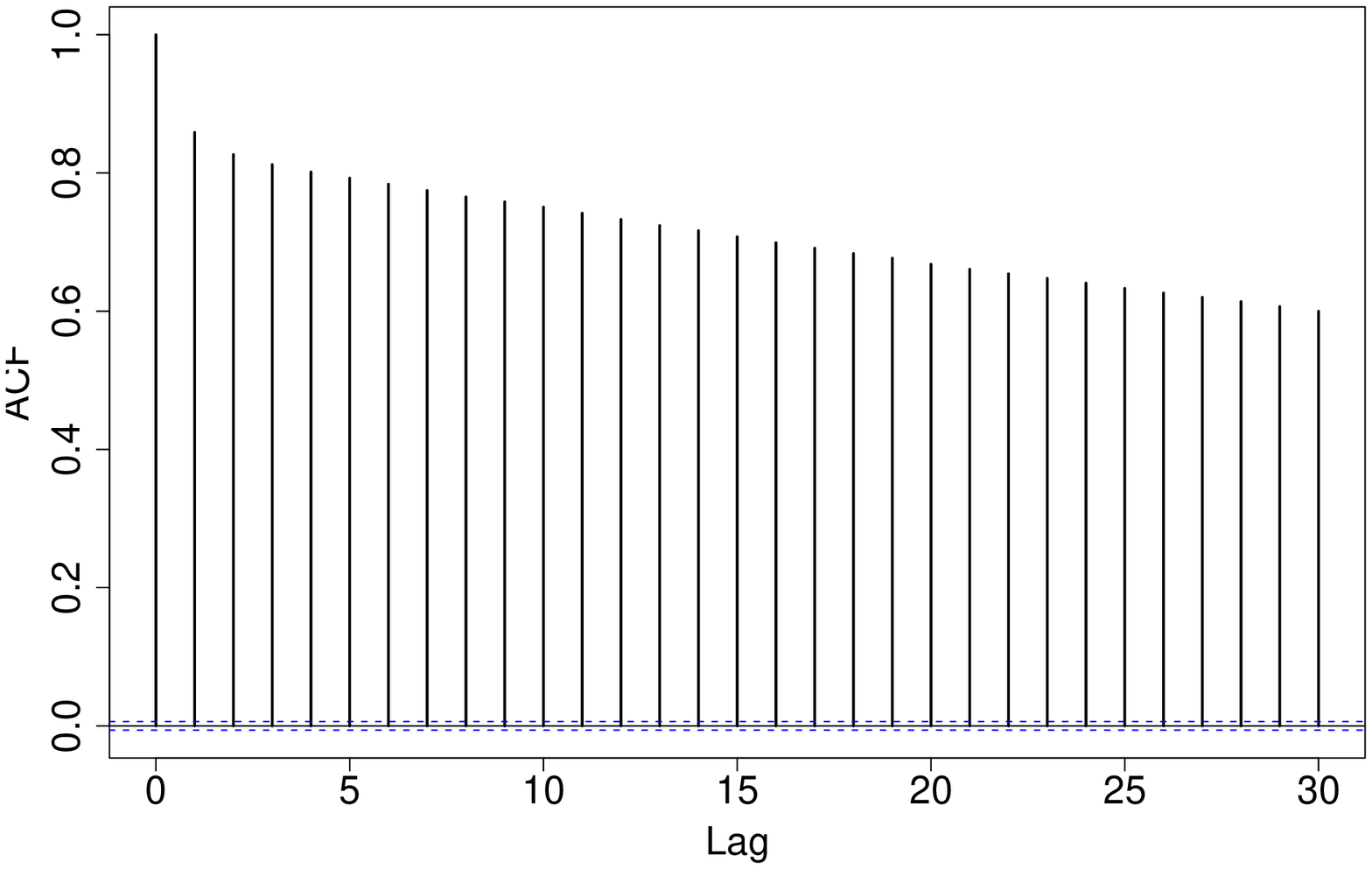}
    \end{minipage}}%
  \subfigure[$\beta_{7}$.]{
    \label{fig:acf-klbeta:07}
    \begin{minipage}[b]{0.23\textwidth}
      \centering \includegraphics[width=\textwidth]{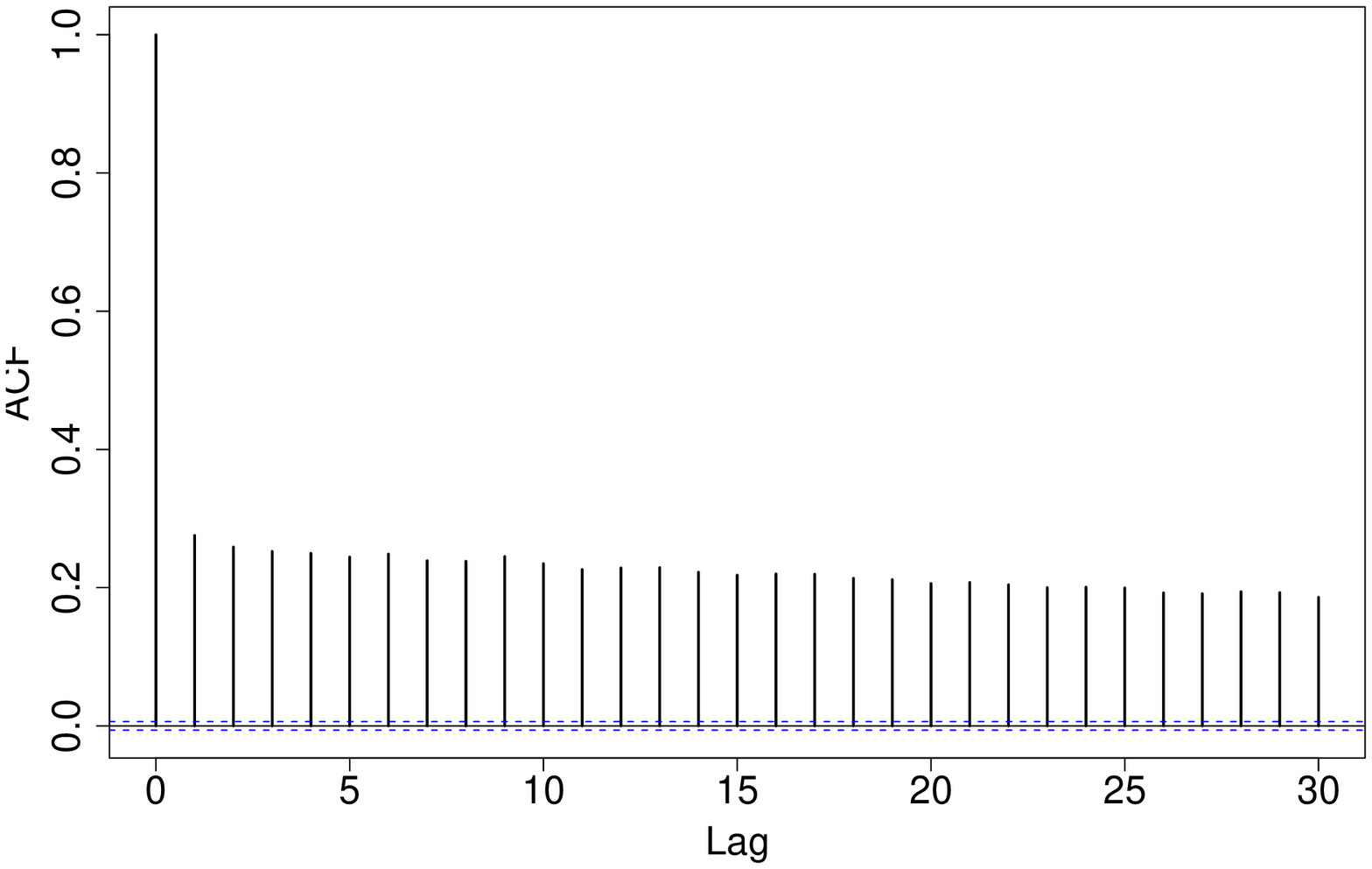}
    \end{minipage}}%
  \subfigure[$\beta_{8}$.]{
    \label{fig:acf-klbeta:08}
    \begin{minipage}[b]{0.23\textwidth}
      \centering \includegraphics[width=\textwidth]{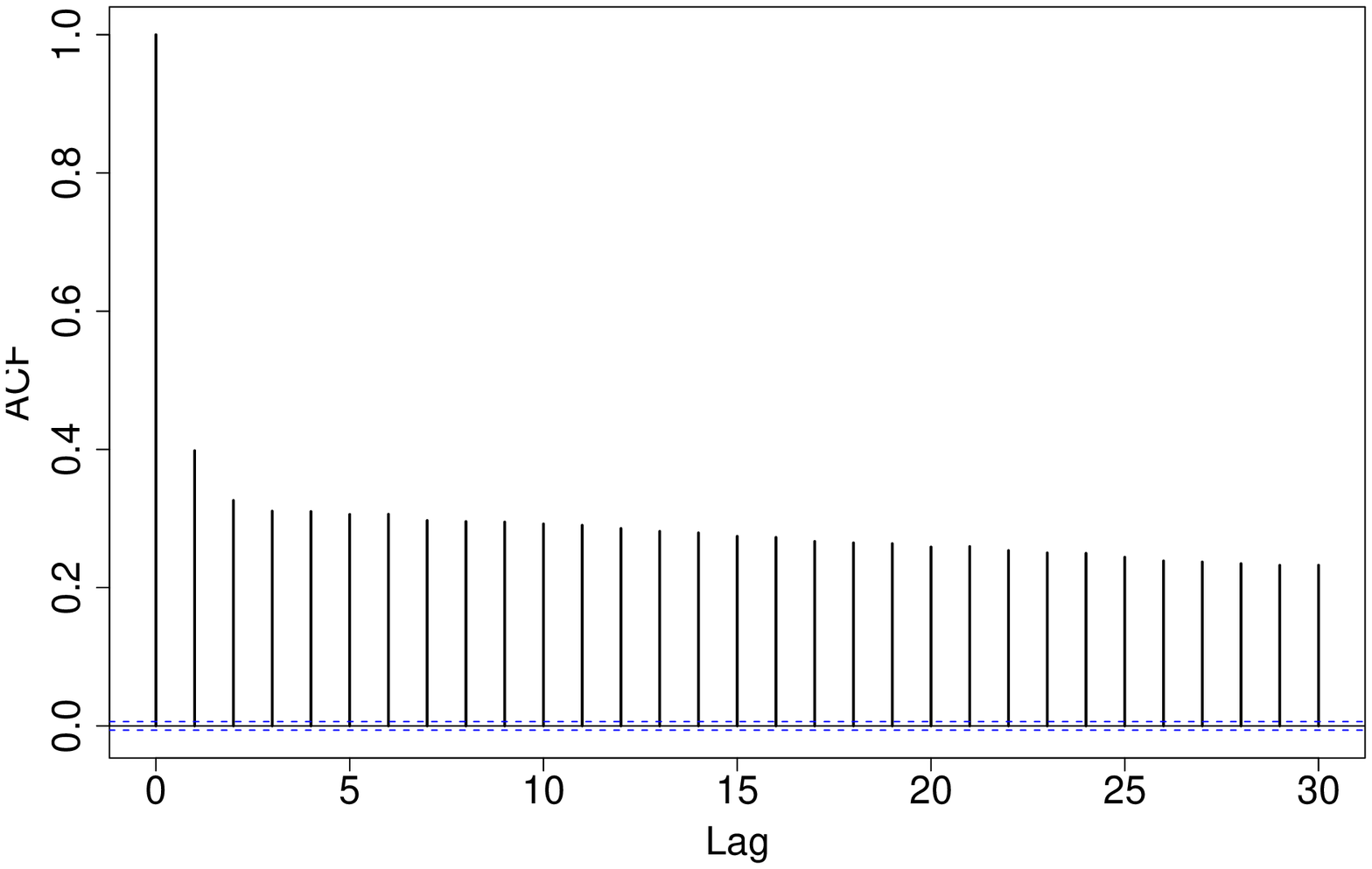}
    \end{minipage}}\\
  \subfigure[$\beta_{9}$.]{
    \label{fig:acf-klbeta:09}
    \begin{minipage}[b]{0.23\textwidth}
      \centering \includegraphics[width=\textwidth]{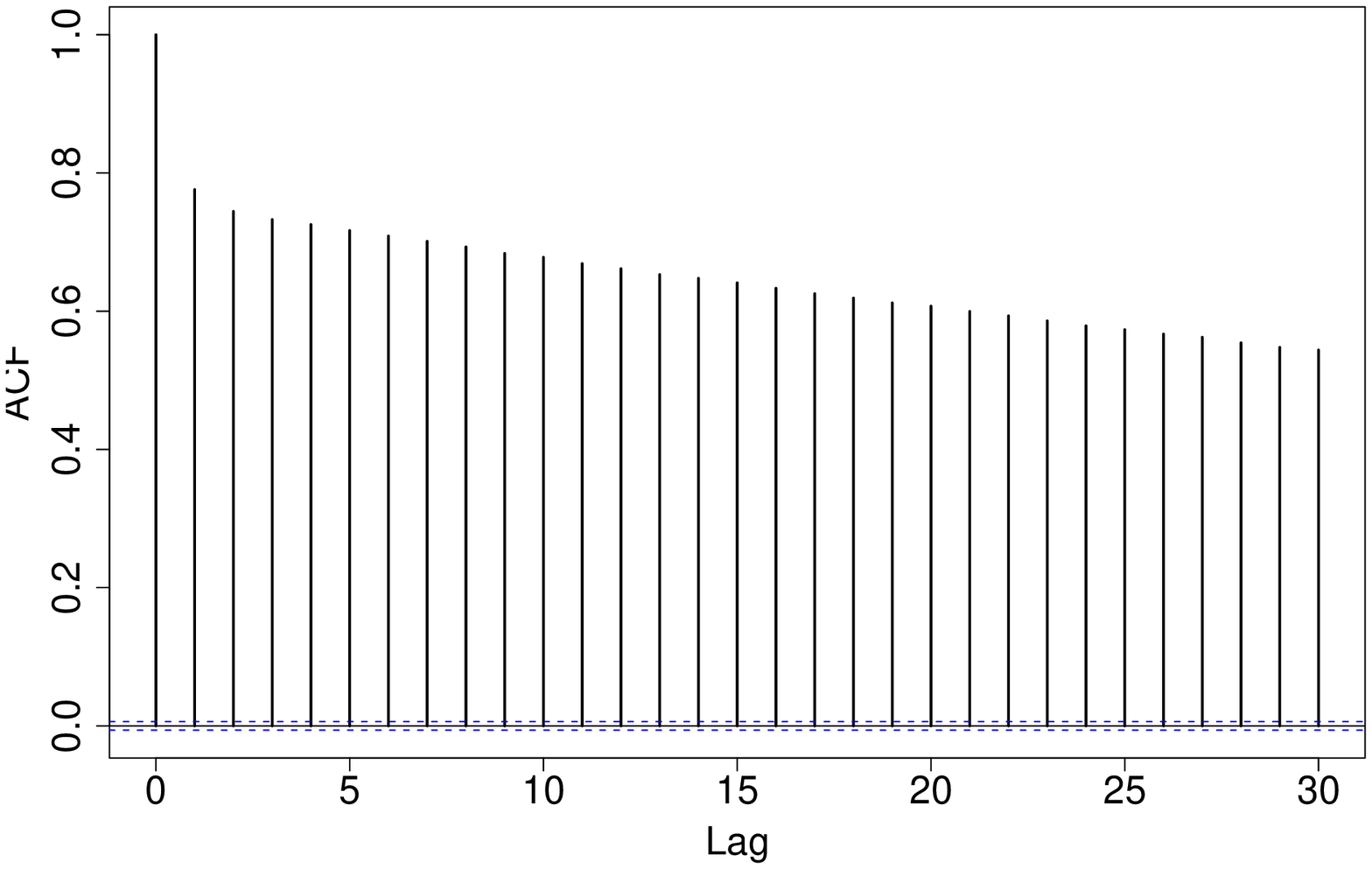}
    \end{minipage}}%
  \subfigure[$\beta_{10}$.]{
    \label{fig:acf-klbeta:10}
    \begin{minipage}[b]{0.23\textwidth}
      \centering \includegraphics[width=\textwidth]{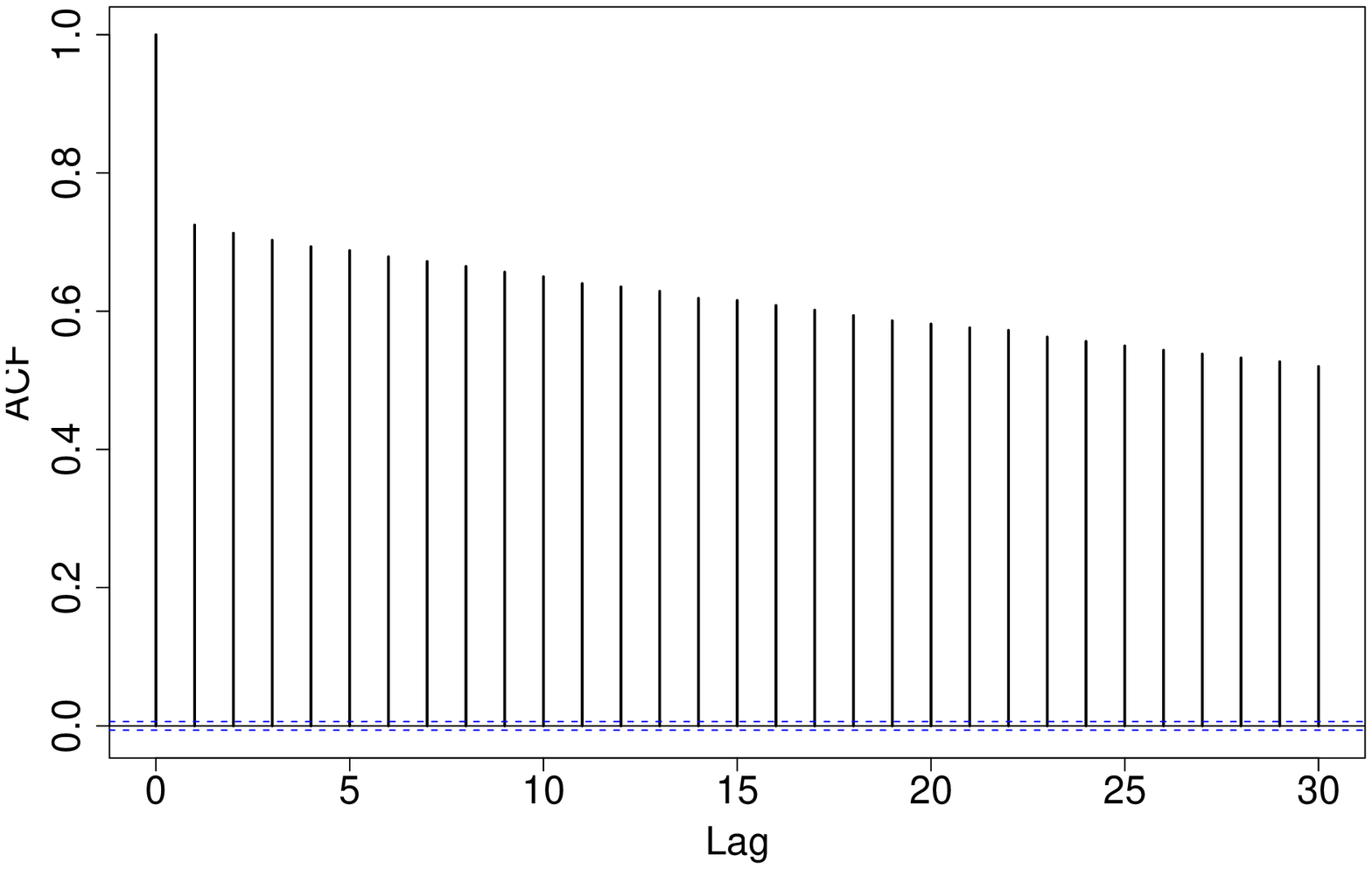}
    \end{minipage}}%
  \subfigure[$\beta_{11}$.]{
    \label{fig:acf-klbeta:11}
    \begin{minipage}[b]{0.23\textwidth}
      \centering \includegraphics[width=\textwidth]{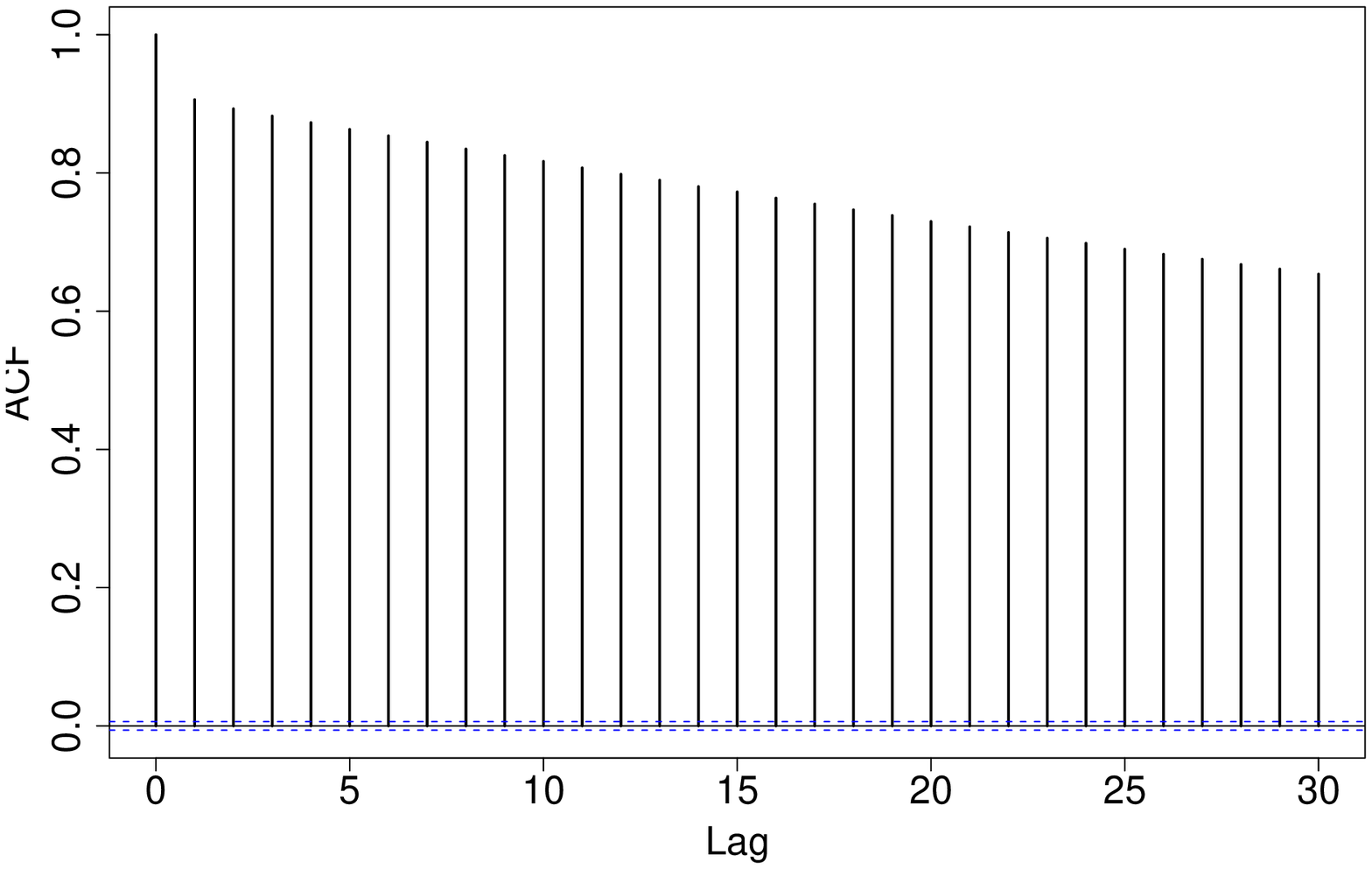}
    \end{minipage}}%
  \subfigure[$\beta_{12}$.]{
    \label{fig:acf-klbeta:12}
    \begin{minipage}[b]{0.23\textwidth}
      \centering \includegraphics[width=\textwidth]{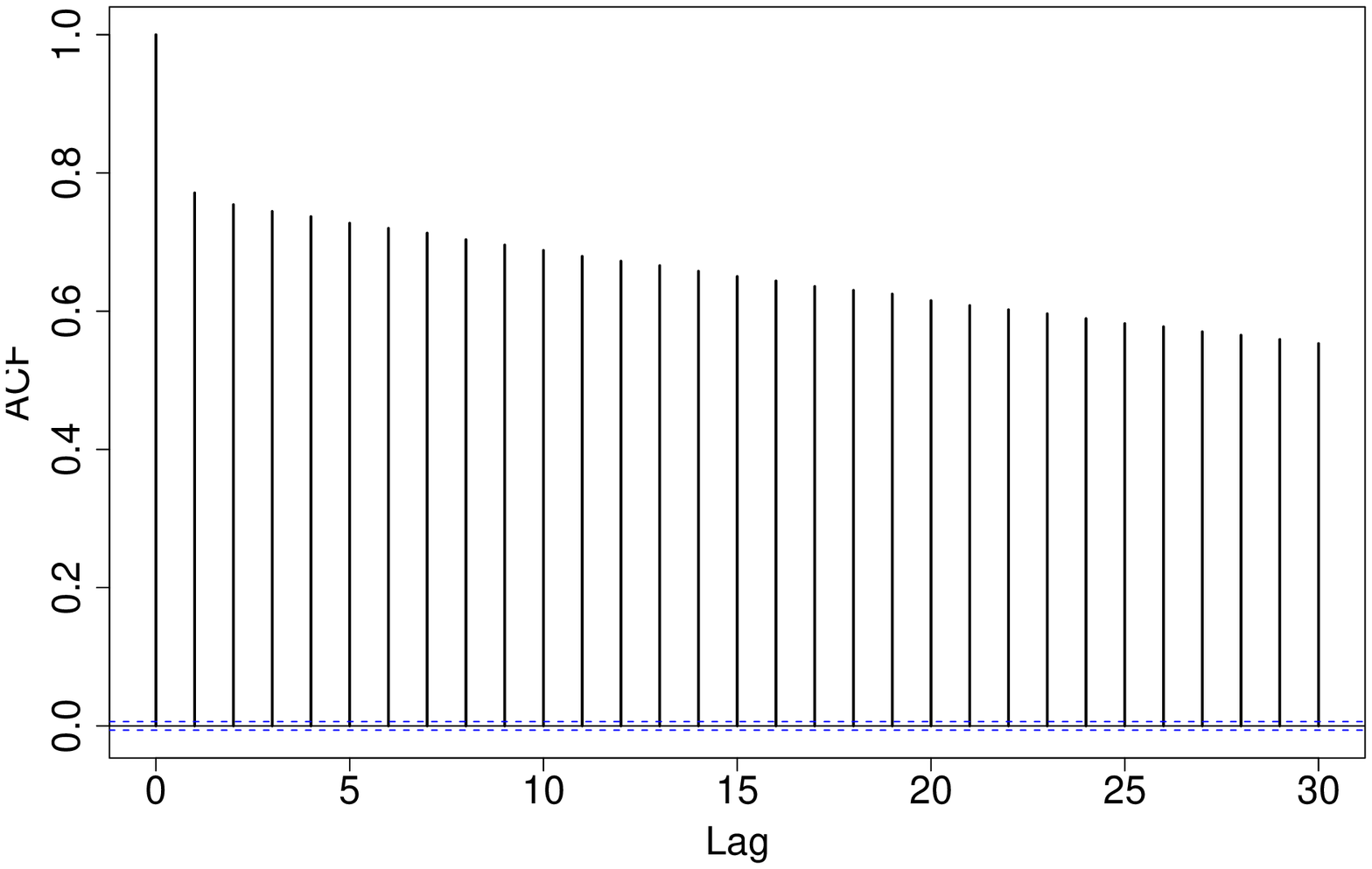}
    \end{minipage}}\\
  \subfigure[$\beta_{13}$.]{
    \label{fig:acf-klbeta:13}
    \begin{minipage}[b]{0.23\textwidth}
      \centering \includegraphics[width=\textwidth]{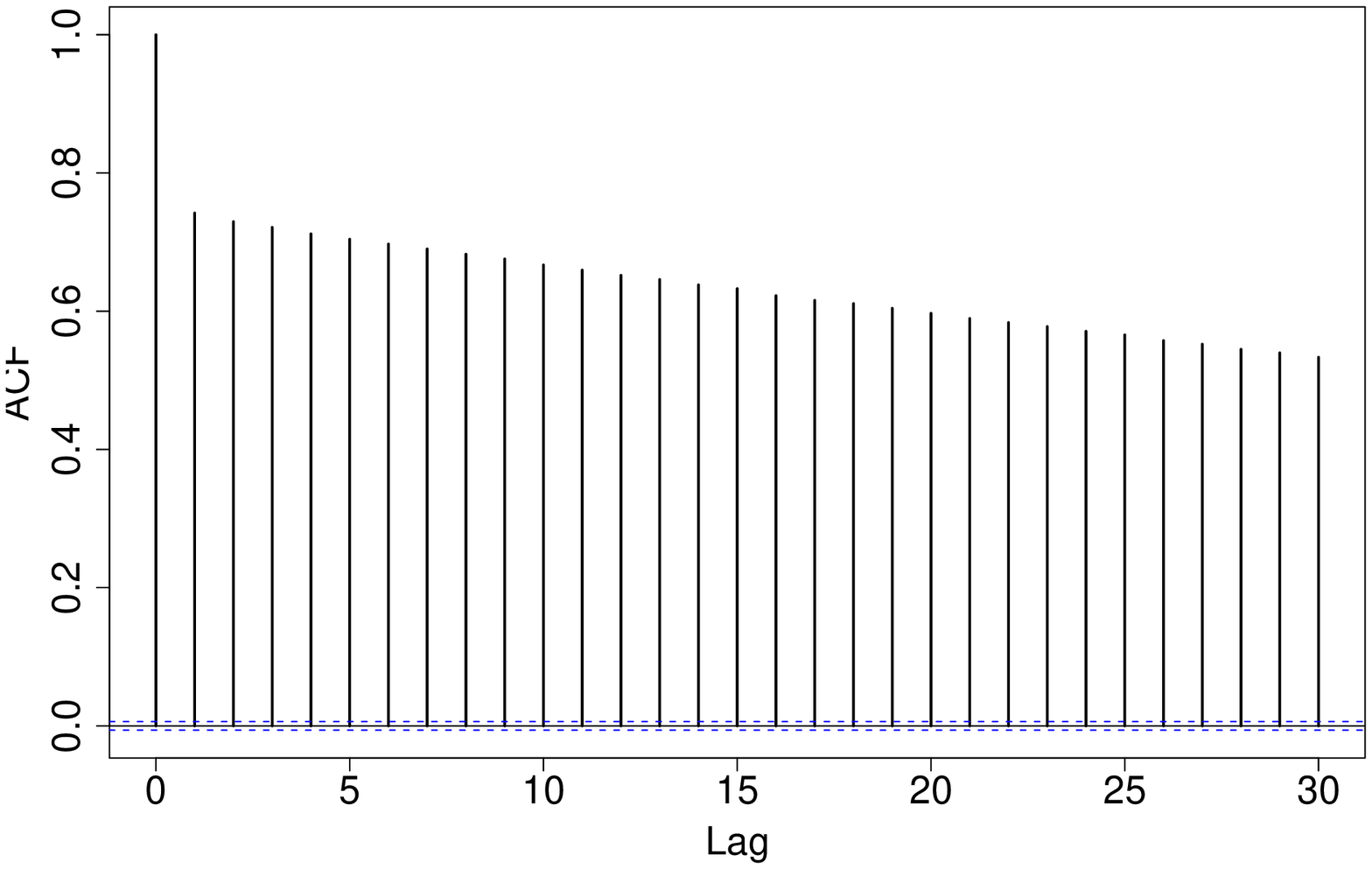}
    \end{minipage}}%
  \subfigure[$\beta_{14}$.]{
    \label{fig:acf-klbeta:14}
    \begin{minipage}[b]{0.23\textwidth}
      \centering \includegraphics[width=\textwidth]{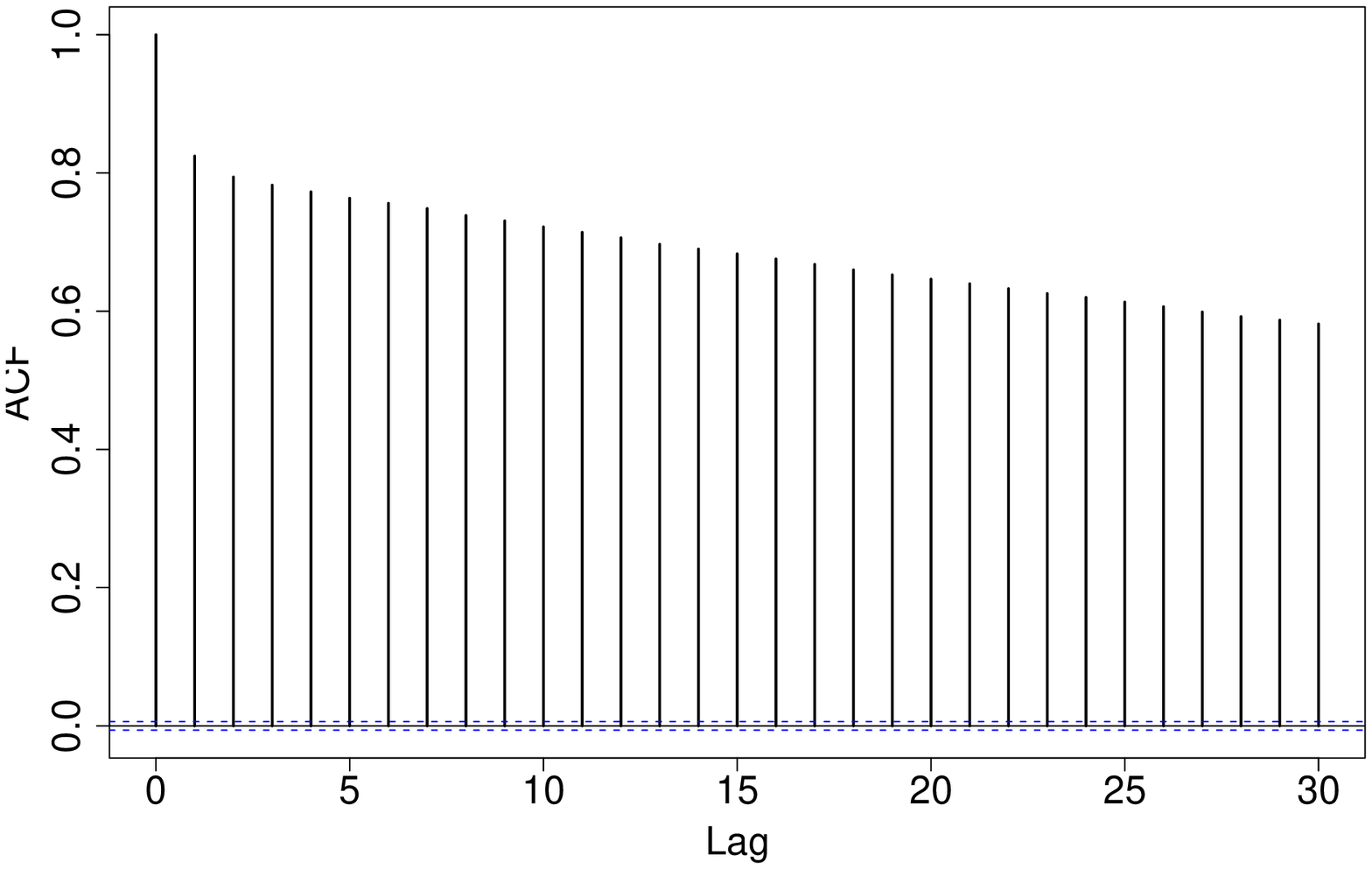}
    \end{minipage}}%
  \subfigure[$\beta_{15}$.]{
    \label{fig:acf-klbeta:15}
    \begin{minipage}[b]{0.23\textwidth}
      \centering \includegraphics[width=\textwidth]{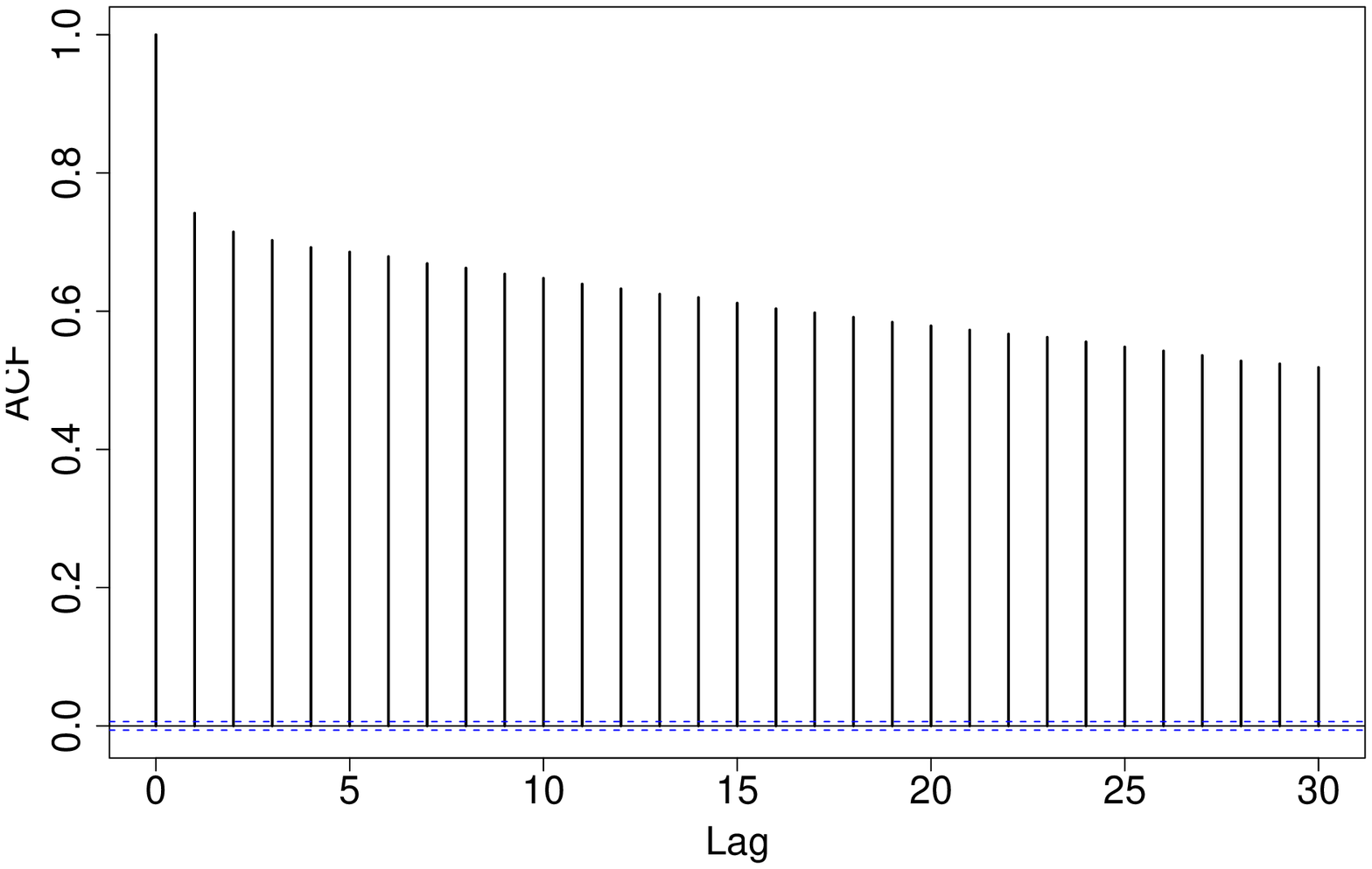}
    \end{minipage}}%
  \subfigure[$\beta_{16}$.]{
    \label{fig:acf-klbeta:16}
    \begin{minipage}[b]{0.23\textwidth}
      \centering \includegraphics[width=\textwidth]{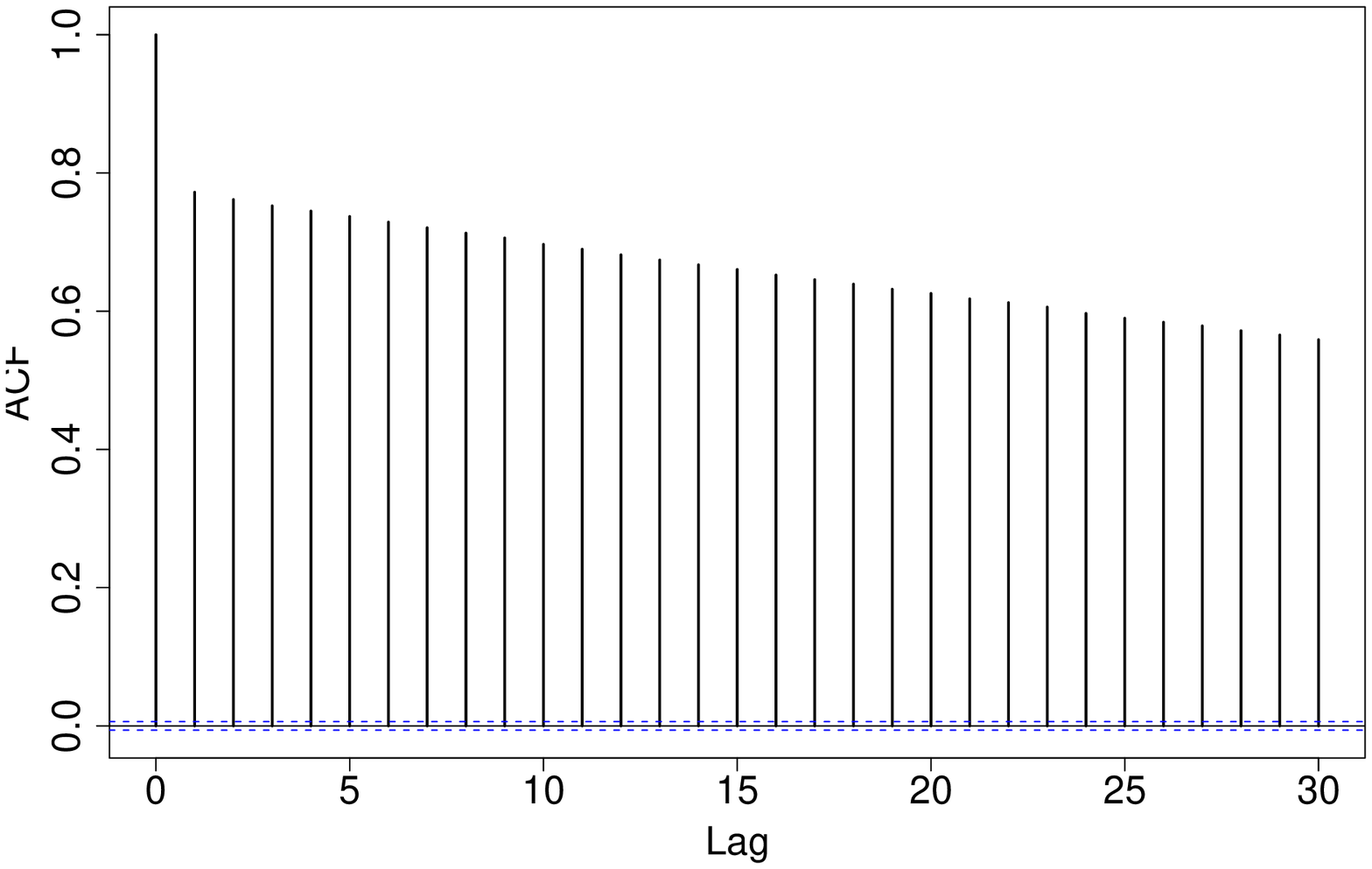}
    \end{minipage}}\\
  \subfigure[$\beta_{17}$.]{
    \label{fig:acf-klbeta:17}
    \begin{minipage}[b]{0.23\textwidth}
      \centering \includegraphics[width=\textwidth]{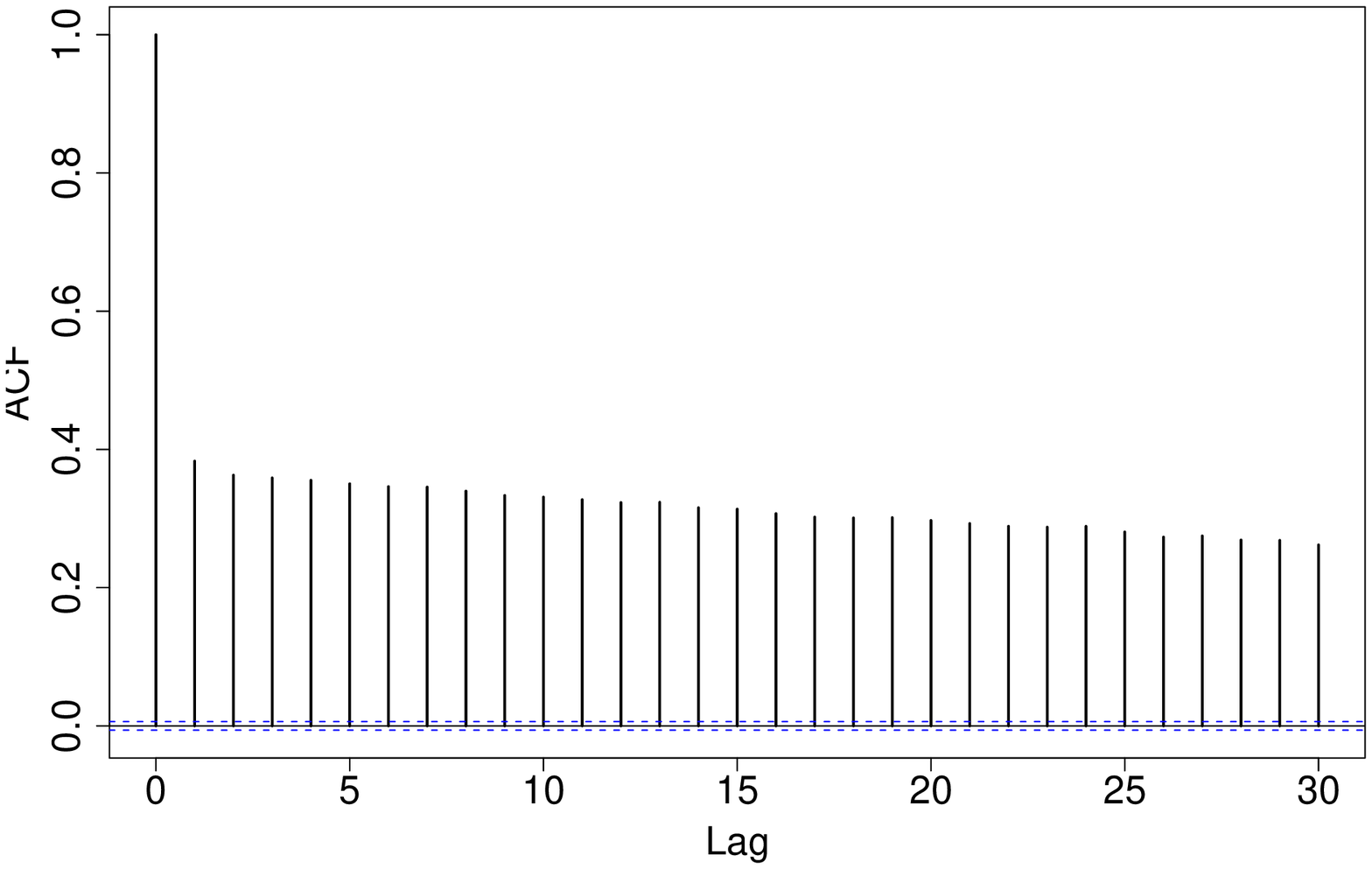}
    \end{minipage}}%
  \subfigure[$\beta_{18}$.]{
    \label{fig:acf-klbeta:18}
    \begin{minipage}[b]{0.23\textwidth}
      \centering \includegraphics[width=\textwidth]{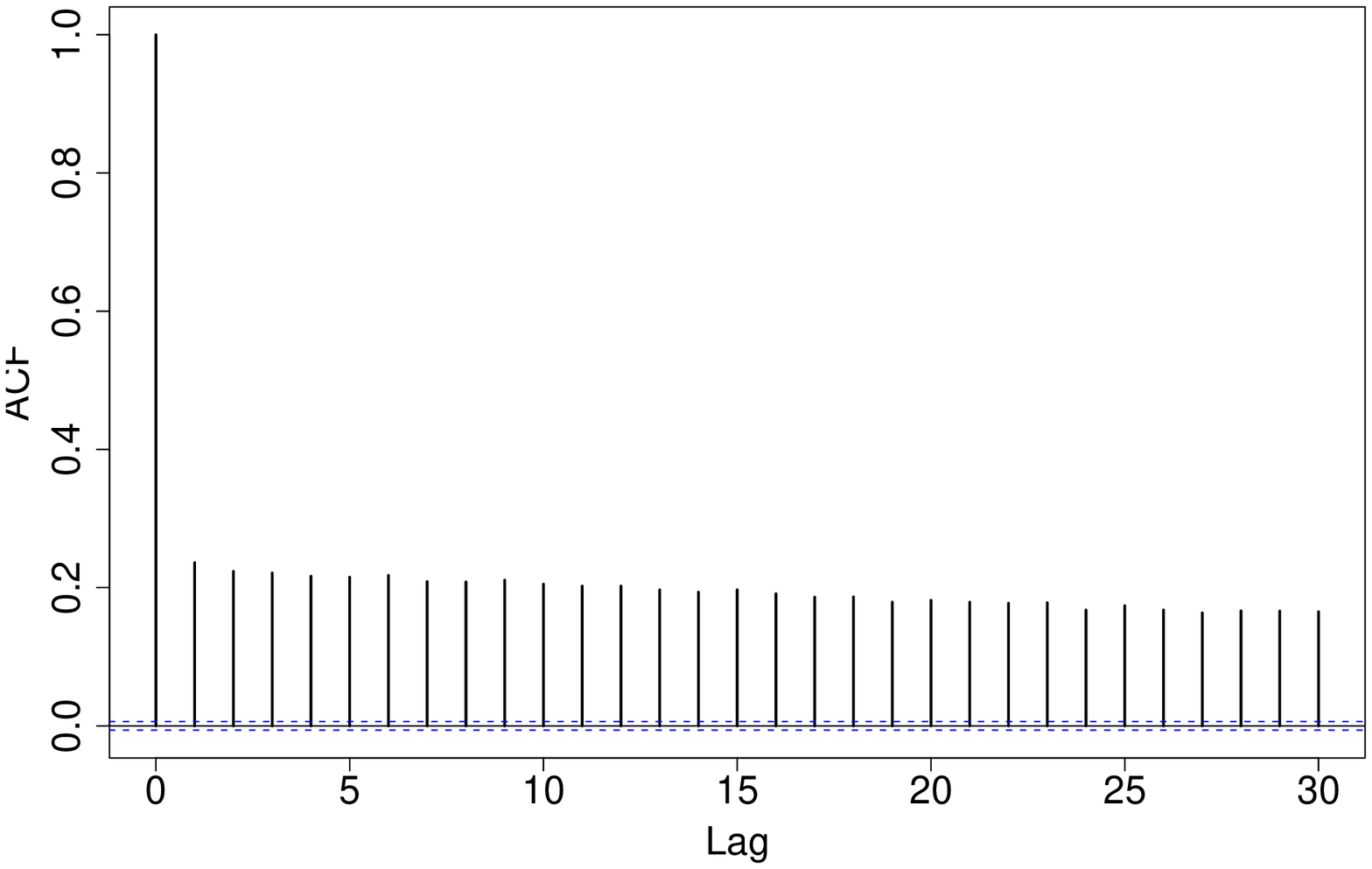}
    \end{minipage}}%
  \subfigure[$\beta_{19}$.]{
    \label{fig:acf-klbeta:19}
    \begin{minipage}[b]{0.23\textwidth}
      \centering \includegraphics[width=\textwidth]{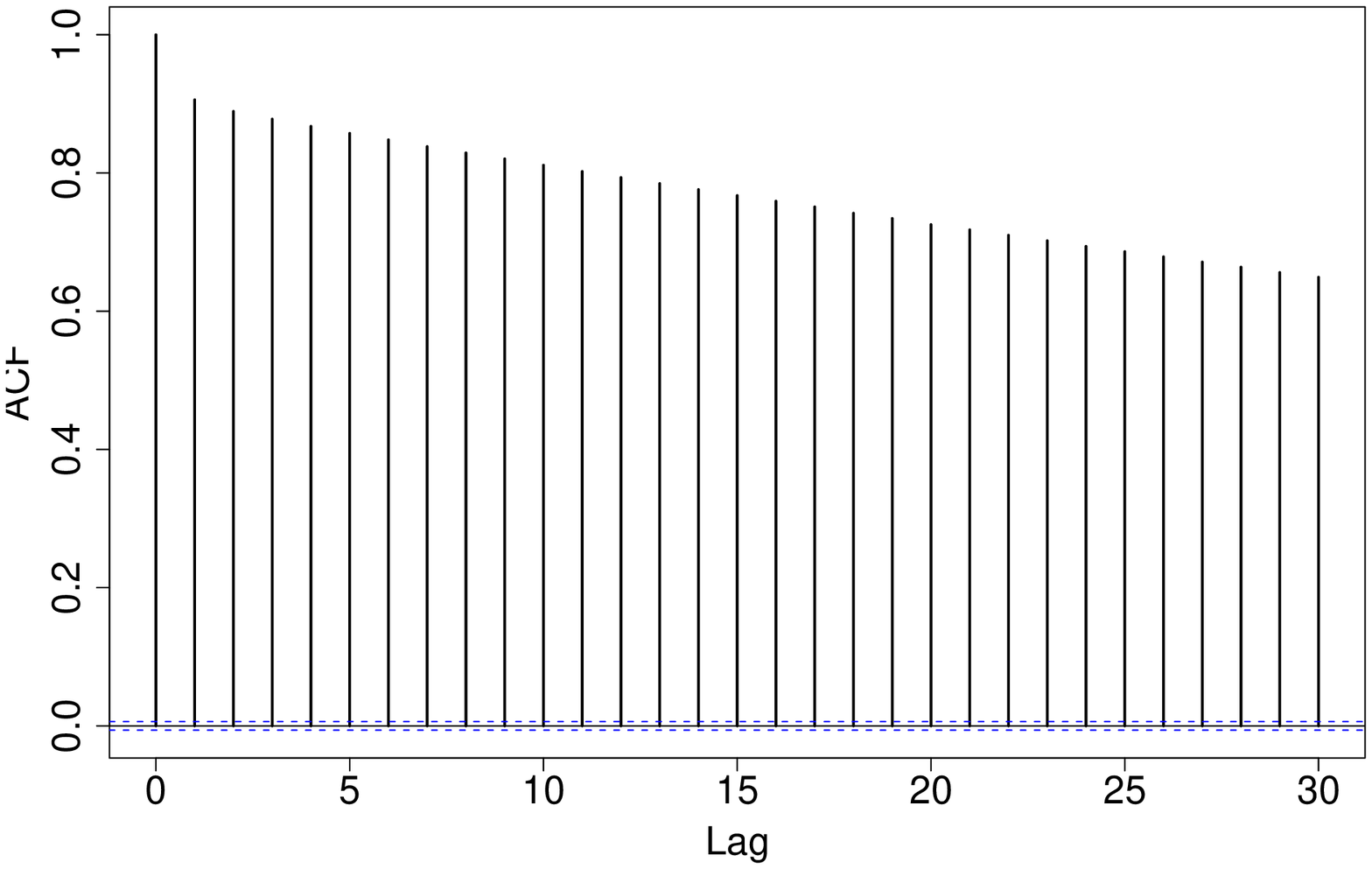}
    \end{minipage}}%
  \subfigure[$\beta_{20}$.]{
    \label{fig:acf-klbeta:20}
    \begin{minipage}[b]{0.23\textwidth}
      \centering \includegraphics[width=\textwidth]{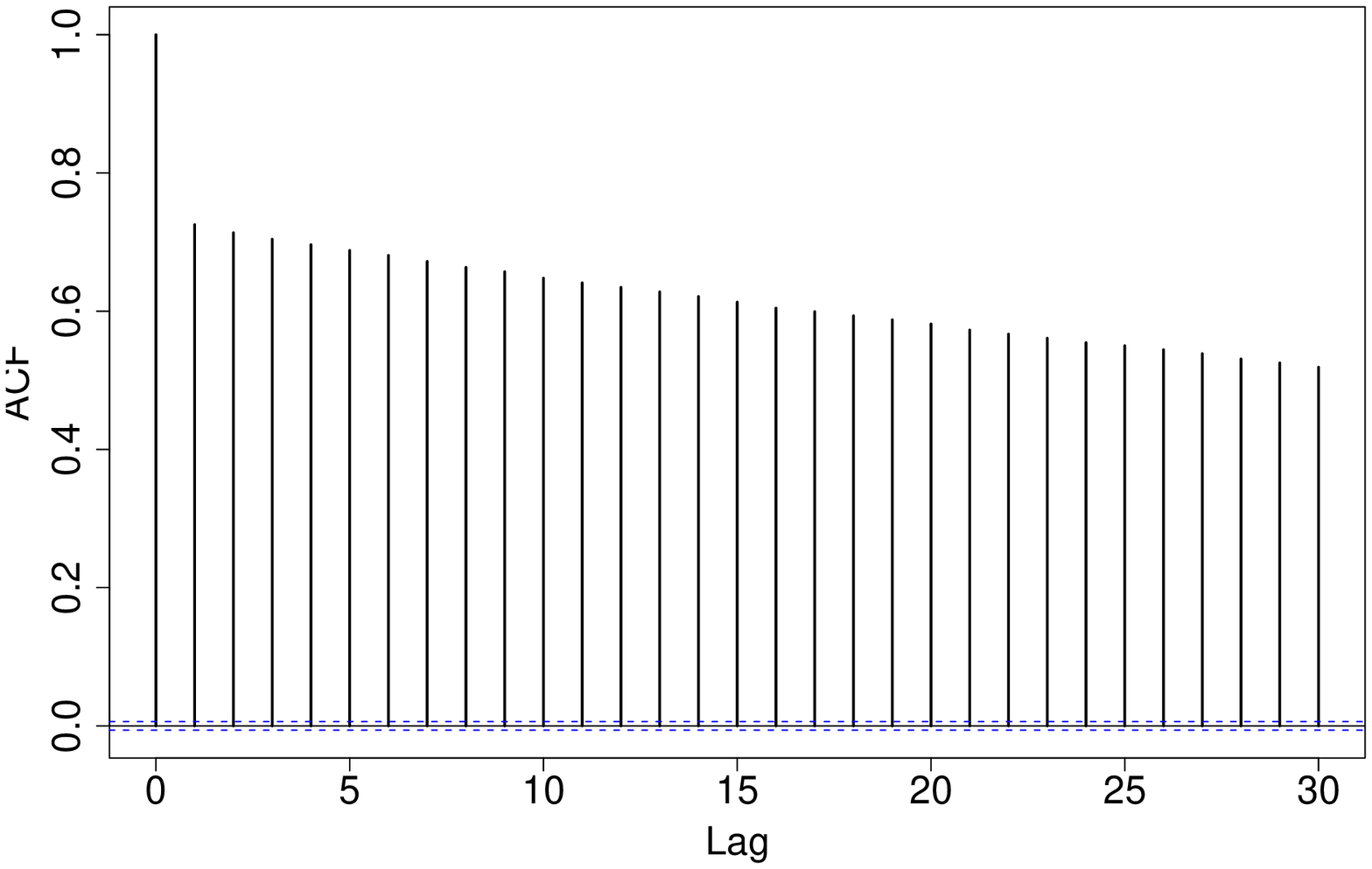}
    \end{minipage}}\\
   \subfigure[$\beta_{21}$.]{
     \label{fig:acf-klbeta:21}
     \begin{minipage}[b]{0.23\textwidth}
       \centering \includegraphics[width=\textwidth]{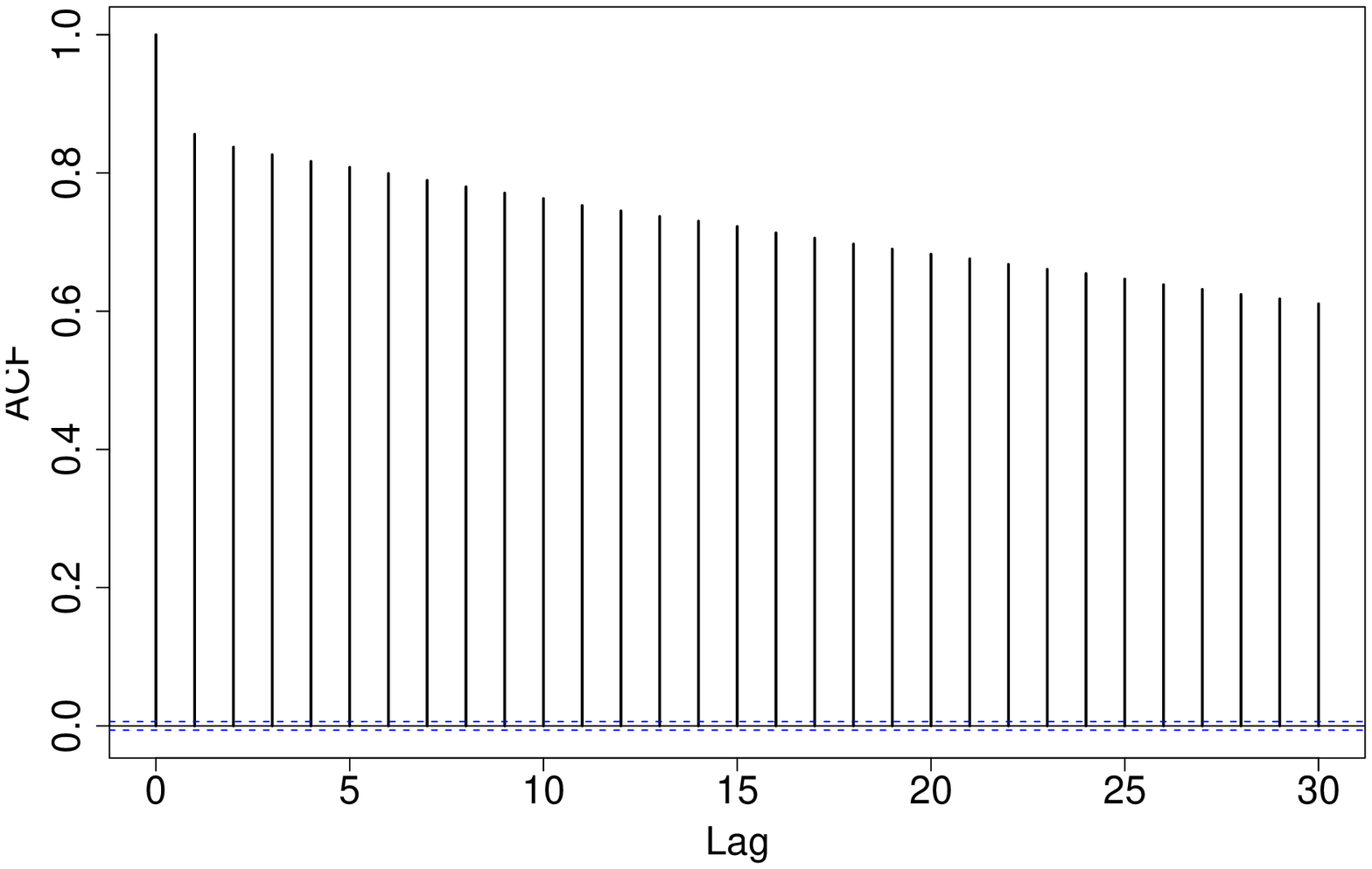}
     \end{minipage}}%
   \subfigure[$\beta_{22}$.]{
     \label{fig:acf-klbeta:22}
     \begin{minipage}[b]{0.23\textwidth}
       \centering \includegraphics[width=\textwidth]{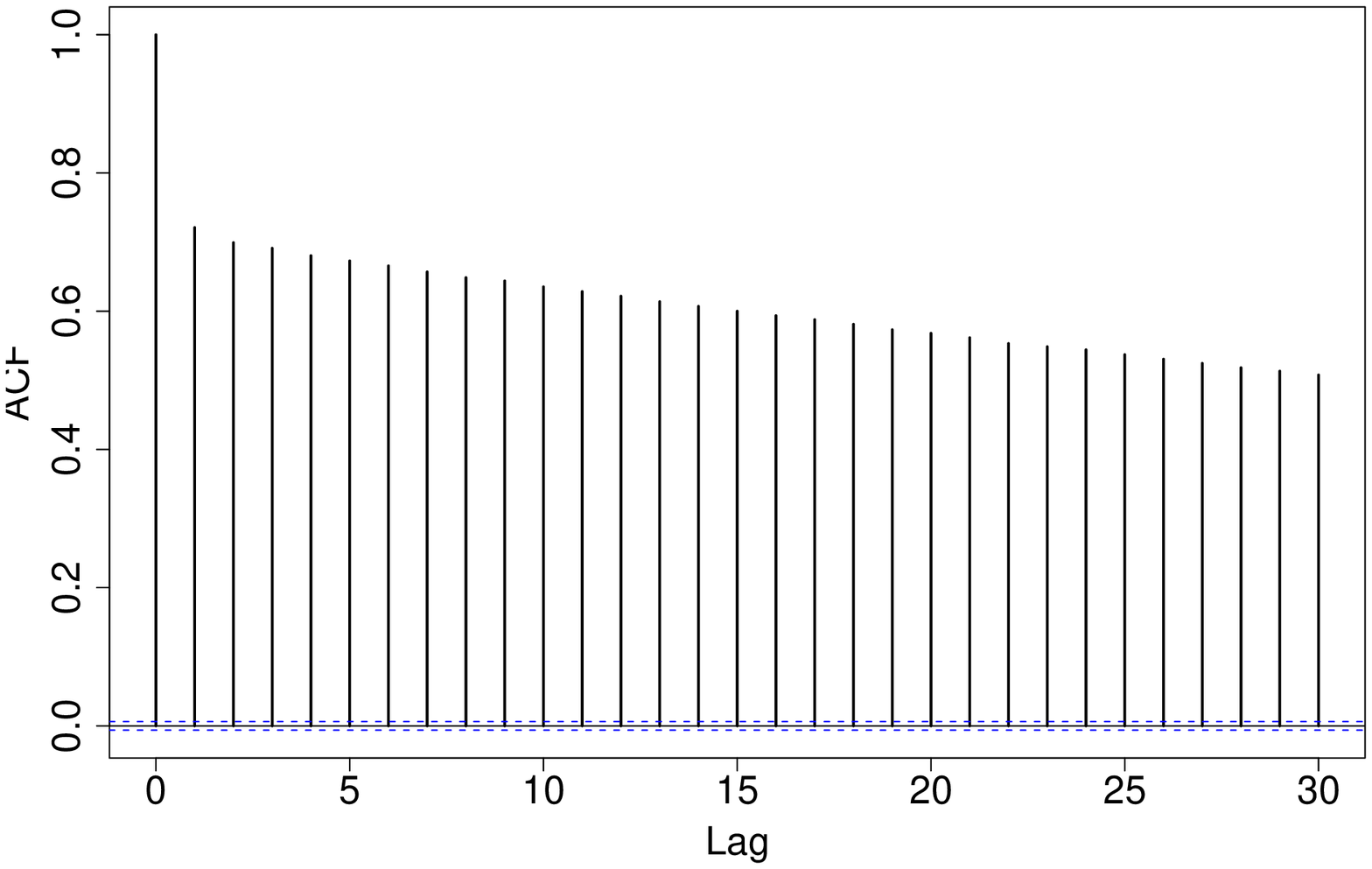}
     \end{minipage}}%
   \subfigure[$\beta_{23}$.]{
     \label{fig:acf-klbeta:23}
     \begin{minipage}[b]{0.23\textwidth}
       \centering \includegraphics[width=\textwidth]{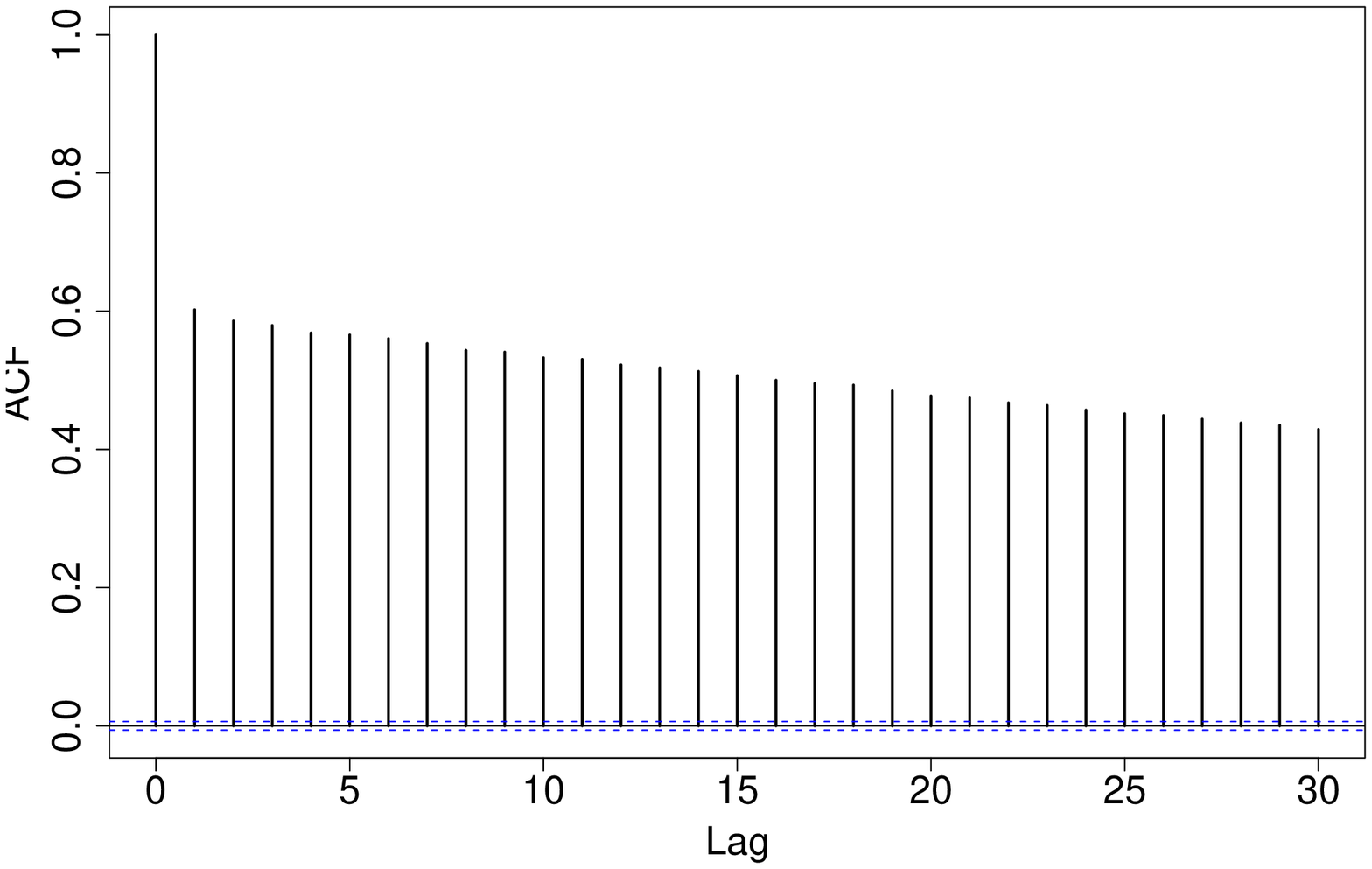}
     \end{minipage}}%
   \subfigure[$\beta_{24}$.]{
     \label{fig:acf-klbeta:24}
     \begin{minipage}[b]{0.23\textwidth}
       \centering \includegraphics[width=\textwidth]{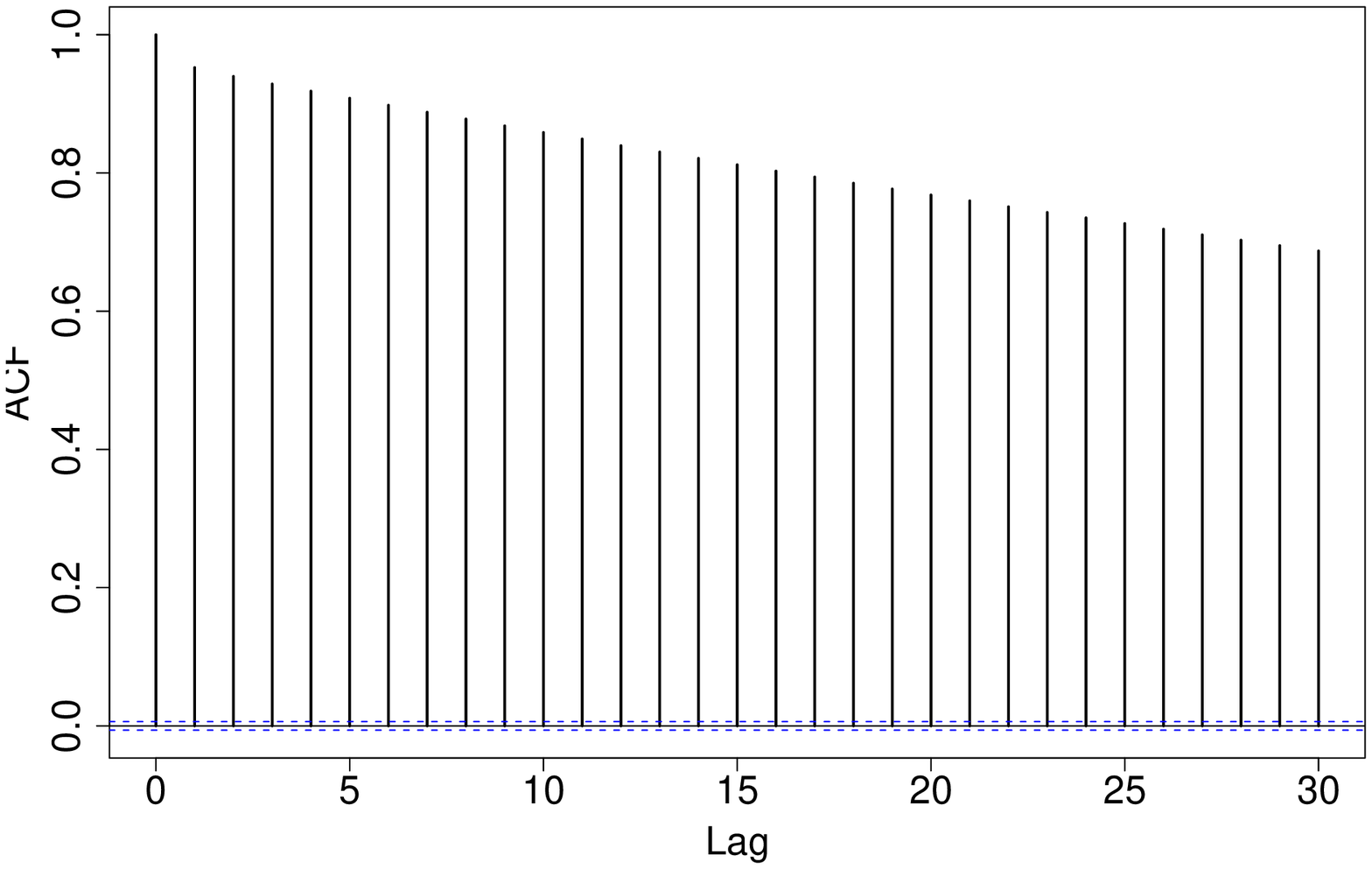}
     \end{minipage}}
  \caption{Autocorrelation plots for $\beta$.}
  \label{fig:acf-klbeta} 
\end{figure}

Tables~\ref{tab:dset4results:alpha} and ~\ref{tab:dset4results:beta} show the
posterior means of the parameters with their corresponding 95\% HPD
intervals.  The autocorrelation plots of the parameters presented in
Figure~\ref{fig:acf-klalpha} and Figure~\ref{fig:acf-klbeta}, show that mixing
is not very good since there is significant dependence even at lags greater
than $30$. The posterior parameter estimates are almost identical to those
obtained by \citeN{klugman1992} even though mixing does not appear to be good.

\begin{figure}
  \subfigure[$\alpha_{2}$.]{
    \label{fig:acf-realpha:02}
    \begin{minipage}[b]{0.33\textwidth}
      \centering \includegraphics[width=\textwidth]{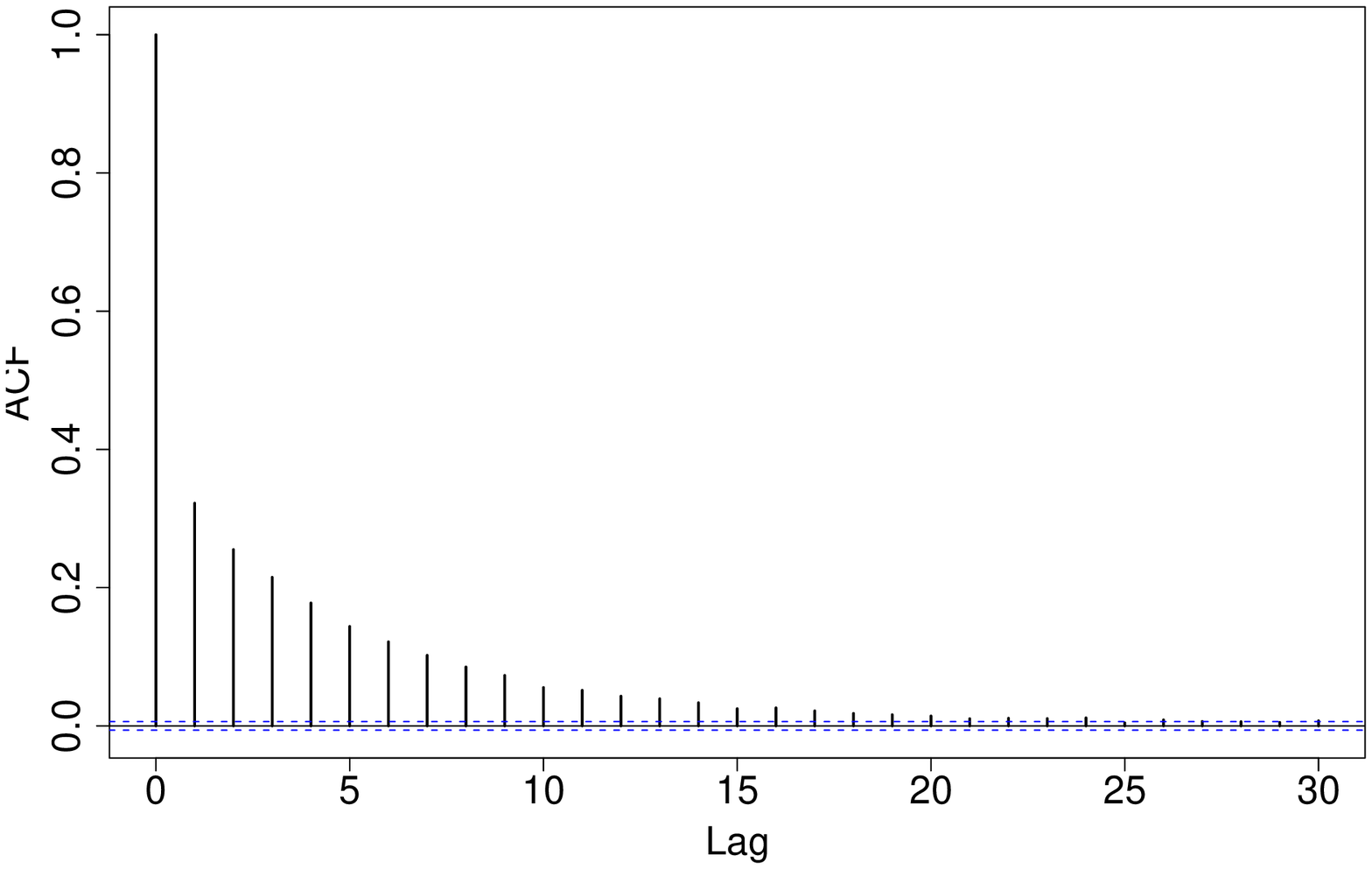}
    \end{minipage}}%
  \subfigure[$\alpha_{3}$.]{
    \label{fig:acf-realpha:03}
    \begin{minipage}[b]{0.33\textwidth}
      \centering \includegraphics[width=\textwidth]{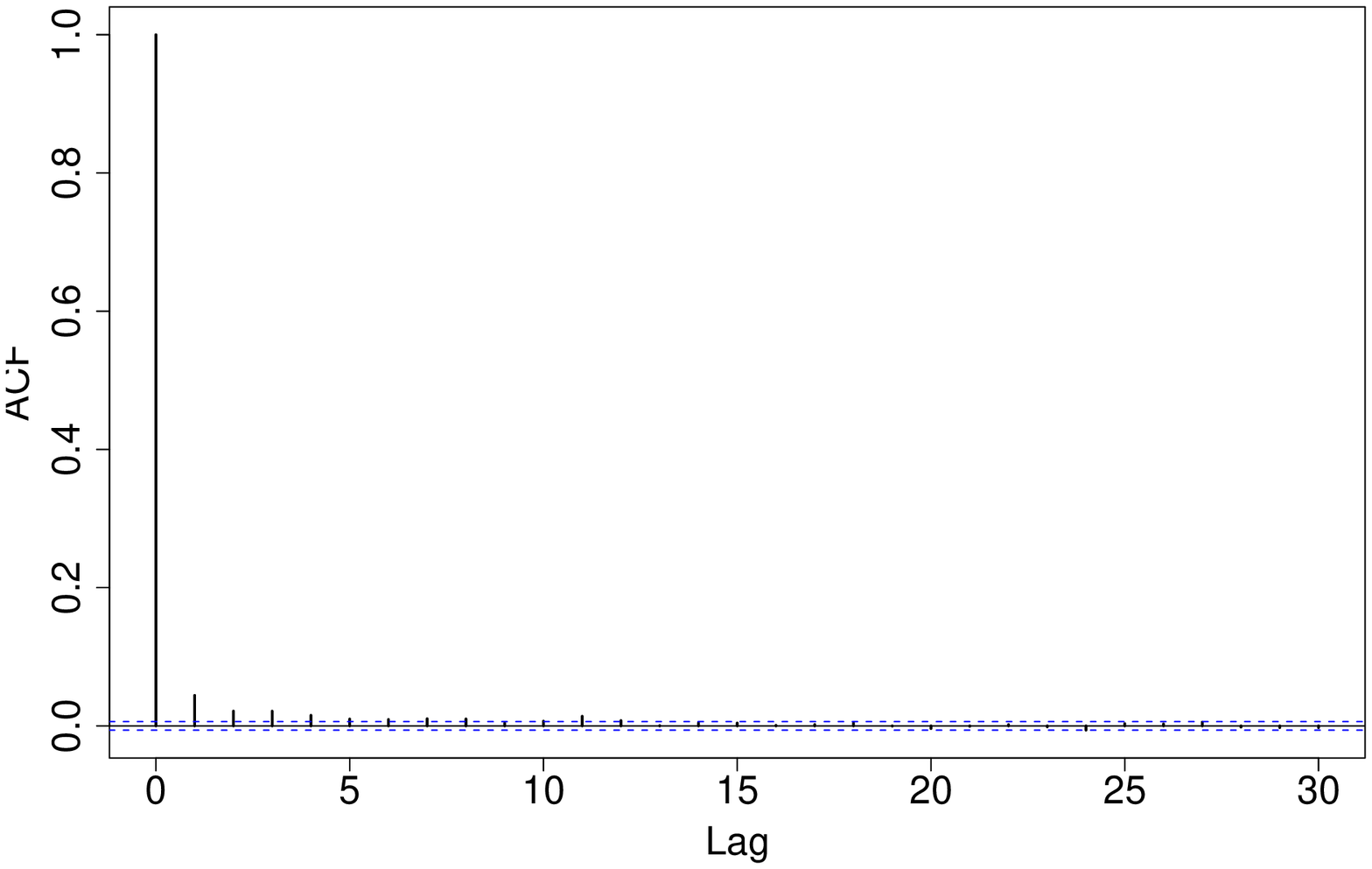}
    \end{minipage}}%
  \subfigure[$\alpha_{4}$.]{
    \label{fig:acf-realpha:04}
    \begin{minipage}[b]{0.33\textwidth}
      \centering \includegraphics[width=\textwidth]{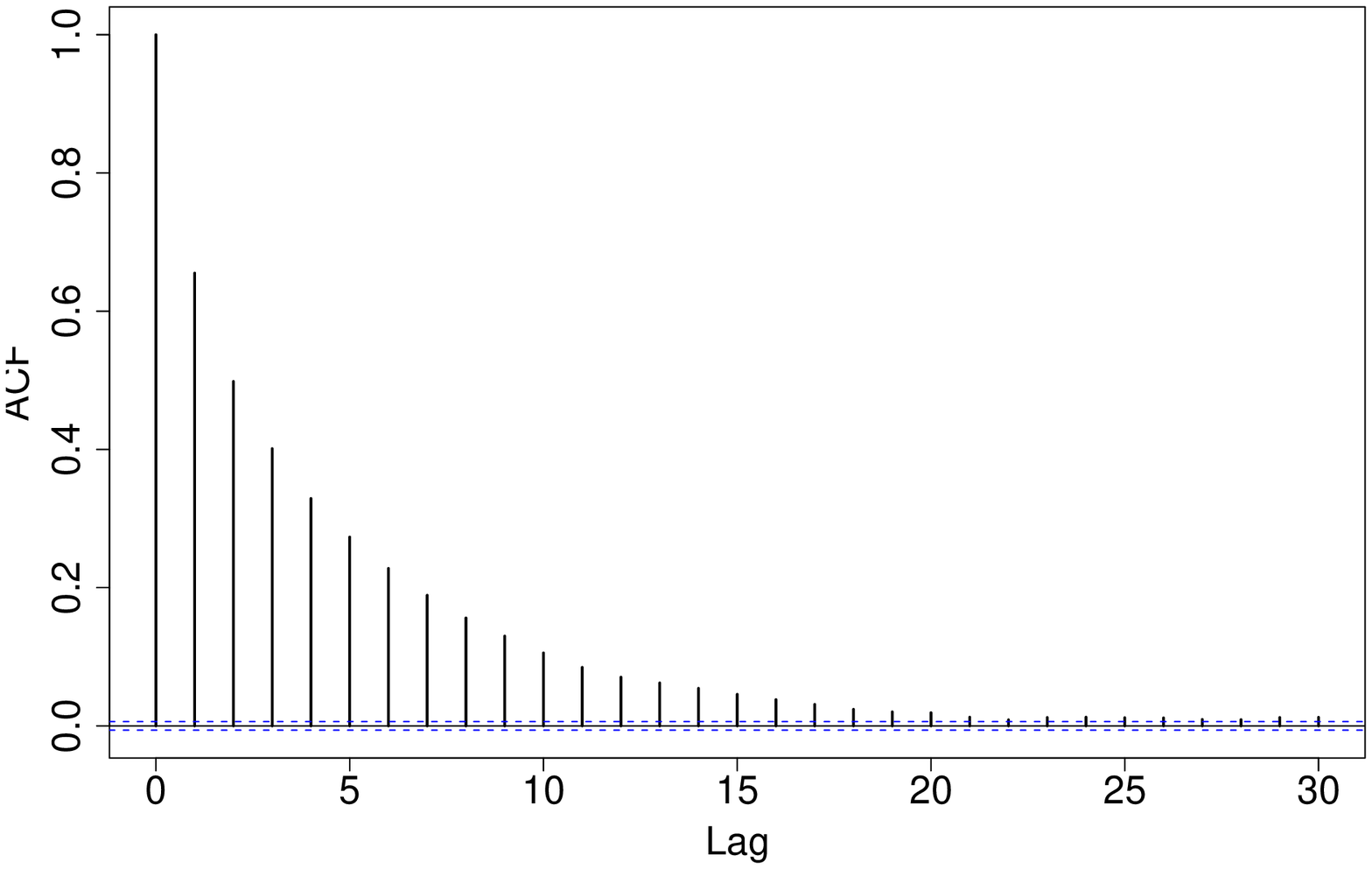}
    \end{minipage}}\\
  \subfigure[$\alpha_{5}$.]{
    \label{fig:acf-realpha:05}
    \begin{minipage}[b]{0.33\textwidth}
      \centering \includegraphics[width=\textwidth]{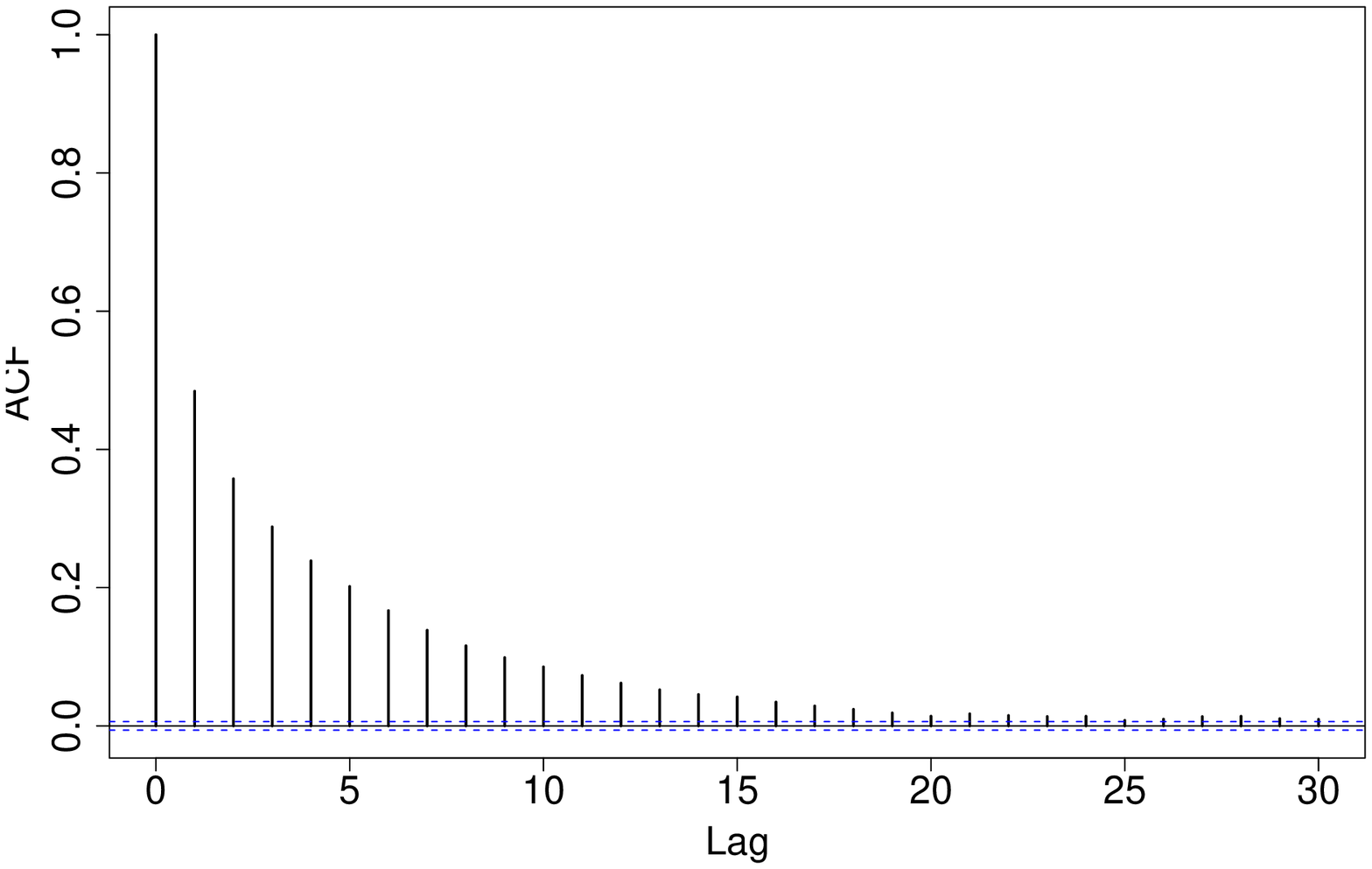}
    \end{minipage}}%
  \subfigure[$\alpha_{6}$.]{
    \label{fig:acf-realpha:06}
    \begin{minipage}[b]{0.33\textwidth}
      \centering \includegraphics[width=\textwidth]{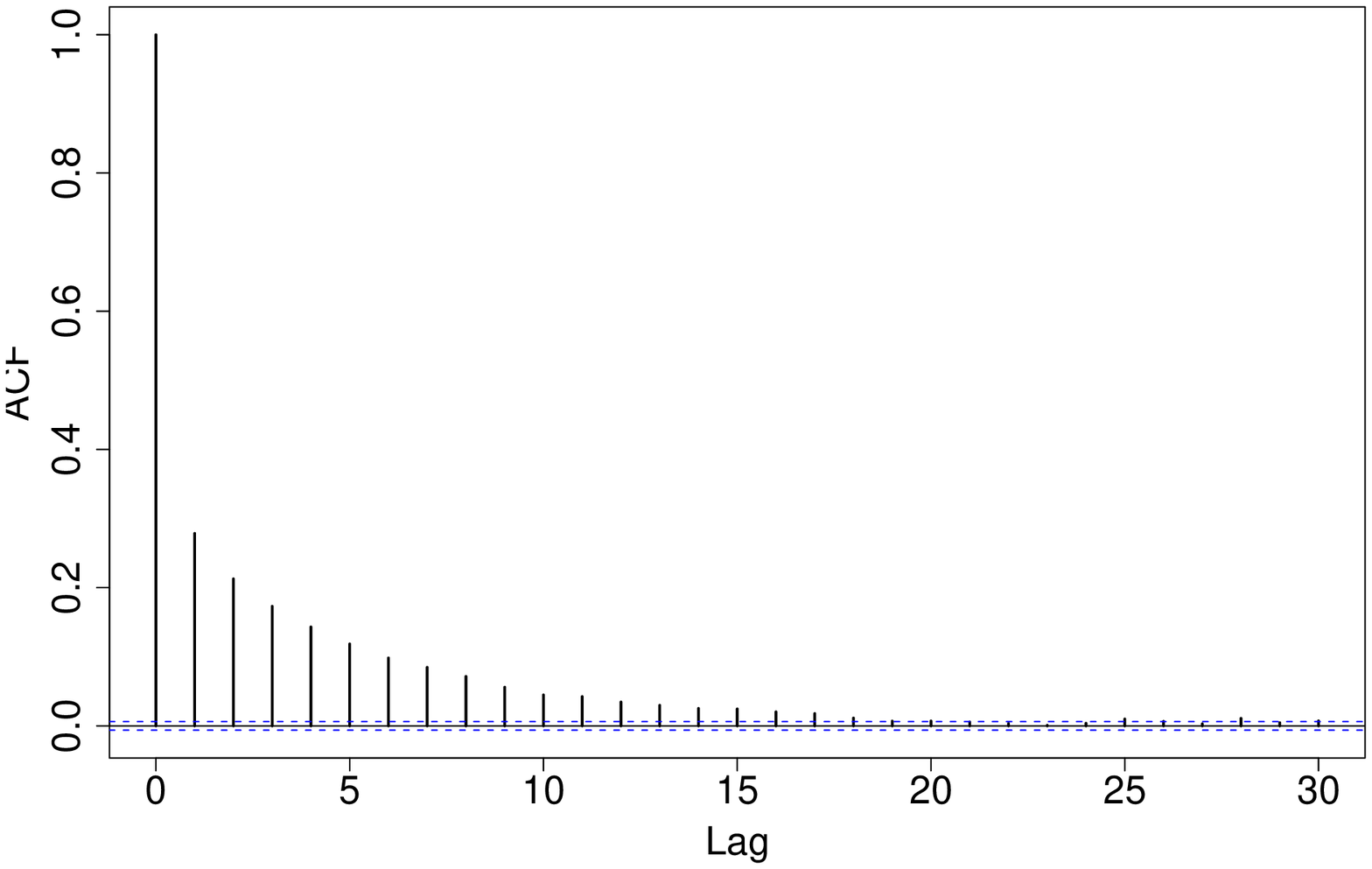}
    \end{minipage}}%
  \subfigure[$\alpha_{7}$.]{
    \label{fig:acf-realpha:07}
    \begin{minipage}[b]{0.33\textwidth}
      \centering \includegraphics[width=\textwidth]{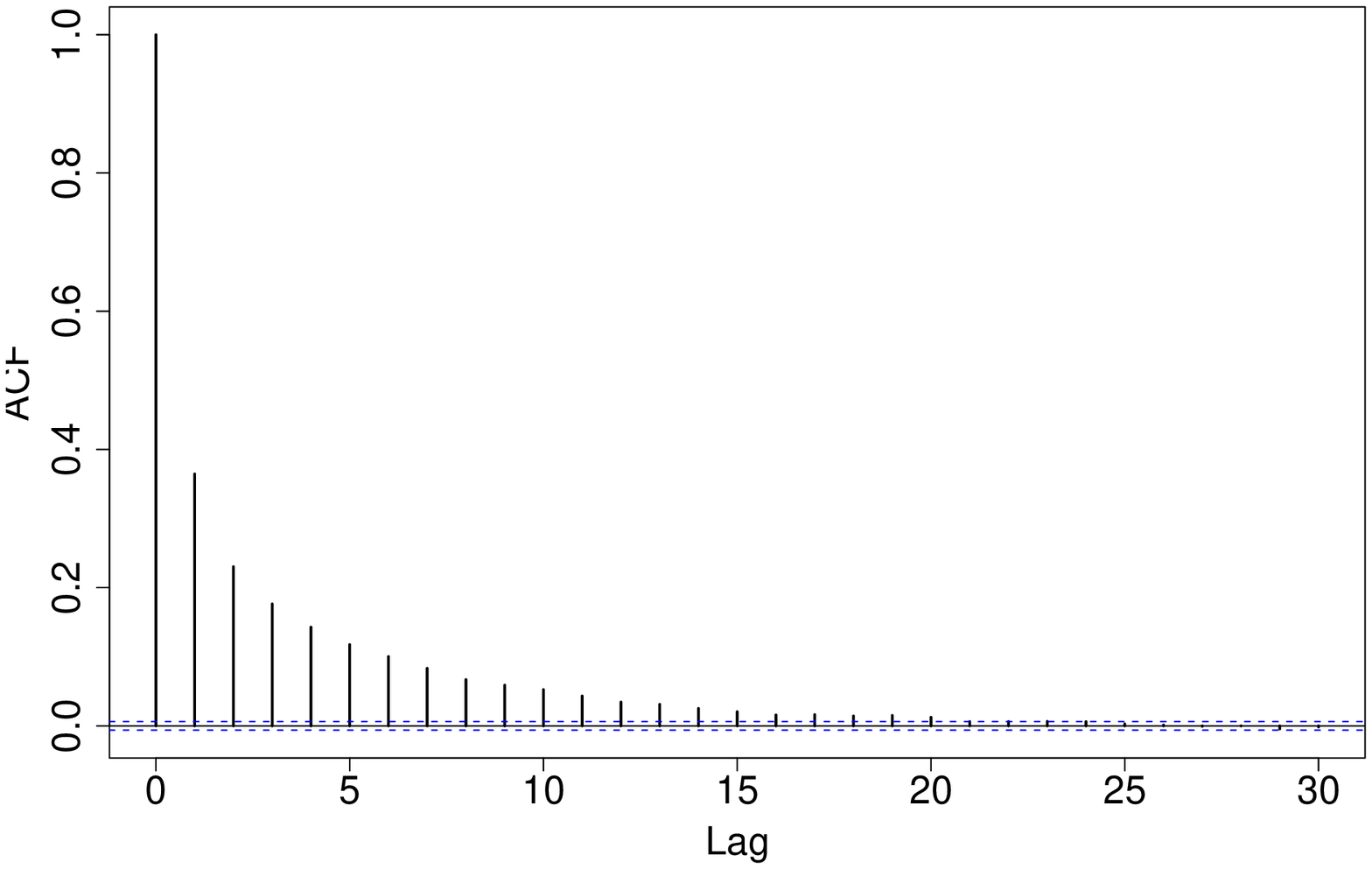}
    \end{minipage}}\\
  \subfigure[$\alpha_{8}$.]{
    \label{fig:acf-realpha:08}
    \begin{minipage}[b]{0.33\textwidth}
      \centering \includegraphics[width=\textwidth]{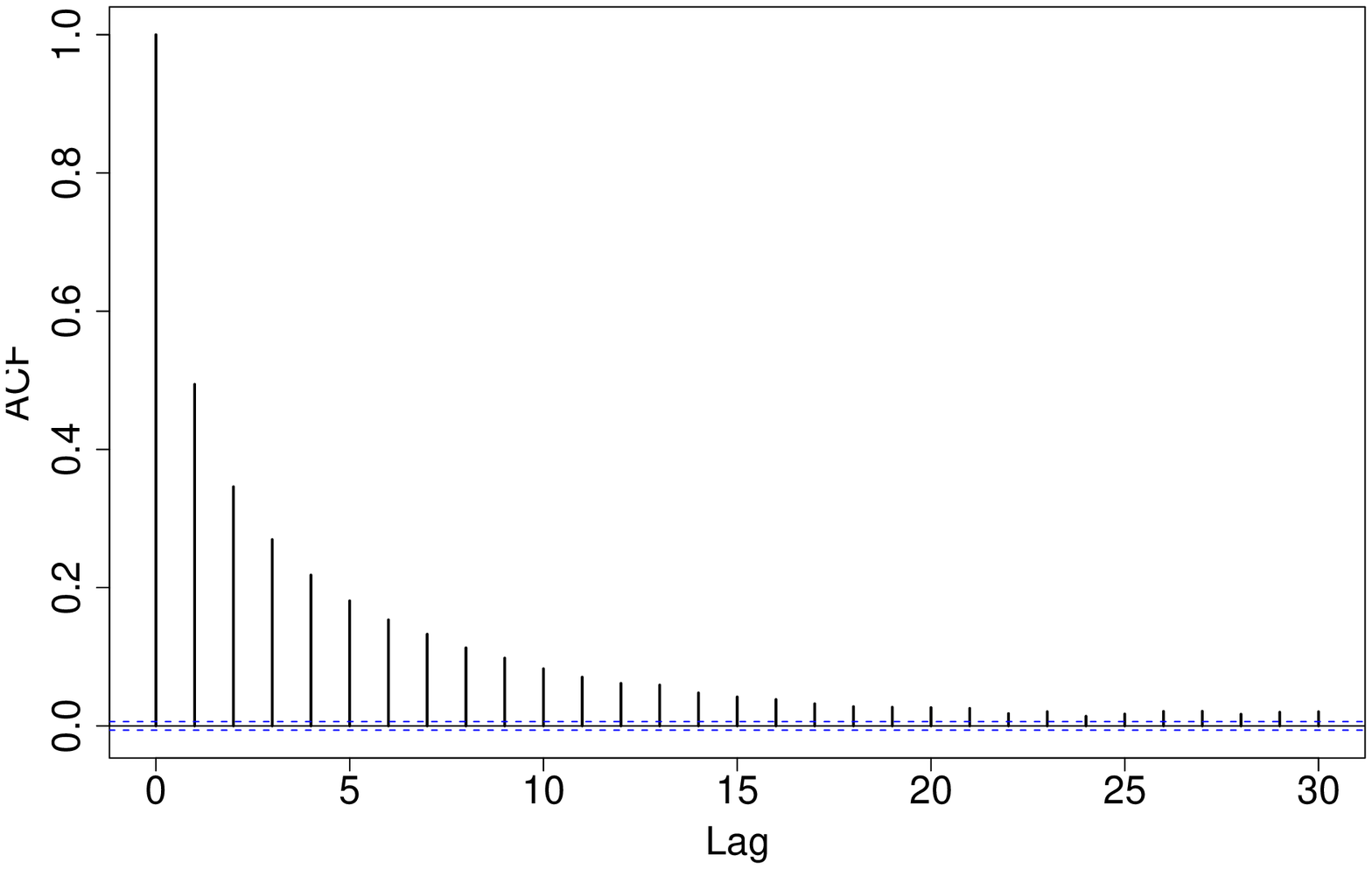}
    \end{minipage}}%
  \subfigure[$\alpha_{9}$.]{
    \label{fig:acf-realpha:09}
    \begin{minipage}[b]{0.33\textwidth}
      \centering \includegraphics[width=\textwidth]{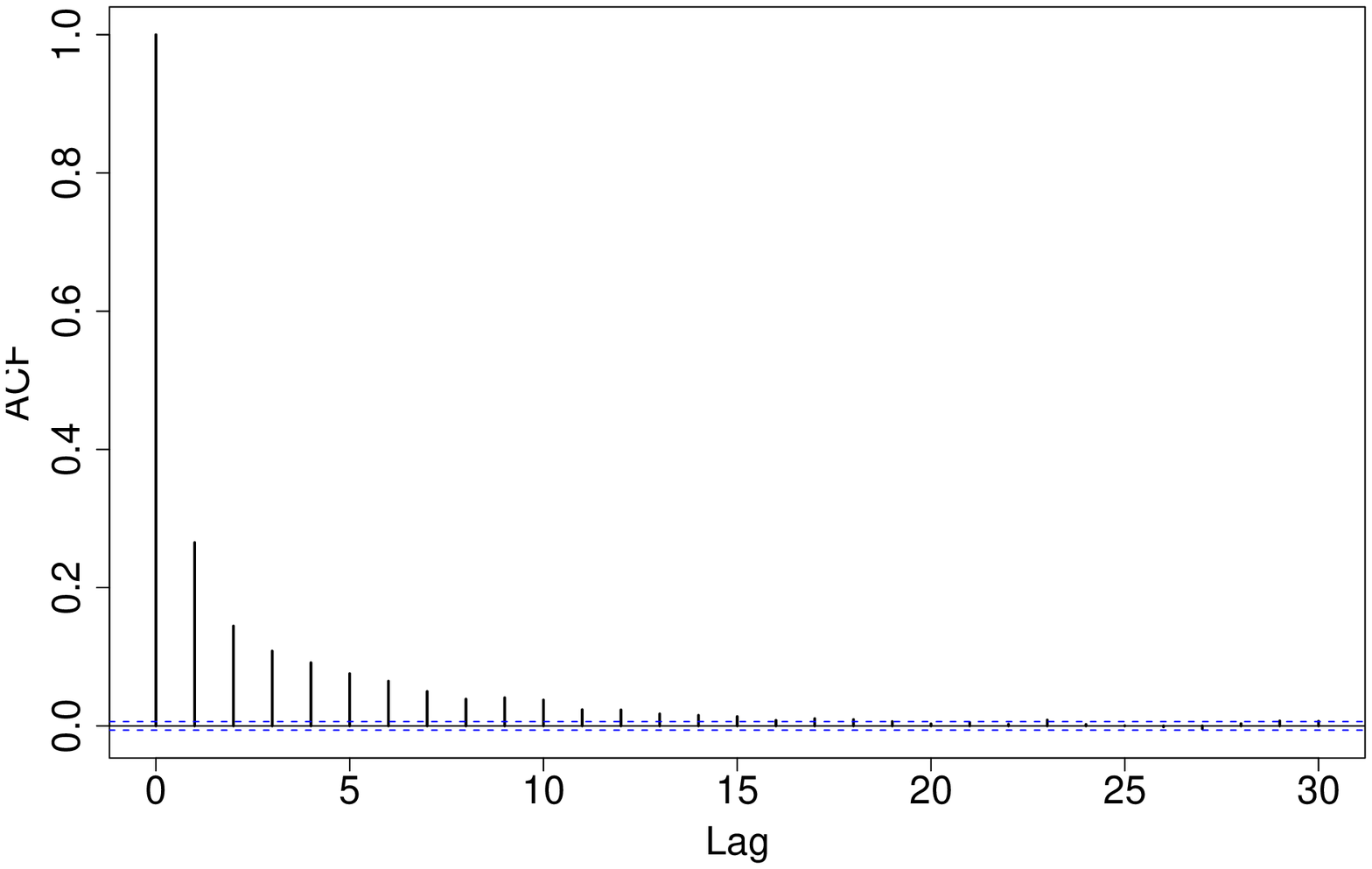}
    \end{minipage}}%
  \subfigure[$\alpha_{10}$.]{
    \label{fig:acf-realpha:10}
    \begin{minipage}[b]{0.33\textwidth}
      \centering \includegraphics[width=\textwidth]{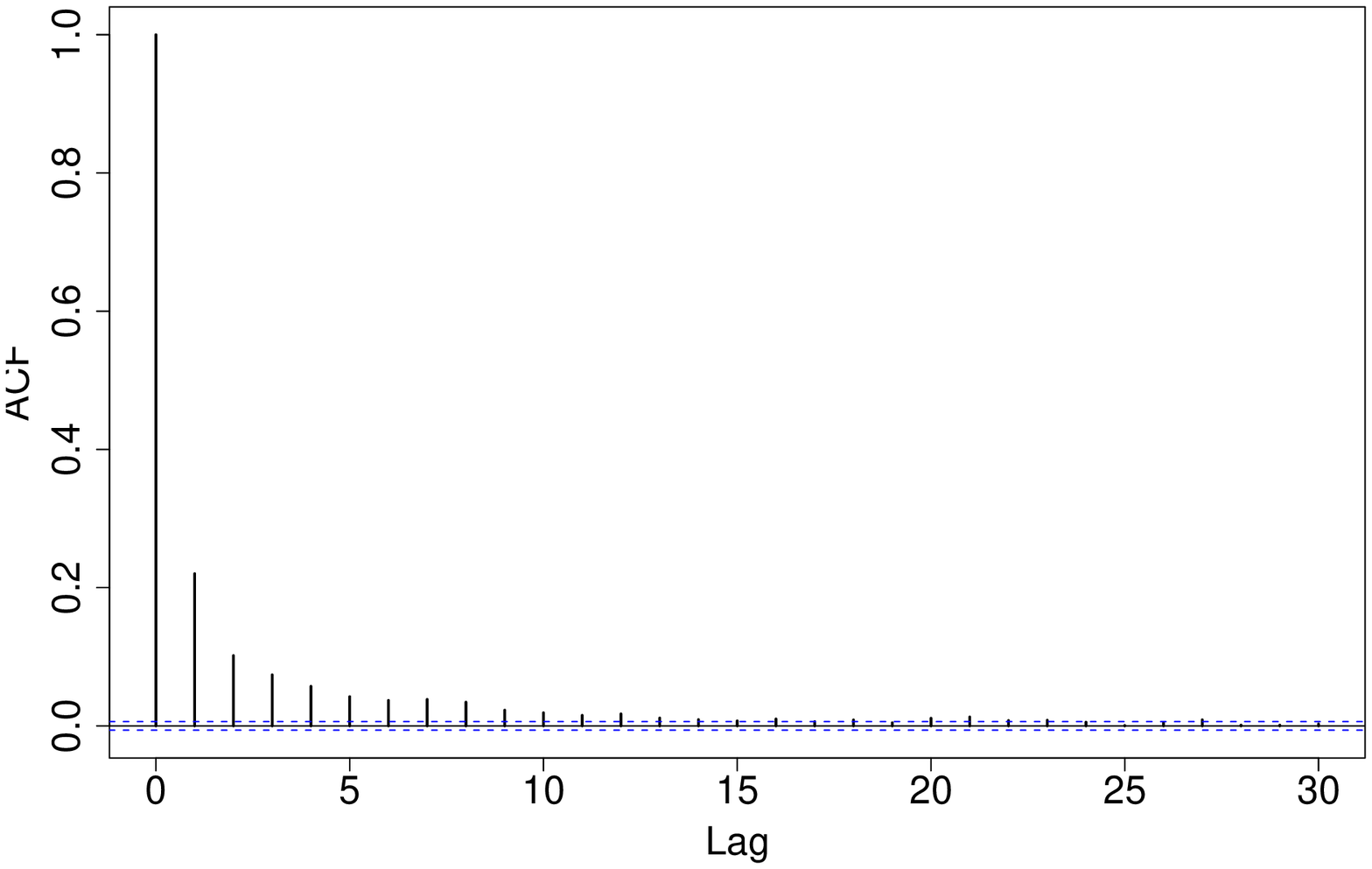}
    \end{minipage}}\\
  \caption{Autocorrelation plots for $\alpha$. The dotted lines are 95\% confidence bands.}  \label{fig:acf-realpha} 
\end{figure}

\begin{figure}
    \subfigure[$\beta_{2}$.]{
    \label{fig:acf-rebeta:02}
    \begin{minipage}[b]{0.24\textwidth}
      \centering \includegraphics[width=\textwidth]{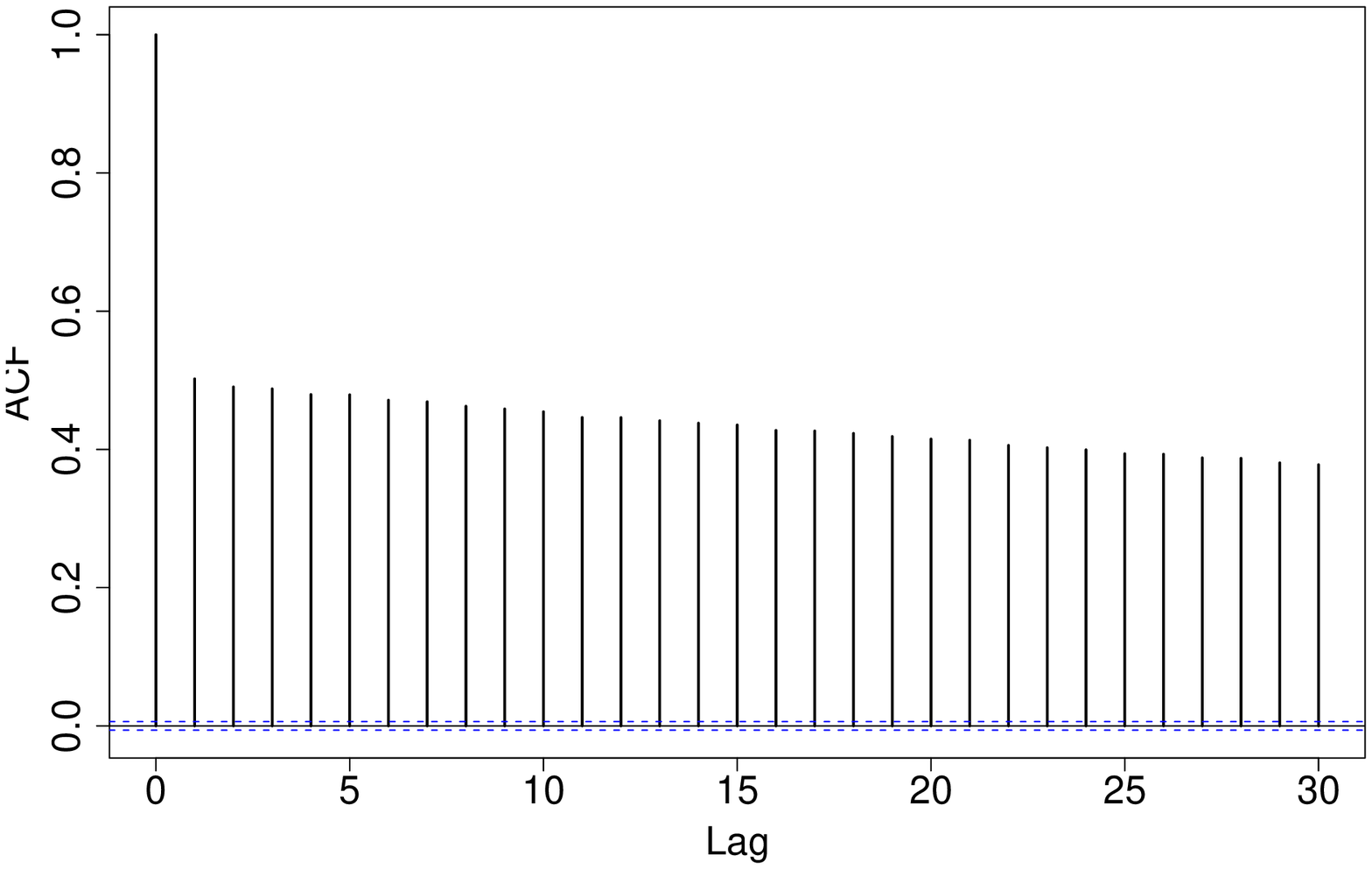}
    \end{minipage}}%
  \subfigure[$\beta_{3}$.]{
    \label{fig:acf-rebeta:03}
    \begin{minipage}[b]{0.24\textwidth}
      \centering \includegraphics[width=\textwidth]{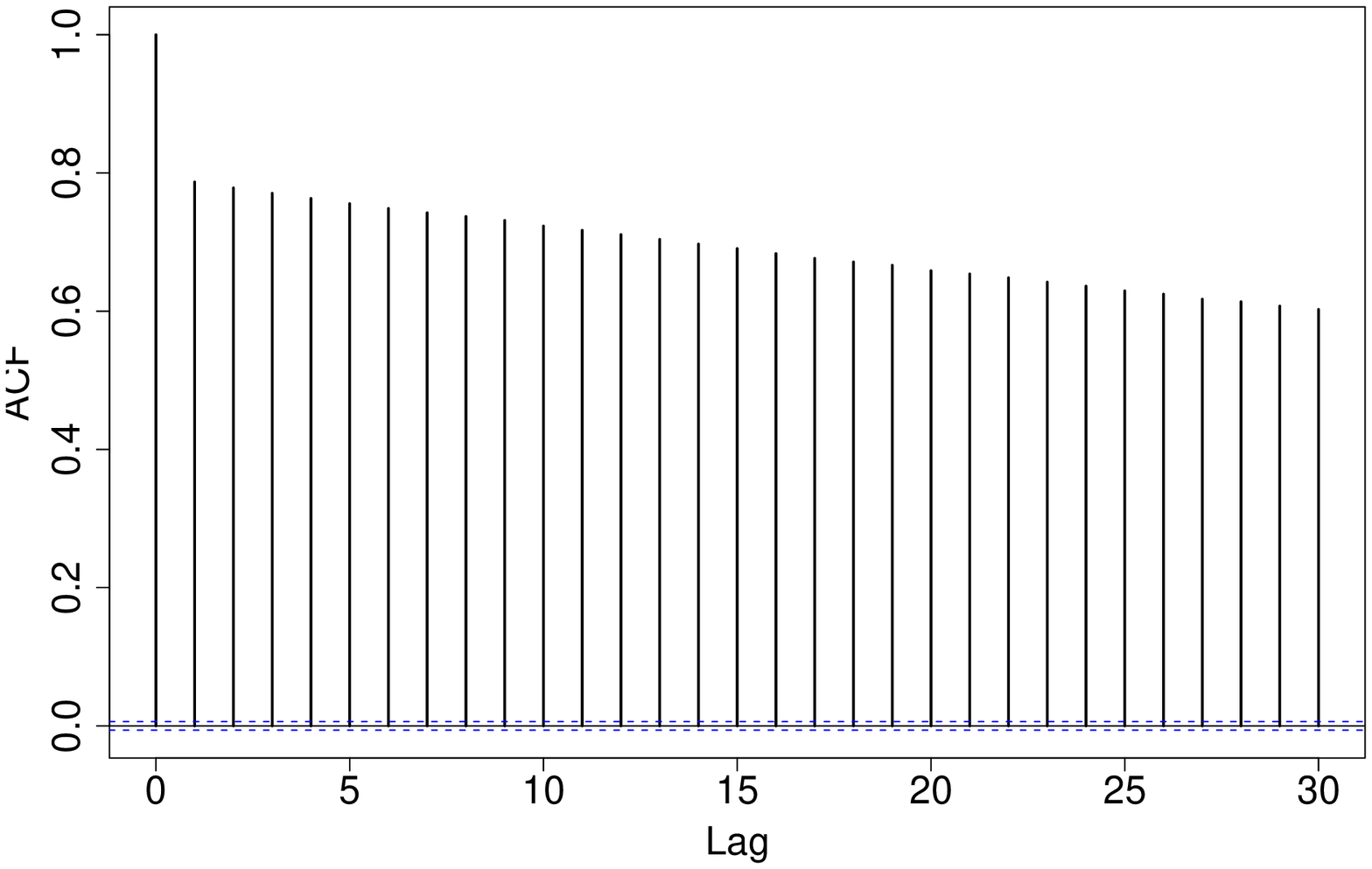}
    \end{minipage}}%
  \subfigure[$\beta_{4}$.]{
    \label{fig:acf-rebeta:04}
    \begin{minipage}[b]{0.24\textwidth}
      \centering \includegraphics[width=\textwidth]{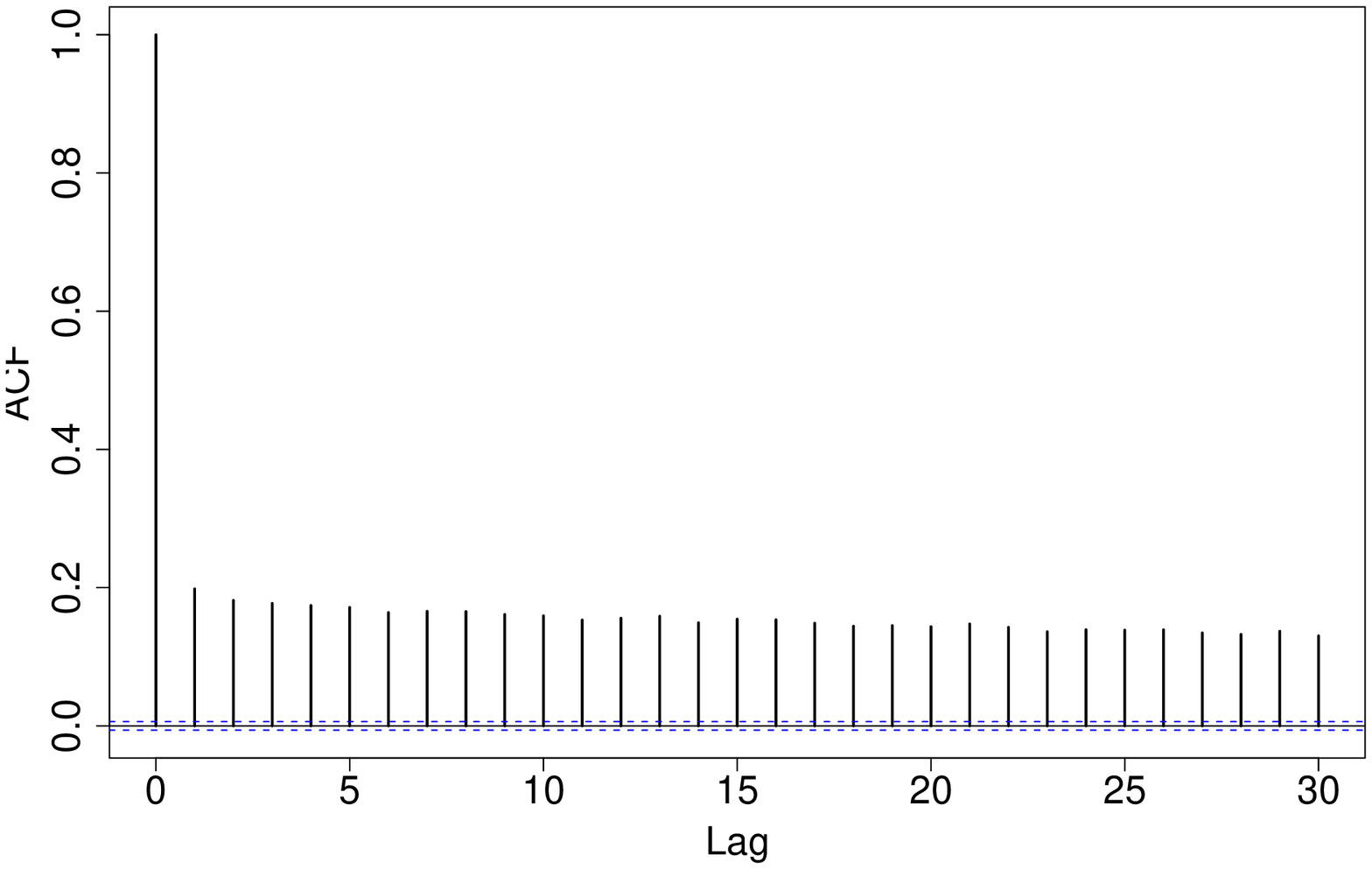}
    \end{minipage}}%
  \subfigure[$\beta_{5}$.]{
    \label{fig:acf-rebeta:05}
    \begin{minipage}[b]{0.24\textwidth}
      \centering \includegraphics[width=\textwidth]{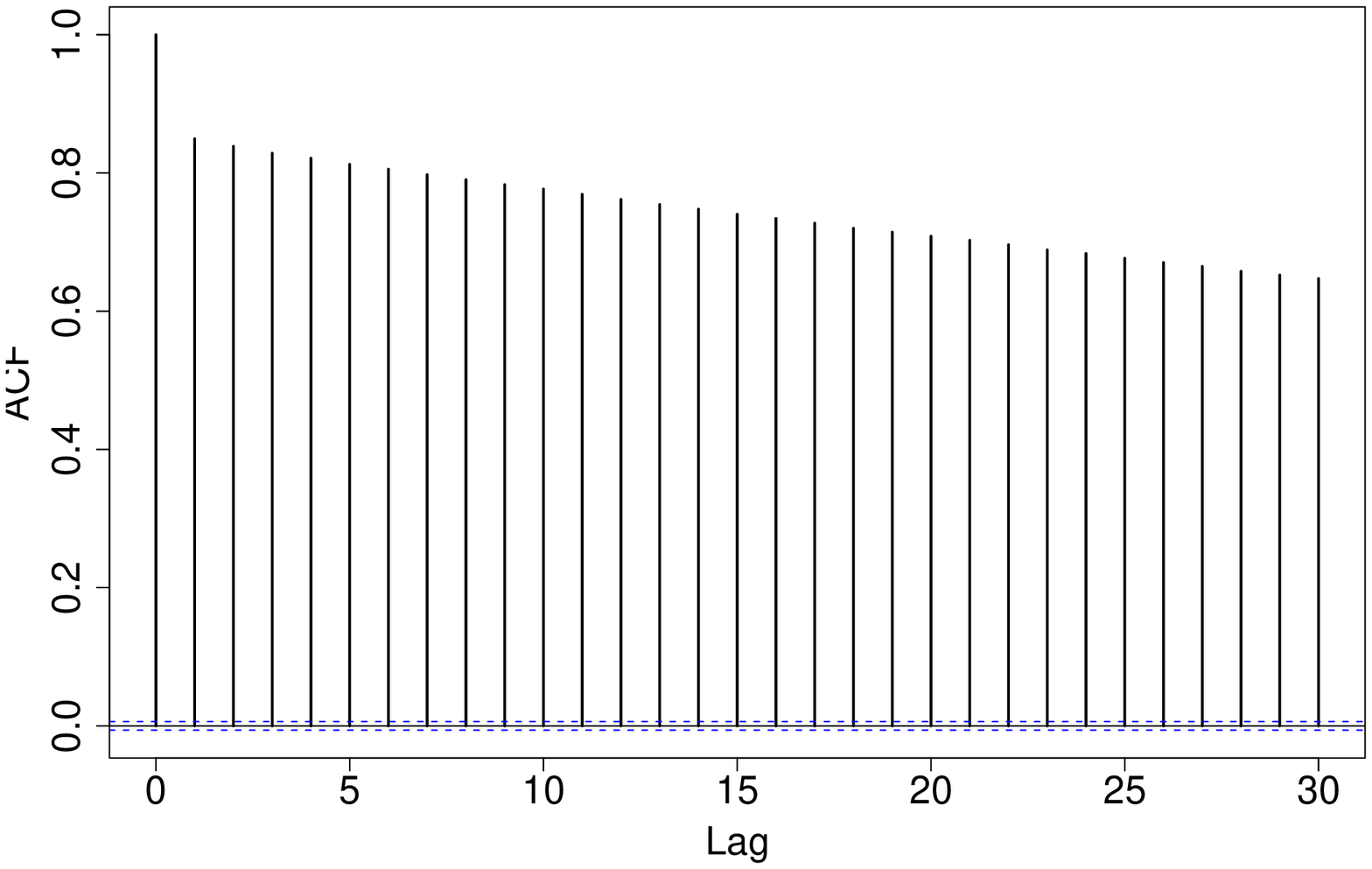}
    \end{minipage}}\\
  \subfigure[$\beta_{6}$.]{
    \label{fig:acf-rebeta:06}
    \begin{minipage}[b]{0.24\textwidth}
      \centering \includegraphics[width=\textwidth]{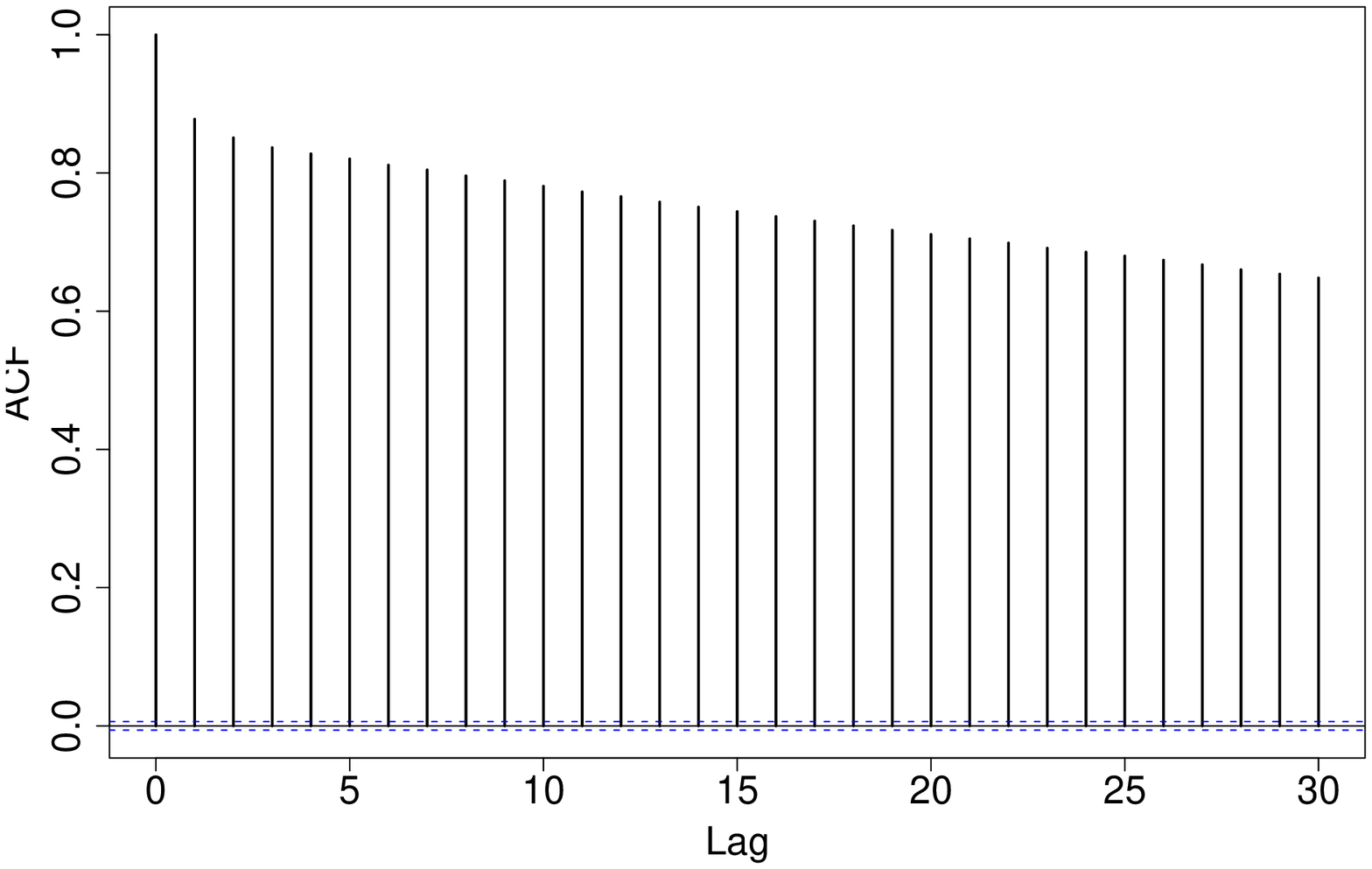}
    \end{minipage}}%
  \subfigure[$\beta_{7}$.]{
    \label{fig:acf-rebeta:07}
    \begin{minipage}[b]{0.24\textwidth}
      \centering \includegraphics[width=\textwidth]{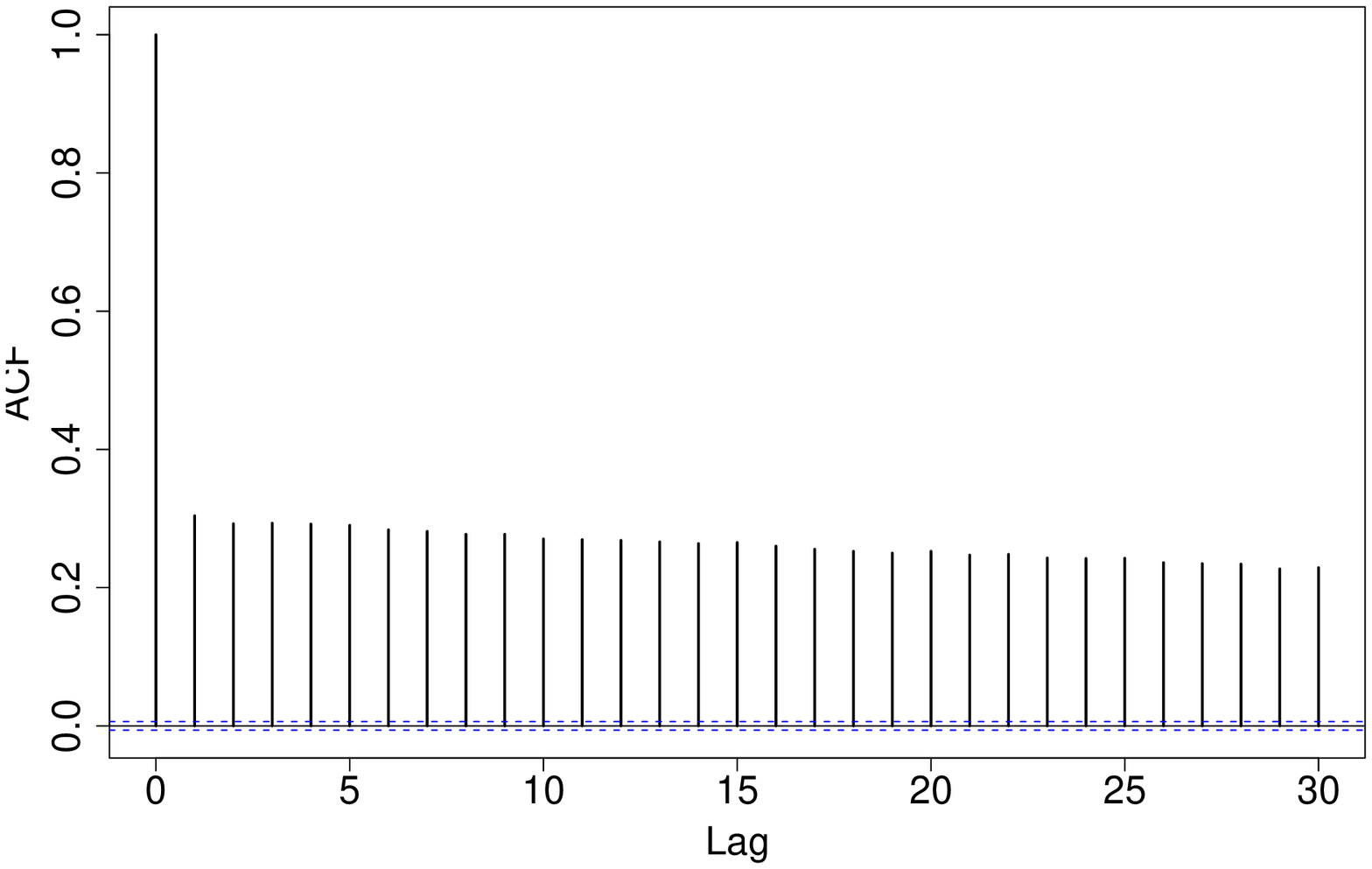}
    \end{minipage}}%
  \subfigure[$\beta_{8}$.]{
    \label{fig:acf-rebeta:08}
    \begin{minipage}[b]{0.24\textwidth}
      \centering \includegraphics[width=\textwidth]{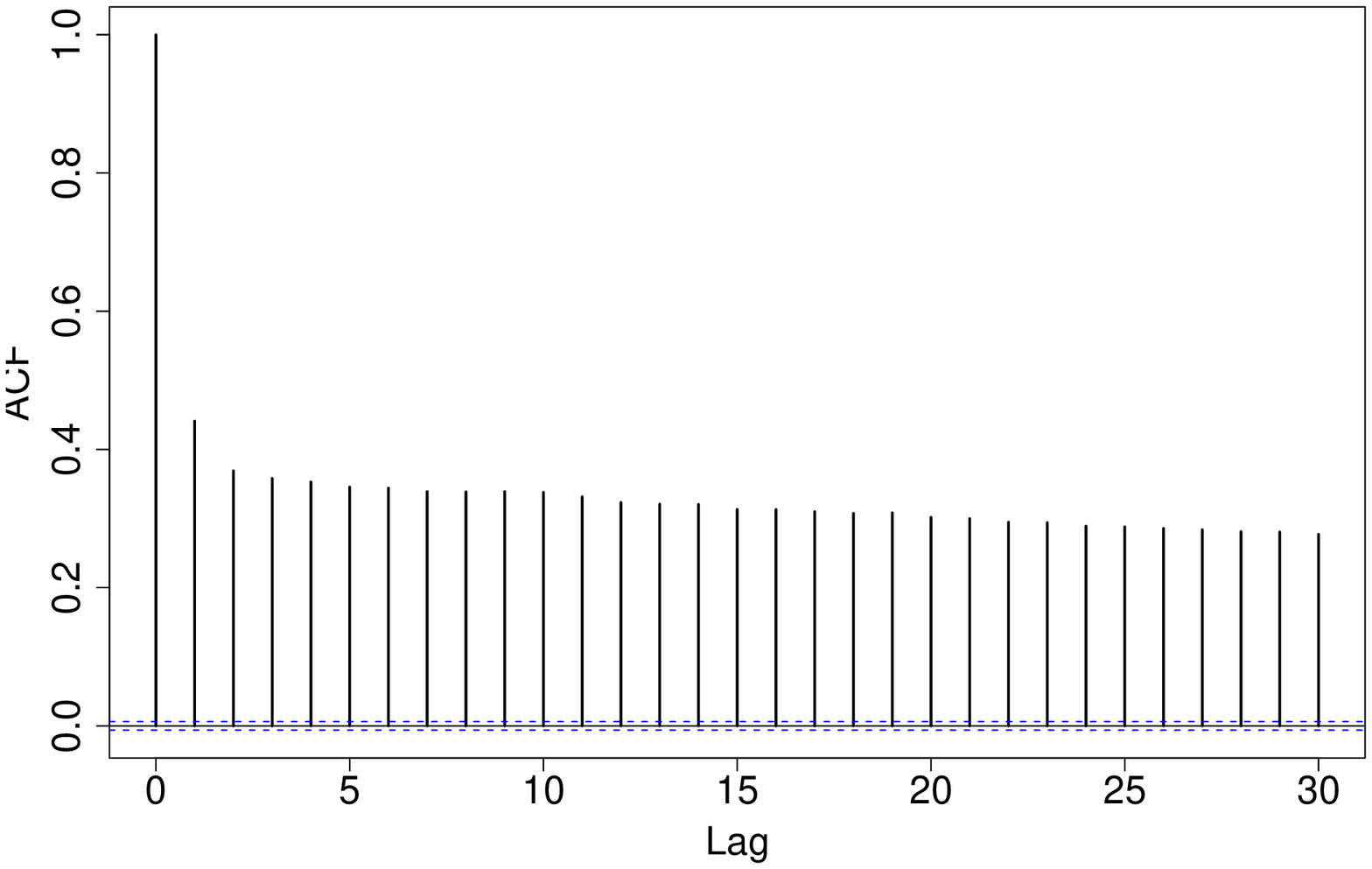}
    \end{minipage}}%
  \subfigure[$\beta_{9}$.]{
    \label{fig:acf-rebeta:09}
    \begin{minipage}[b]{0.24\textwidth}
      \centering \includegraphics[width=\textwidth]{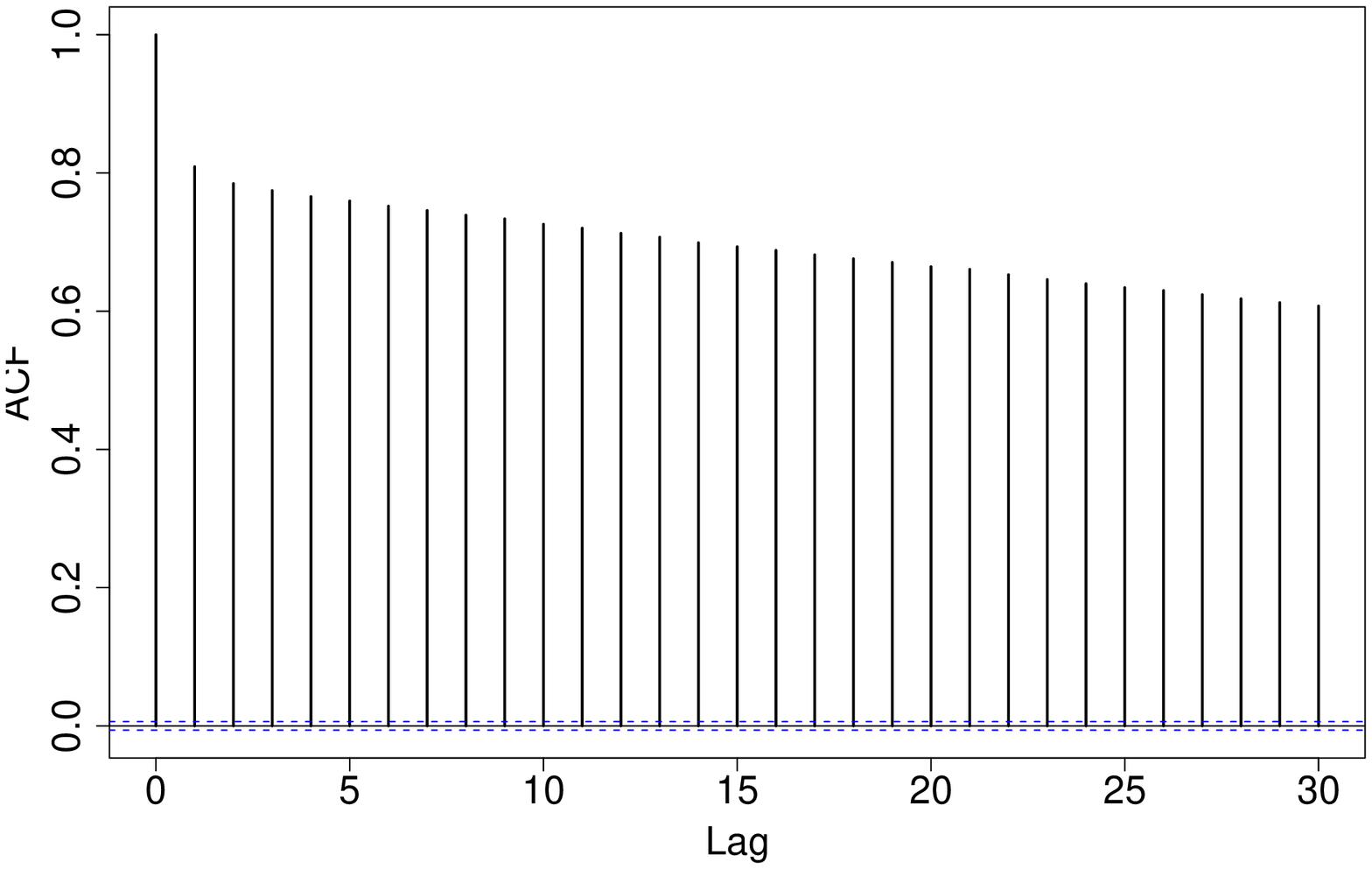}
    \end{minipage}}\\
  \subfigure[$\beta_{10}$.]{
    \label{fig:acf-rebeta:10}
    \begin{minipage}[b]{0.24\textwidth}
      \centering \includegraphics[width=\textwidth]{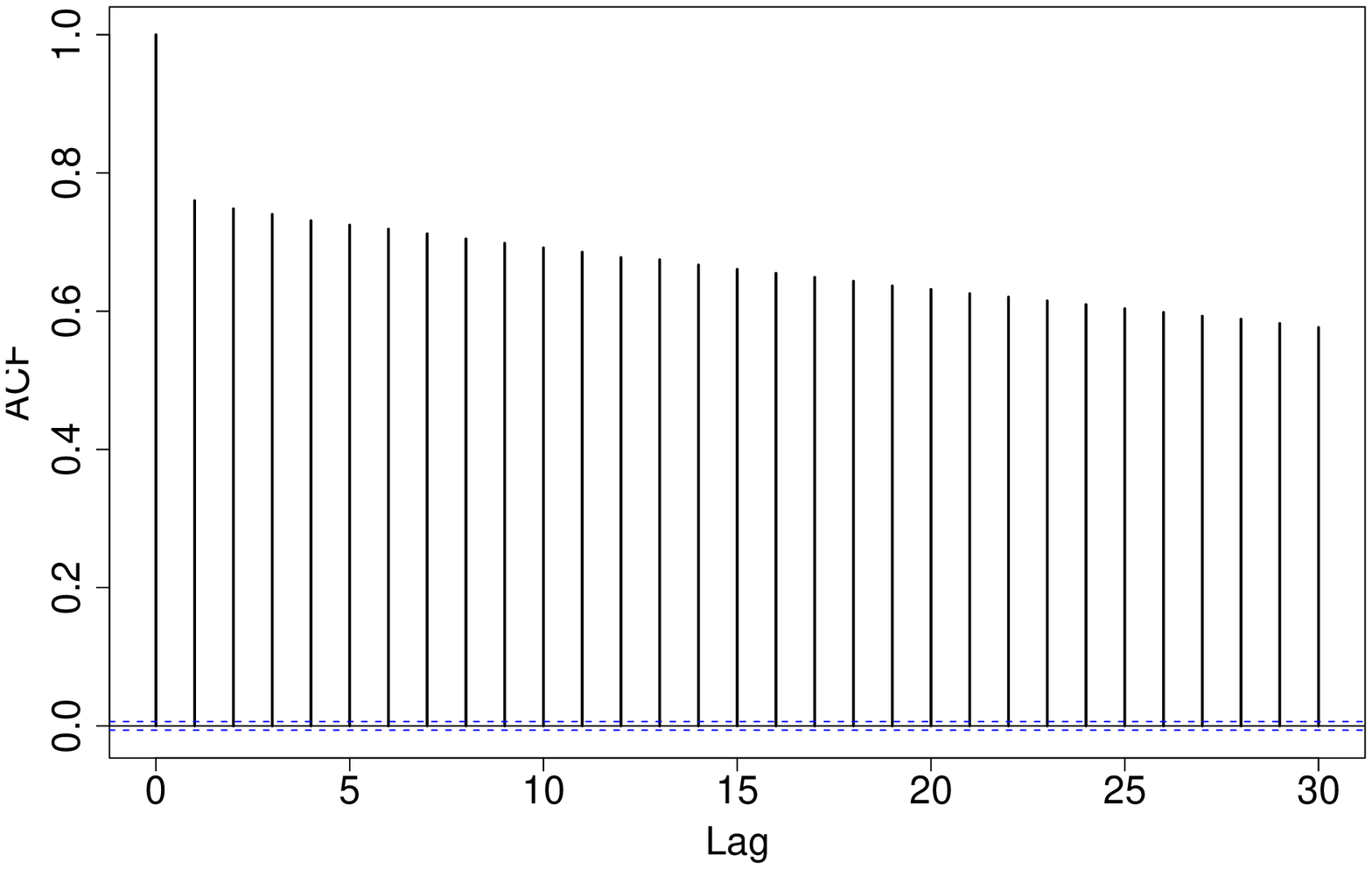}
    \end{minipage}}%
  \subfigure[$\beta_{11}$.]{
    \label{fig:acf-rebeta:11}
    \begin{minipage}[b]{0.24\textwidth}
      \centering \includegraphics[width=\textwidth]{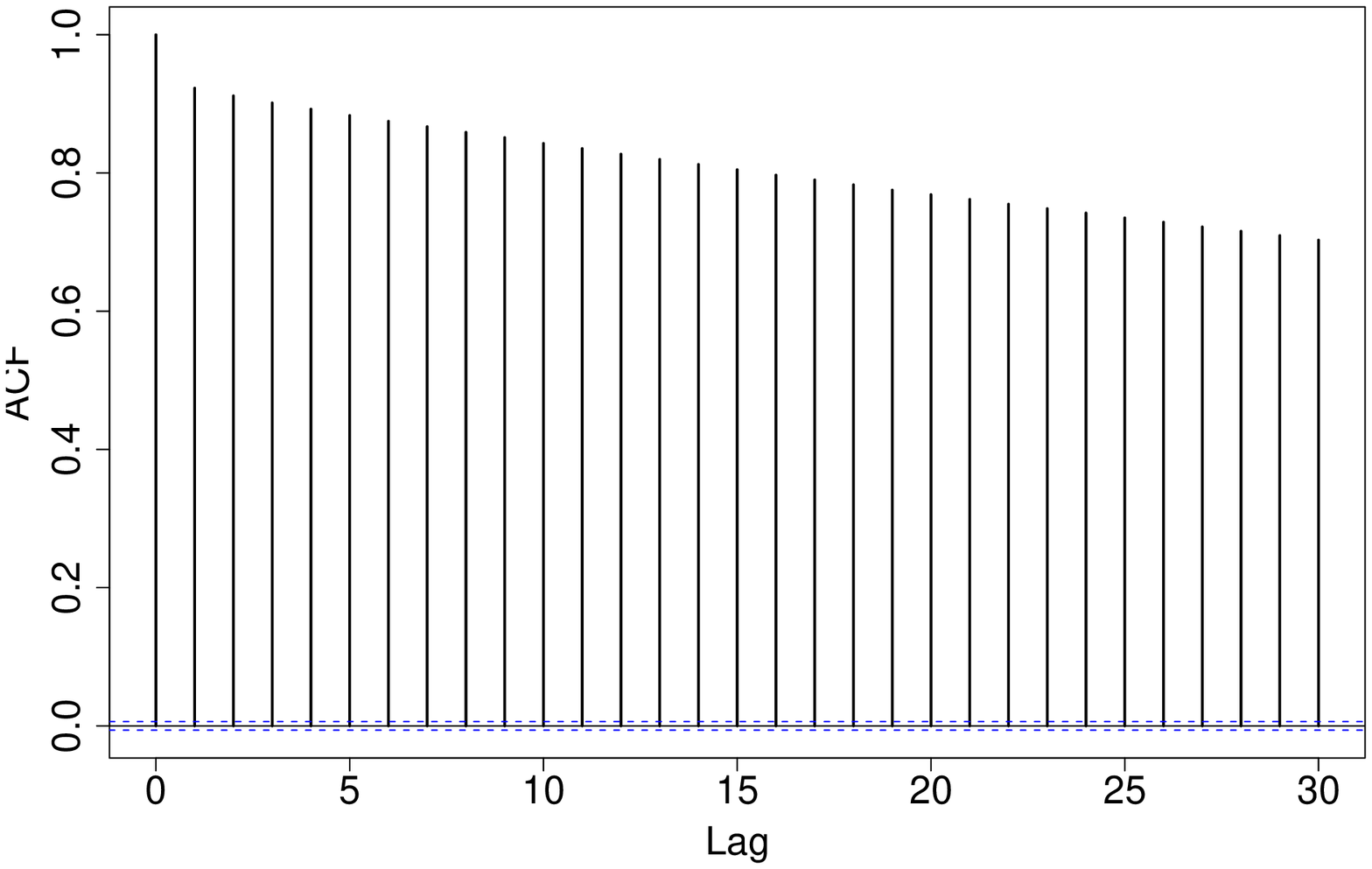}
    \end{minipage}}%
  \subfigure[$\beta_{12}$.]{
    \label{fig:acf-rebeta:12}
    \begin{minipage}[b]{0.24\textwidth}
      \centering \includegraphics[width=\textwidth]{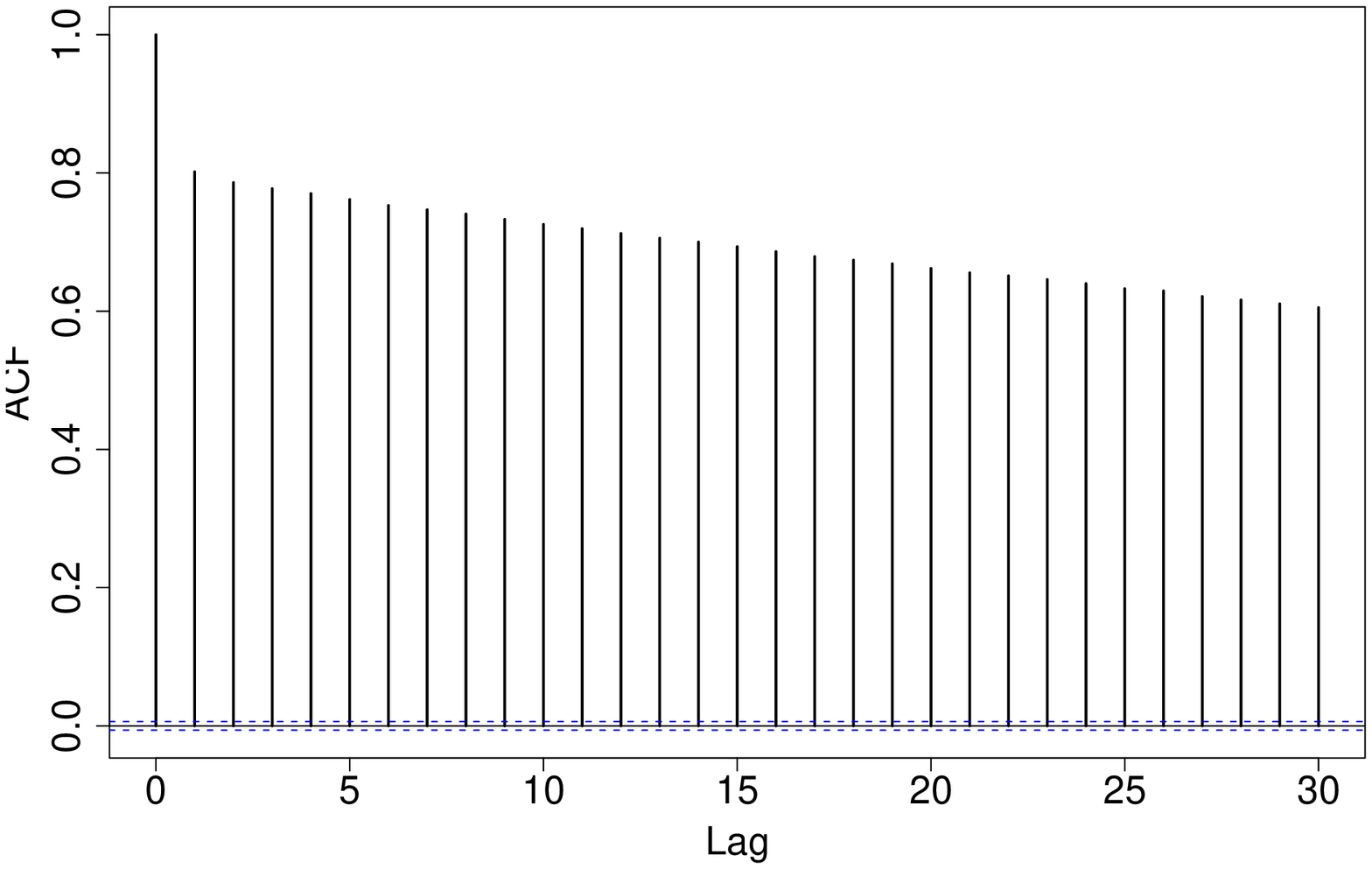}
    \end{minipage}}%
  \subfigure[$\beta_{13}$.]{
    \label{fig:acf-rebeta:13}
    \begin{minipage}[b]{0.24\textwidth}
      \centering \includegraphics[width=\textwidth]{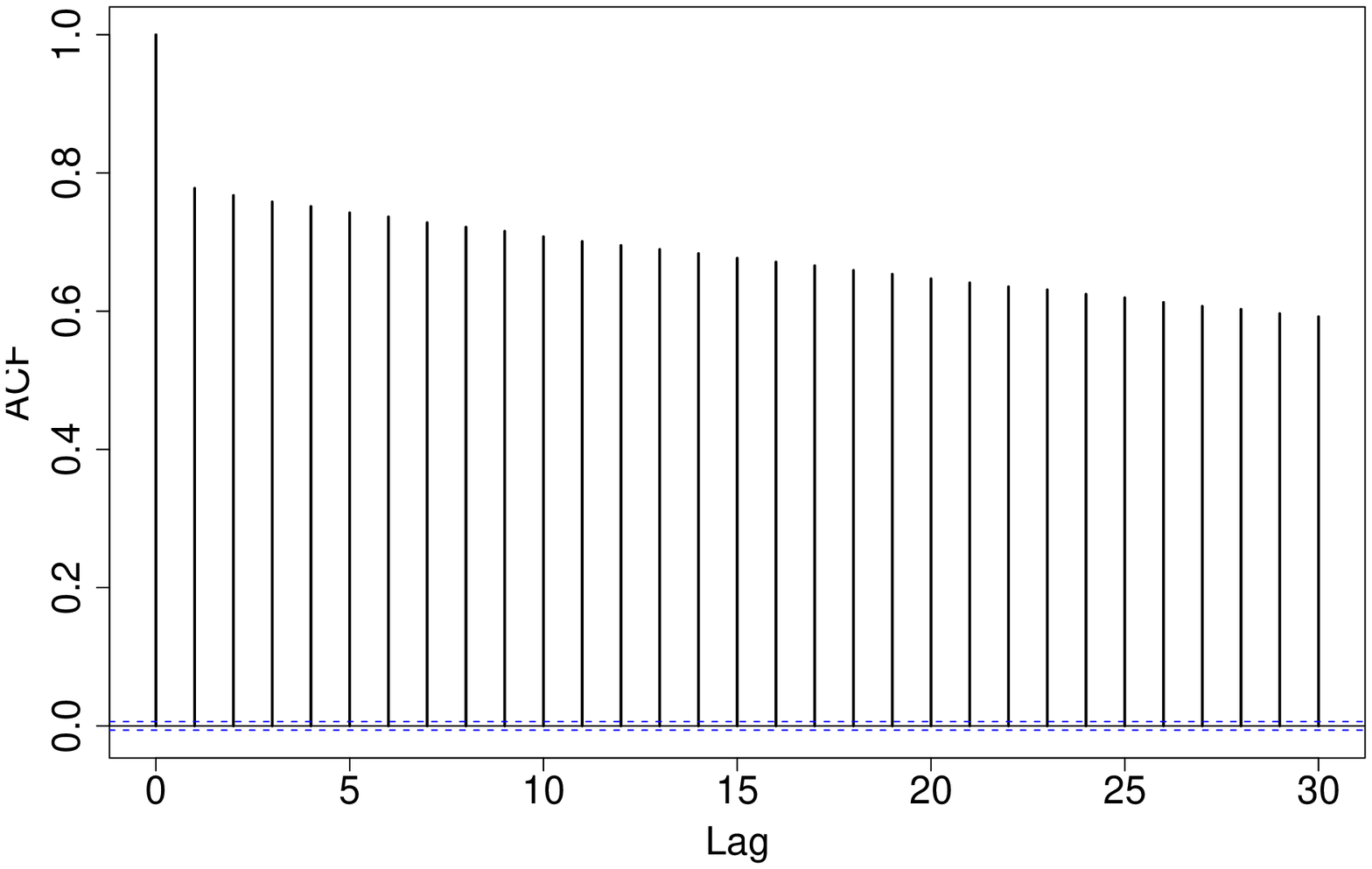}
    \end{minipage}}\\
  \subfigure[$\beta_{14}$.]{
    \label{fig:acf-rebeta:14}
    \begin{minipage}[b]{0.24\textwidth}
      \centering \includegraphics[width=\textwidth]{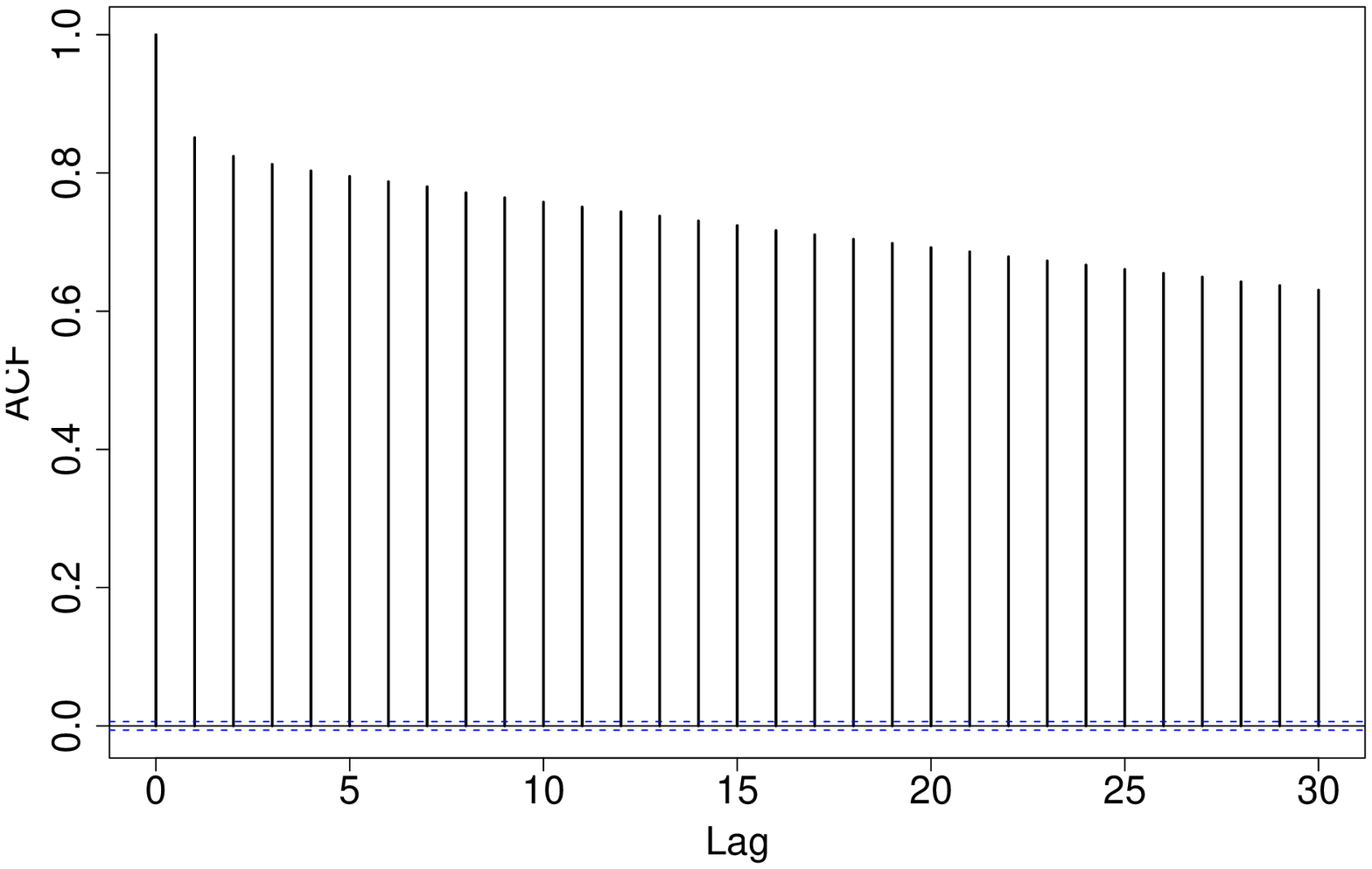}
    \end{minipage}}%
  \subfigure[$\beta_{15}$.]{
    \label{fig:acf-rebeta:15}
    \begin{minipage}[b]{0.24\textwidth}
      \centering \includegraphics[width=\textwidth]{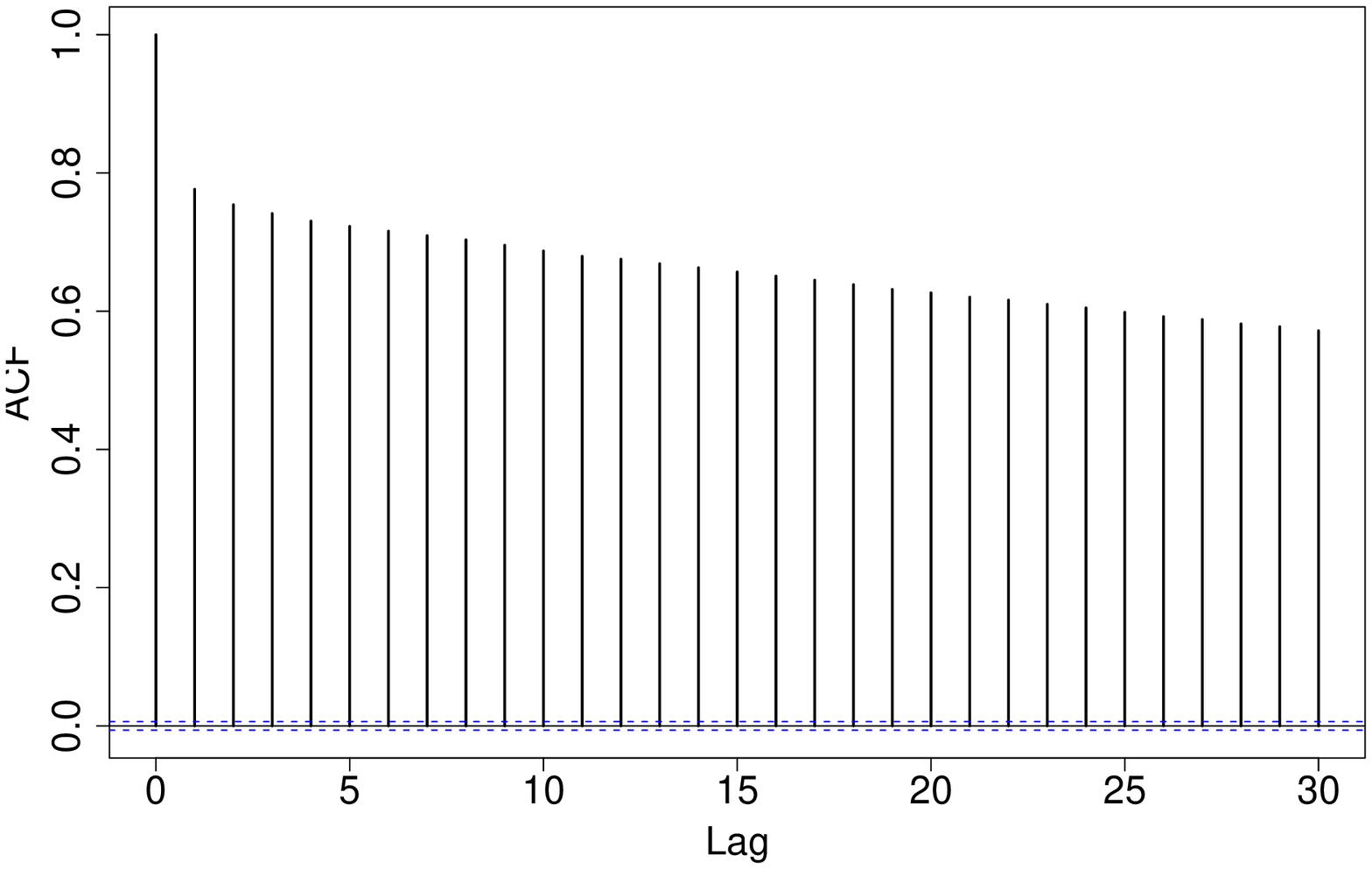}
    \end{minipage}}%
  \subfigure[$\beta_{16}$.]{
    \label{fig:acf-rebeta:16}
    \begin{minipage}[b]{0.24\textwidth}
      \centering \includegraphics[width=\textwidth]{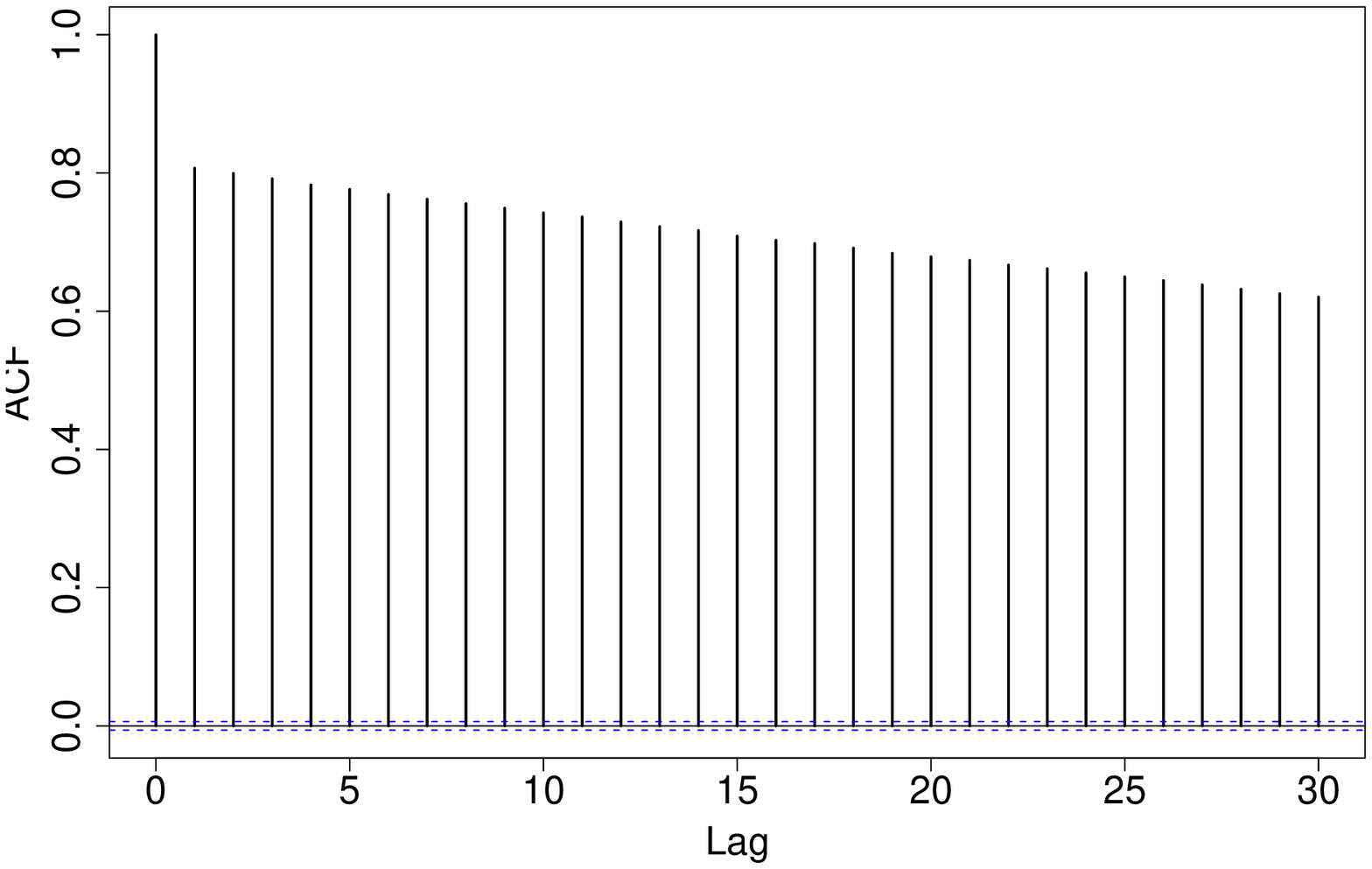}
    \end{minipage}}%
  \subfigure[$\beta_{17}$.]{
    \label{fig:acf-rebeta:17}
    \begin{minipage}[b]{0.24\textwidth}
      \centering \includegraphics[width=\textwidth]{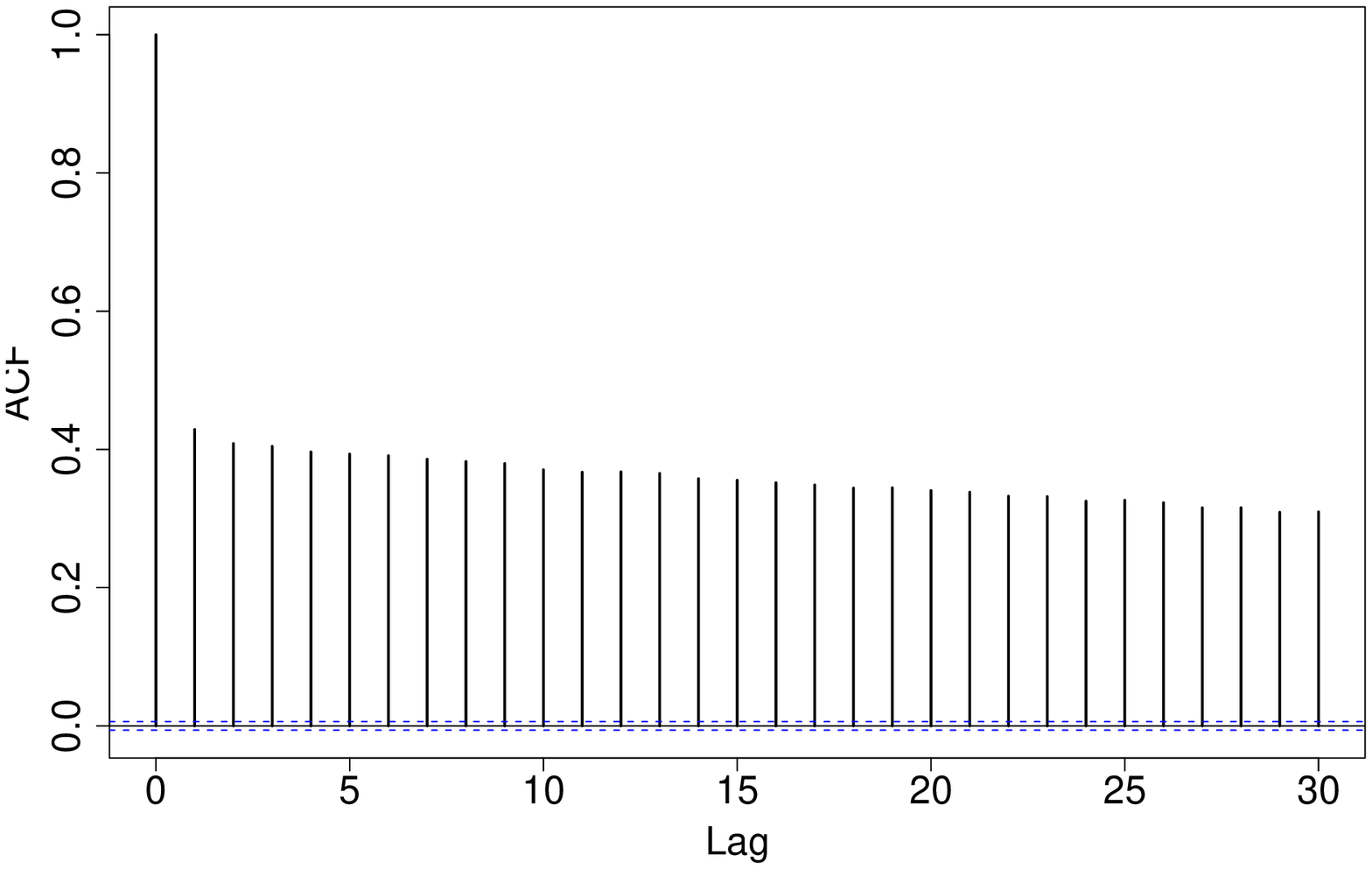}
    \end{minipage}}\\
  \subfigure[$\beta_{18}$.]{
    \label{fig:acf-rebeta:18}
    \begin{minipage}[b]{0.24\textwidth}
      \centering \includegraphics[width=\textwidth]{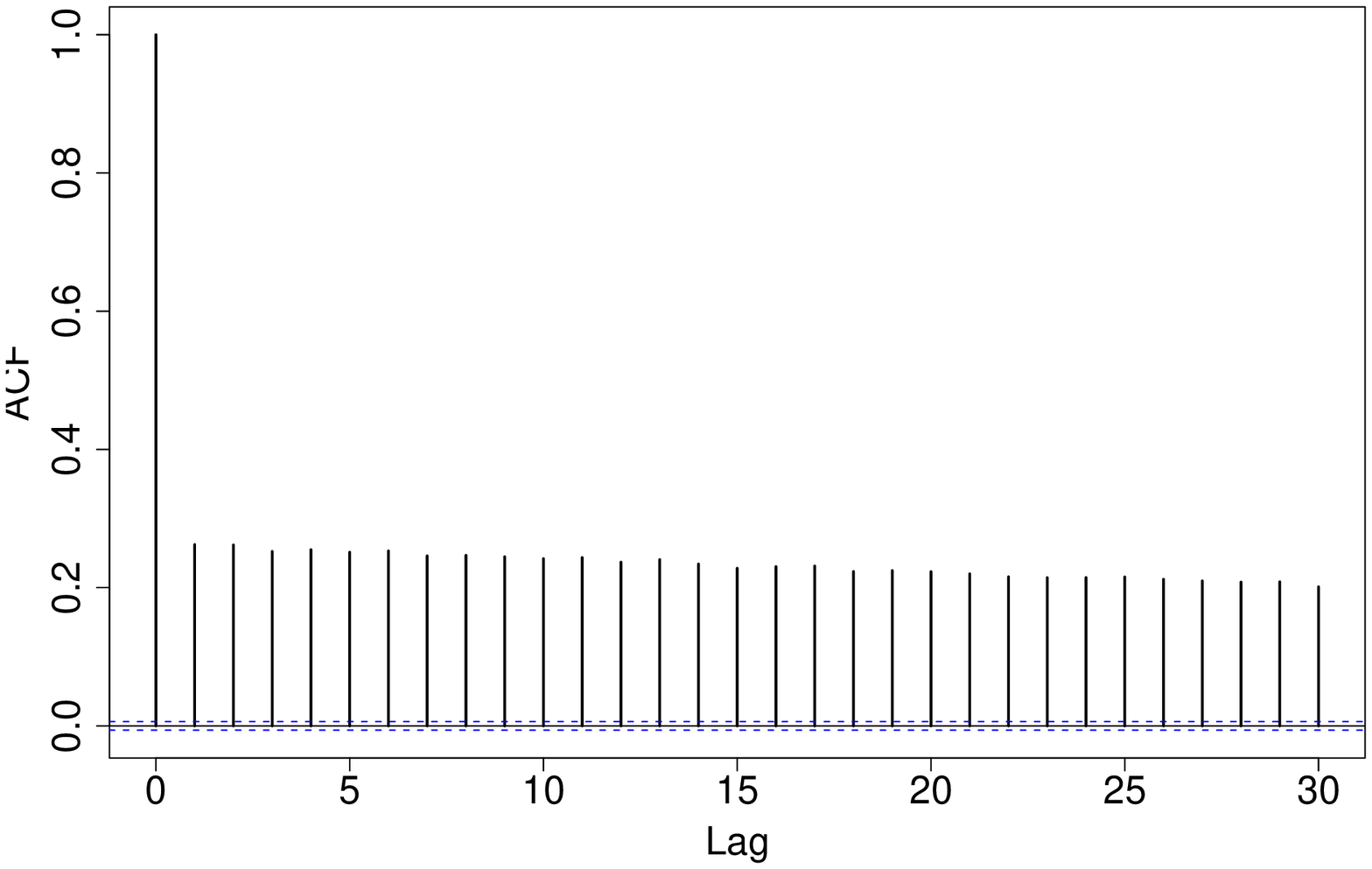}
    \end{minipage}}%
  \subfigure[$\beta_{19}$.]{
    \label{fig:acf-rebeta:19}
    \begin{minipage}[b]{0.24\textwidth}
      \centering \includegraphics[width=\textwidth]{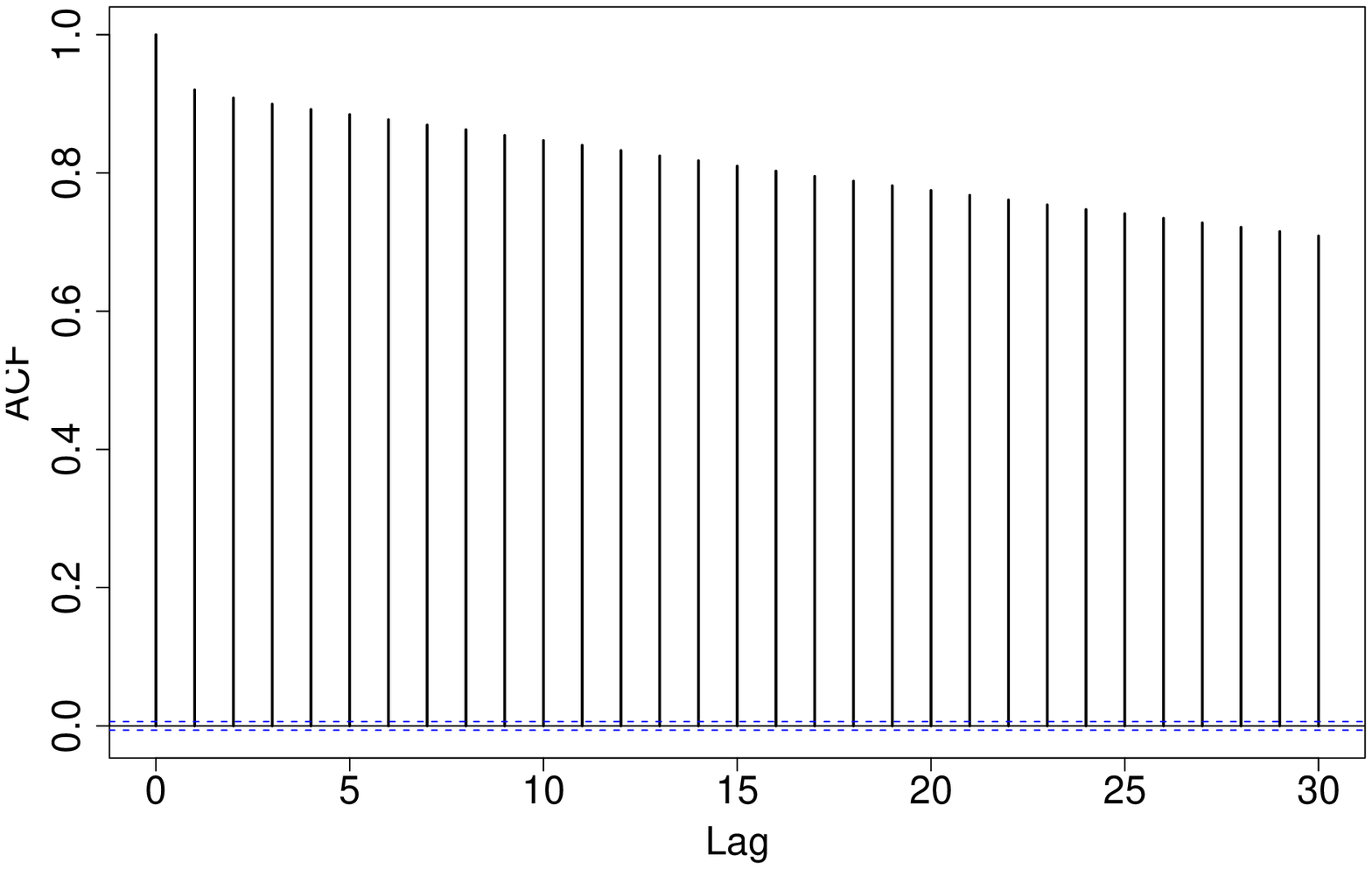}
    \end{minipage}}%
  \subfigure[$\beta_{20}$.]{
    \label{fig:acf-rebeta:20}
    \begin{minipage}[b]{0.24\textwidth}
      \centering \includegraphics[width=\textwidth]{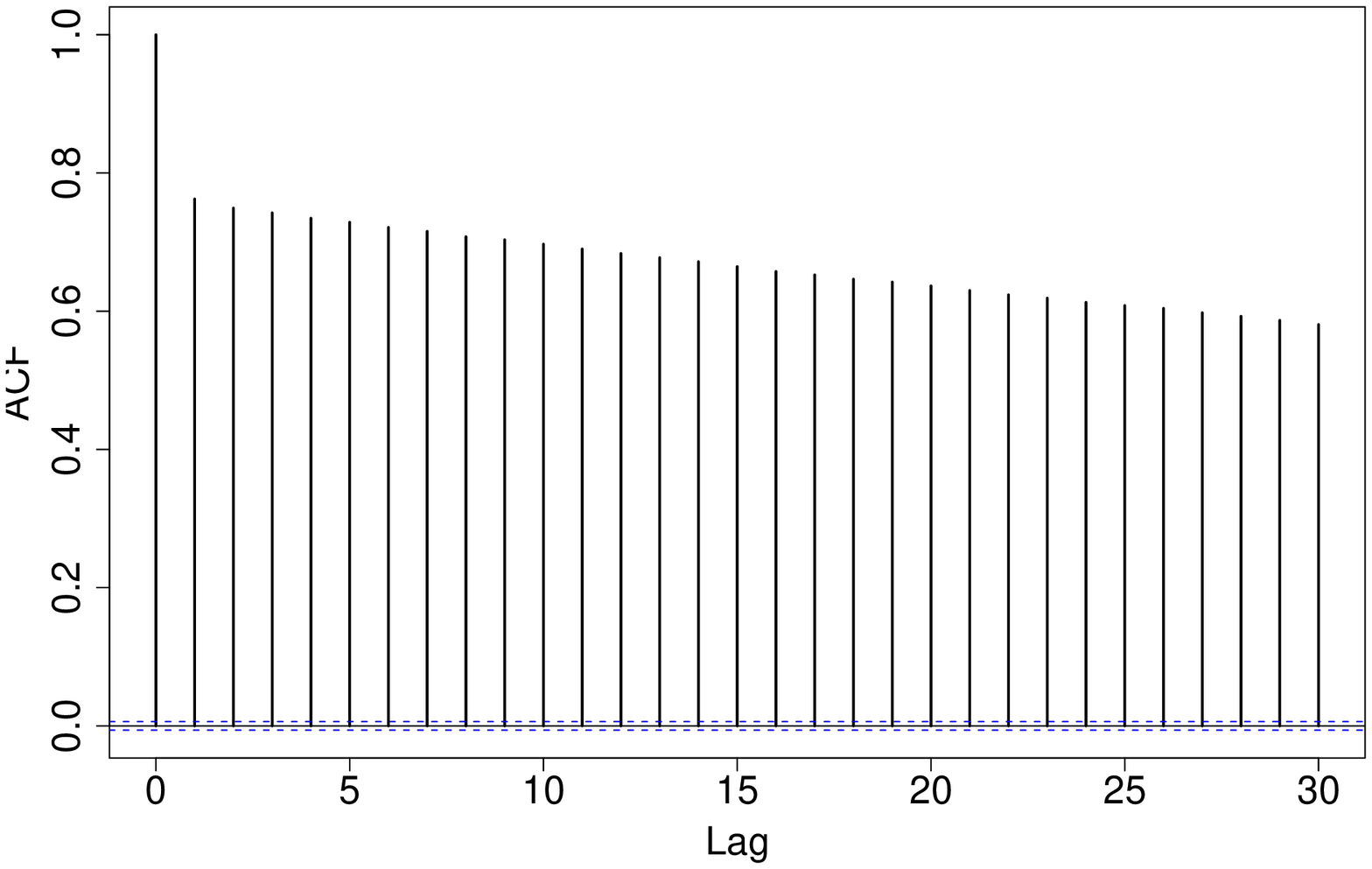}
    \end{minipage}}%
  \subfigure[$\beta_{21}$.]{
    \label{fig:acf-rebeta:21}
    \begin{minipage}[b]{0.24\textwidth}
      \centering \includegraphics[width=\textwidth]{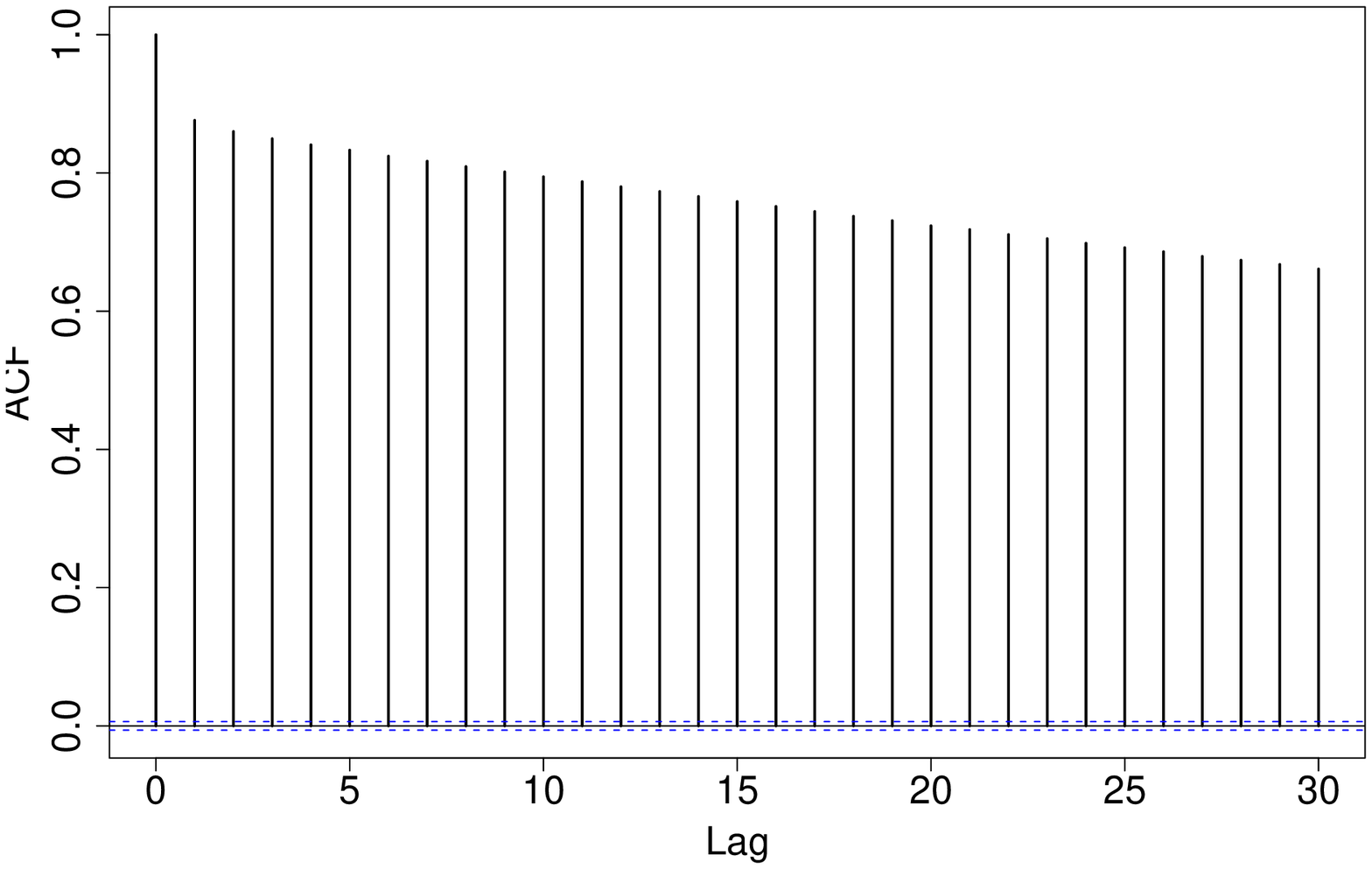}
    \end{minipage}}\\
  \subfigure[$\beta_{22}$.]{
    \label{fig:acf-rebeta:22}
    \begin{minipage}[b]{0.24\textwidth}
      \centering \includegraphics[width=\textwidth]{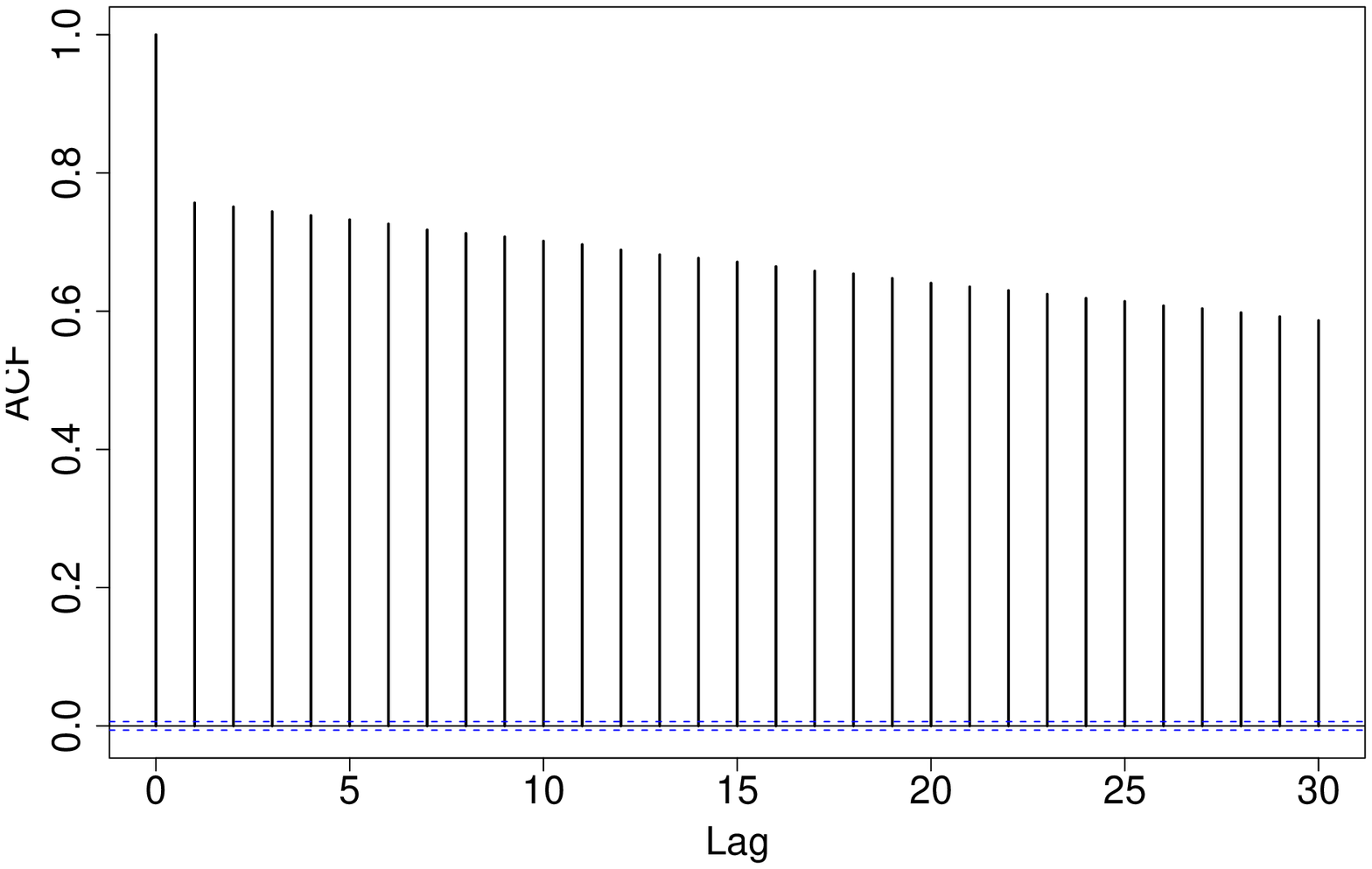}
    \end{minipage}}%
  \subfigure[$\beta_{23}$.]{
    \label{fig:acf-rebeta:23}
    \begin{minipage}[b]{0.24\textwidth}
      \centering \includegraphics[width=\textwidth]{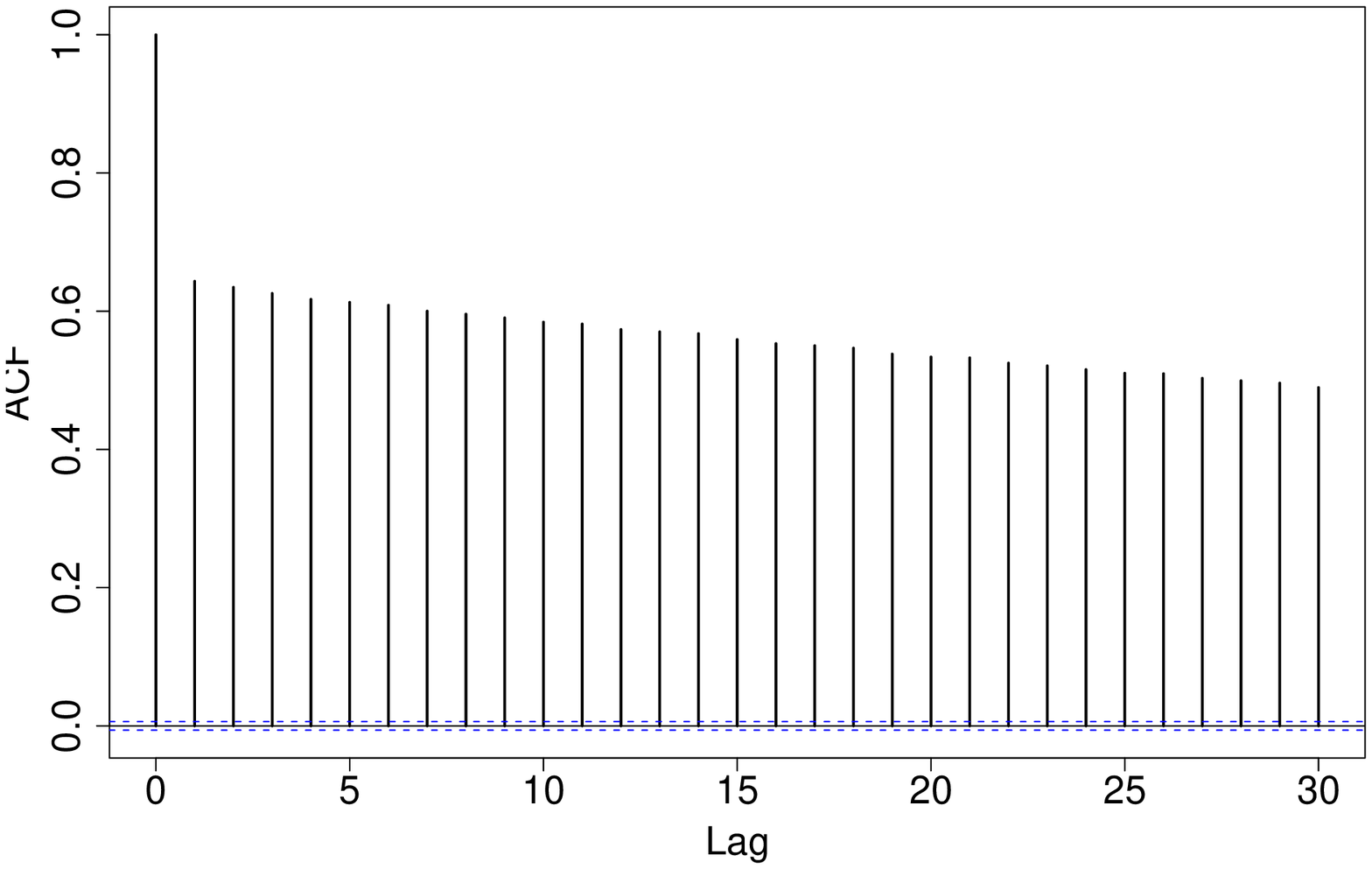}
    \end{minipage}}%
  \subfigure[$\beta_{24}$.]{
    \label{fig:acf-rebeta:24}
    \begin{minipage}[b]{0.24\textwidth}
      \centering \includegraphics[width=\textwidth]{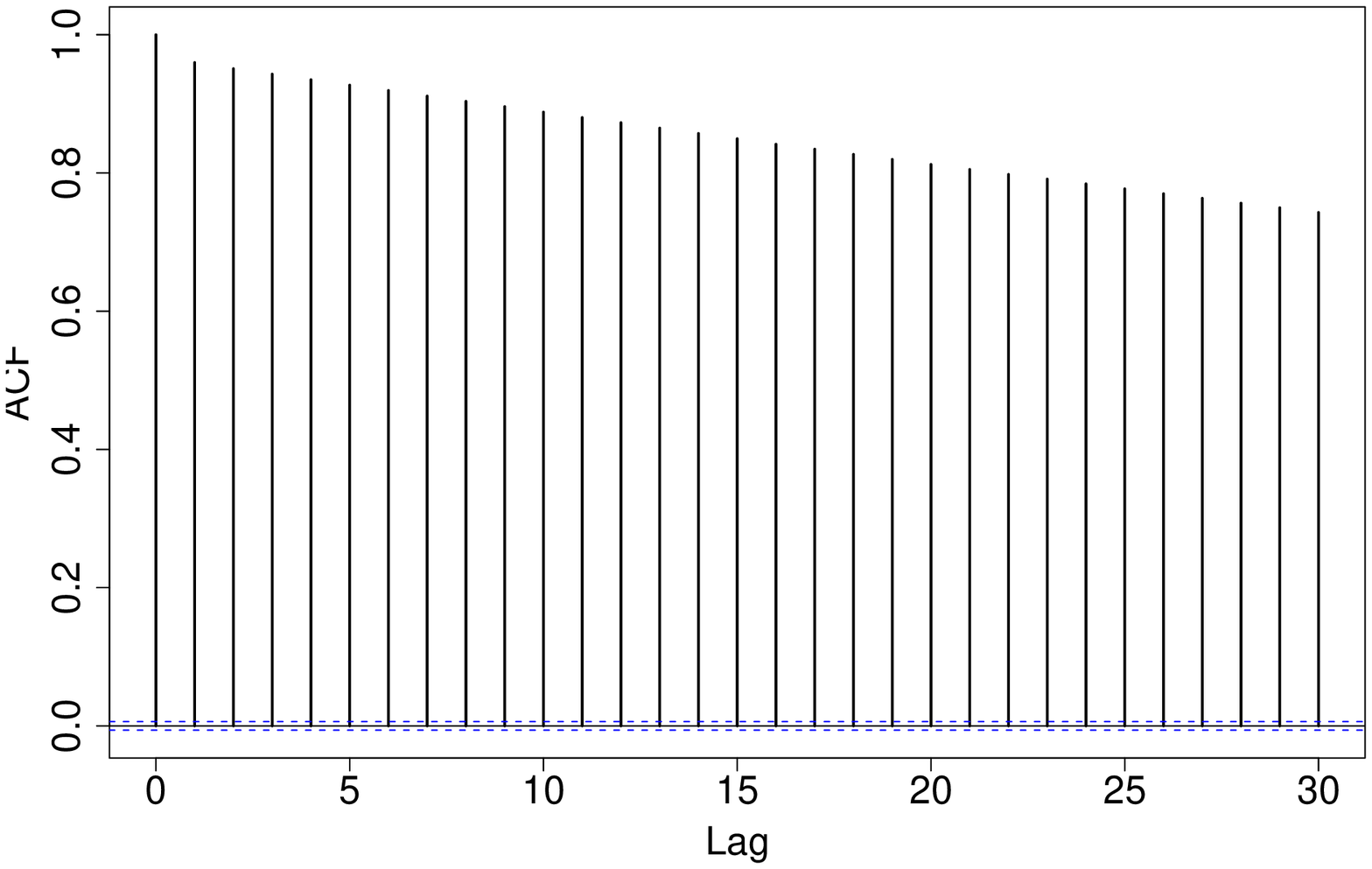}
    \end{minipage}}
  \caption{Autocorrelation plots for $\beta$.}
  \label{fig:acf-rebeta} 
\end{figure}

In the next section, we reparameterise the model given in
Section~\ref{sec:klugmanreview}. This reparameterisation results in improved
mixing for $\bfalpha$ and the results were marginally better for $\bfbeta$.
 Figures~\ref{fig:acf-klalpha} and \ref{fig:acf-klbeta} are the
autocorrelation plots for the first implementation,  while Figures
\ref{fig:acf-realpha} and \ref{fig:acf-rebeta} show the corresponding plots
for the reparameterised implementation.  The new parameterisation used is
the corner point constraint, where one of the state or occupation effects is
fixed at zero.  Without loss of generality, we assume that $\alpha_1$ and
$\beta_1$ are both identically 0. The reparameterised model is presented in
Section~\ref{sec:noncentred}.


We also investigate whether the state effects ($i$) are substantial and also
whether the occupation effects ($j$) are non--trivial. For this, we employ the
reversible jump algorithm.  In the absence of a natural jump function, we
propose to change all model parameters when moving between models. This is
not strictly necessary. However,  if parameter values change substantially
between models, then keeping such values fixed will result in high reject
rates at the accept/reject stage of the Reversible jump algorithm. We discuss
the reversible jump algorithm for this example in Section~\ref{sec:ds4-rj}.
\begin{table}[h]
  \centering
  \caption{Estimates of the model parameters, with 95\% HPD Intervals.}
  \label{tab:dset4results:alpha}
  \begin{tabular}{rrr}
    \hline\hline
    & mean & 95\% HPD Interval  \\
    \hline
    $\alpha_1$ &  0.0095   & (-0.0085, 0.0270)  \\
    $\alpha_2$ &  0.0492   &  (0.0310, 0.0676)  \\
    $\alpha_3$ &  0.0993   &  (0.0687, 0.1305)  \\
    $\alpha_4$ &  0.0310   &  (0.0134, 0.0487)  \\
    $\alpha_5$ &  0.0068   & (-0.0113, 0.0246) \\
    $\alpha_6$ &  0.0322   &  (0.0131, 0.0507) \\
    $\alpha_7$ &  0.0100   & (-0.0090, 0.0290) \\
    $\alpha_8$ &  0.1109   &  (0.0926, 0.1291) \\
    $\alpha_9$ &  0.0067   & (-0.0131, 0.0263) \\
    $\alpha_{10}$ &  0.0525&  (0.0312, 0.0738)\\
    \hline
  \end{tabular}\vskip1cm
\end{table}
\begin{table}[h]
  \centering
  \caption{Estimates of the model parameters, with 95\% HPD Intervals.}
  \label{tab:dset4results:beta}
  \begin{tabular}{rrr}
    \hline\hline
    & mean & 95\% HPD Interval  \\
    \hline
    $\beta_1$    & 0.0711 &   (0.0458, 0.0967) \\
    $\beta_2$    & 0.0792 &   (0.0545, 0.1042) \\  
    $\beta_3$    & 0.0206 &   (0.0009, 0.0403) \\ 
    $\beta_4$    &-0.0269 &  (-0.0667, 0.0131) \\
    $\beta_5$    & 0.0539 &   (0.0351, 0.0729)  \\ 
    $\beta_6$    & 0.1873 &   (0.1684, 0.2060) \\ 
    $\beta_7$    & 0.0924 &   (0.0606, 0.1243) \\ 
    $\beta_8$    & 0.0532 &   (0.0246, 0.0822) \\ 
    $\beta_9$    & 0.0120 &  (-0.0074, 0.0317) \\ 
    $\beta_{10}$ & 0.0360 &   (0.0160, 0.0561) \\ 
    $\beta_{11}$ & 0.0206 &   (0.0026, 0.0387) \\ 
    $\beta_{12}$ & 0.0308 &   (0.0111, 0.0503) \\ 
    $\beta_{13}$ & 0.0392 &   (0.0194, 0.0592) \\ 
    $\beta_{14}$ & 0.0515 &   (0.0326, 0.0708) \\ 
    $\beta_{15}$ & 0.0816 &   (0.0617, 0.1017) \\ 
    $\beta_{16}$ & 0.0306 &   (0.0114, 0.0503) \\ 
    $\beta_{17}$ & 0.0222 &  (-0.0051, 0.0494) \\
    $\beta_{18}$ & 0.0178 &  (-0.0162, 0.0516)  \\
    $\beta_{19}$ & 0.0256 &   (0.0074, 0.0440) \\
    $\beta_{20}$ & 0.0143 &  (-0.0058, 0.0346) \\ 
    $\beta_{21}$ & 0.0307 &   (0.0119, 0.0494) \\ 
    $\beta_{22}$ & 0.0034 &  (-0.0170, 0.0237) \\
    $\beta_{23}$ & 0.0357 &   (0.0137, 0.0578) \\
    $\beta_{24}$ &-0.0003 &  (-0.0178, 0.0176)\\
    \hline
  \end{tabular}
\end{table}

\section{Reparameterisation Issues}\label{sec:reparam}
We reparameterise the model to allow for more freedom of the first level
parameters. In model ~\eqref{eq:dset4first}, both $\alpha_i$ and $\beta_j$ are
restricted to have mean $\mu/2$. There is, however, no direct interpretation
of the model parameters,  given this parameterisation

In addition, we reparameterise the model so that each state effect and occupation effect
are compared with the first level. For this, we set both $\alpha_1$ and
$\beta_1$ equal to zero. This is the well known corner point
constraint~\cite[Section 6.2]{venables1999}. The reparameterised model is
\begin{equation}\label{eq:reparam}
  R_{ijt}| \mu,\bfalpha,\bfbeta,\sigma \sim
  \dnorm{\mu+\alpha_i+\beta_j}{(\sigma E_{ijt})^{-1}},\quad 
  \alpha_1=0,\,\, \beta_1=0.
\end{equation}
We use multivariate normal priors for $\bfalpha=(\alpha_2,\ldots,\alpha_m)'$
and $\bfbeta=(\beta_2,\ldots,\beta_n)'$. The derivation of the posterior
conditional distributions are given in the next section.

The reparameterisation does not affect the final interpretation of the model.
It could, however, affect the convergence rate of the Gibbs sampler. Problems
such as these are considered by ~\shortciteN{papaspilioppoulos2003} who show
that the centered parameterisation is not uniformly superior to the
non--centered parameterisation, and indeed that a partially centered
parameterisation might give the fastest convergence rate of the Gibbs
sampler.  We also note that the autocorrelations have decreased because of
the parameterisations chosen. Generally the autocorrelation for
hierarchical models depends on the parameterisation chosen. See for example
\citeN{gilks1996:mcmcip}, \shortciteN{gelfand1995}, and \citeN{gelfand1996}.

\subsection{Using Non--Centred Block Updates} \label{sec:noncentred}
Let $\bfalpha=(\alpha_2,\ldots,\alpha_m)'$,
$\bfbeta=(\beta_2,\ldots,\beta_n)'$ and, as before, let $\bfR=\{R_{ijt}\}$
denote the loss ratios. The first level is described as
\begin{equation*}
  R_{ijt}\sim \dnorm{\mu+\alpha_i+\beta_j}{(\sigma E_{ijt})^{-1}}, \quad
  \alpha_1=0,\,\,\, \beta_1 = 0.
\end{equation*}
Given the first above, we  choose the following prior
distributions for the model parameters
\begin{equation*}
  \mu\sim\dnorm{0}{\tau_{\mu}^{-1}},\;
  \bfalpha\sim\dnorm{\bfzero}{\tau_{\alpha}^{-1}\bfI}, \;
  \bfbeta\sim\dnorm{\bfzero}{\tau_{\beta}^{-1}\bfI}, \text{ and }
  \sigma\sim\dgamma{a}{b},
\end{equation*}
where $\tau_{\mu}=\tau_{\alpha}=\tau_{\beta}=0.001$, and $a=b=0.001$, so that
these prior distributions are vague and flat.  The law of
$(\bfalpha,\bfbeta,\mu,\sigma)$, given the data is
\begin{equation*}
  \pi(\mu,\bfalpha,\bfbeta,\sigma|\bfR)\propto
  p(\sigma) p(\mu)p(\bfalpha)p(\bfbeta)\bfL(\bfR|\mu,\bfalpha,\bfbeta,\sigma).
\end{equation*}
The updating scheme used is the deterministic updating strategy with
components $(\sigma$, $\mu$, $\bfalpha$, $\bfbeta)$.  Other updating schemes
are possible such as grouping two or more of the parameters given above
\cite{roberts1997}.  An initial implementation using a random walk metropolis
algorithm, which updates all model parameters at once, using a covariance matrix
estimated from a trial run,
did not perform well. The acceptance rate for that
algorithm was $0.857\%$ and the autocorrelations were large even at lags
greater than $50$.

\subsection{Posterior Conditionals}
The posterior conditional for $\mu$ is
\begin{equation*}
  \pi(\mu|\bfalpha, \bfbeta, \sigma) \propto 
  p(\mu) \likelihood{\bfR|\mu,\bfalpha,\bfbeta,\sigma}.
\end{equation*}
We determine that $\mu$ has a normal distribution with mean
\begin{equation*}
\left(\tau_{\mu} + \sigma\sum_{ijt}E_{ijt}\right)^{-1}
\left(\sigma\sum_{ijt}E_{ijt}(R_{ijt}-\alpha_i-\beta_j)\right),
\end{equation*}
and variance 
\begin{equation*}
\left(\tau_{\mu} + \sigma\sum_{ijt}E_{ijt}\right)^{-1}.
\end{equation*}
The posterior conditional of $\bfalpha$ is
\begin{align*}
  \pi(\bfalpha|\mu, \bfbeta,\sigma) &\propto
  p(\bfalpha) \likelihood{\bfR|\mu,\bfalpha,\bfbeta,\sigma} \\
  &\propto\exp\left\{-\tfrac{\tau_{\alpha}}{2}\sum_{i=2}^m\alpha_i^2\right\}
  \exp\left\{
    -\tfrac{\sigma}{2}\sum_{ijt}E_{ijt}(R_{ijt} - \mu-\alpha_i-\beta_j)^2
  \right\} .
\end{align*}
After simplification, we observe that the posterior conditional of $\bfalpha$,
given the other model parameters, is  multivariate normal with mean
vector given by
\begin{equation*}
  \begin{pmatrix}
    \tau_{\alpha} + \sigma\sum_{jt}E_{2jt} & 0 &\dots & 0\\
    0 & \tau_{\alpha} + \sigma\sum_{jt} E_{3jt} &\dots& 0\\
    \hdotsfor[2]{4}\\
    0 & 0 &\dots& \tau_{\alpha} + \sigma\sum_{jt}E_{m,jt}
\end{pmatrix}^{-1}
\begin{pmatrix}
  \sigma\sum_{jt} E_{2jt}(R_{2jt} - \mu-\beta_j) \\
  \sigma\sum_{jt} E_{3jt}(R_{3jt} - \mu -\beta_j)\\
  \vdots \\
  \sigma\sum_{jt} E_{m,jt}(R_{m,jt} - \mu -\beta_j)\\
\end{pmatrix}
\end{equation*}
and variance matrix
\begin{equation*}
  \begin{pmatrix}
    \tau_{\alpha} + \sigma\sum_{jt}E_{2jt} & 0 &\dots & 0\\
    0 & \tau_{\alpha} + \sigma\sum_{jt} E_{3jt} &\dots& 0\\
    \hdotsfor[2]{4}\\
    0 & 0 &\dots& \tau_{\alpha} + \sigma\sum_{jt}E_{m,jt}
\end{pmatrix}^{-1}
\end{equation*}
The posterior conditional distribution for $\bfbeta$ is given by
\begin{align*}
  \pi(\bfbeta|\mu, \bfalpha, \sigma) &\propto p(\bfbeta)
  \likelihood{\bfR|\mu,\bfalpha,\bfbeta,\sigma}\\
  &\propto \exp\left\{-\tfrac{\tau_{\beta}}{2}\sum_{j=2}^n\beta_j \right\}
  \exp\left\{ 
    -\tfrac{\sigma}{2}\sum_{ijt} E_{ijt}(R_{ijt}-\mu-\alpha_i-\beta_j)^2
  \right\}
\end{align*}
from which we determine that $\bfbeta$ follows a multivariate normal
distribution with mean vector
 \begin{equation*}
   \begin{pmatrix}
     \tau_{\beta}+ \sigma\sum_{it}E_{i2t} & 0 &\dots & 0\\
     0 & \tau_{\beta}+ \sigma\sum_{it} E_{i3t} &\dots& 0\\
     \hdotsfor[2]{4}\\
     0 & 0 &\dots& \tau_{\beta}+ \sigma\sum_{it}E_{i,n,t}
   \end{pmatrix}^{-1}
 \begin{pmatrix}
 \sigma  \sum_{it} E_{i2t} (R_{i2t} - \mu - \alpha_i) \\
 \sigma  \sum_{it} E_{i3t} (R_{i3t} - \mu - \alpha_i) \\
   \vdots \\
 \sigma  \sum_{it} E_{i,n,t} (R_{i,n,t} - \mu - \alpha_i) \\
 \end{pmatrix}
 \end{equation*}
and variance matrix
 \begin{equation*}
   \begin{pmatrix}
     \tau_{\beta}+ \sigma\sum_{it}E_{i2t} & 0 &\dots & 0\\
     0 & \tau_{\beta}+ \sigma\sum_{it} E_{i3t} &\dots& 0\\
     \hdotsfor[2]{4}\\
     0 & 0 &\dots& \tau_{\beta}+ \sigma\sum_{it}E_{i,n,t}
   \end{pmatrix}^{-1}
\end{equation*}
The precision parameter $\sigma$,  has posterior conditional distribution which
is a gamma distribution with shape and scale parameters, respectively, given by
\begin{equation*}
  a+\tfrac{mns}{2} \text{ and }
  b + \tfrac1{2}\sum_{ijt}E_{ijt}(R_{ijt} - \mu-\alpha_i-\beta_j )^2.
\end{equation*}

\subsection{Results and Model Interpretation}
We implemented the Gibbs algorithm for the model described in Equation
\eqref{eq:reparam} using the conditional distributions derived above. The
resulting mean of the posterior parameter distributions are given in
Tables~\ref{tab:ds4fullresults:alpha} \,and ~\ref{tab:ds4fullresults:beta}.
Posterior 95\% HPD intervals for each parameter are also given.  Recalling
that the quantity of interest is $\mu + \alpha_i + \beta_j$, the negative values
of some of the parameters are irrelevant, as these are off-set by the
value of $\mu$.  Therefore, the $\alpha$ values are all relative to $\alpha_1$
which was fixed at 0. If instead, we fixed $\alpha_1$ at some other value, the
other $\alpha$ values would have compensated for this by changing. In
particular, if we fix $\alpha_1$ at the minimum observed value of $0.0035$
($\alpha_9$) then all the $\alpha$'s would be positive. Likewise, fixing
$\alpha_1$ at $0.1015$ ($\alpha_8$) would result in all other $\alpha$'s
being negative. In each case ,$\mu$ would also change so that $\mu + \alpha_i
+ \beta_j$ remains constant. A similar discussion holds for the $\beta_j$
values observed.

In later sections, we allow for models which try to explain the data using
only the assumptions of dependence on state (index $i$) or on occupation
(index $j$) only. Our results will show that such models are unlikely to give
adequate description of the underlying process generating the data.

\begin{table}[h]
  \centering
  \caption{Parameter Estimates for the Non centred Parameterisation.}
  \label{tab:ds4fullresults:alpha}
  \begin{tabular}{rrr}
    \hline\hline
    Parameter & Posterior mean & 95\% HPD Interval\\
    \hline
    $\alpha_2$&   0.0394&  (0.0316,   0.0473) \\
    $\alpha_3$&   0.0966&  (0.0680,   0.1245) \\
    $\alpha_4$&   0.0215&  (0.0161,   0.0268) \\
    $\alpha_5$&  -0.0027&  (-0.0099,  0.0041) \\
    $\alpha_6$&   0.0223&  (0.0131,   0.0314) \\
    $\alpha_7$&   0.0001&  (-0.0095,  0.0099) \\
    $\alpha_8$&   0.1015&  (0.0932,   0.1093) \\
    $\alpha_9$&  -0.0035&  (-0.0149,   0.0076)\\
    $\alpha_{10}$ &0.0430& (0.0284,   0.0571) \\
    \hline
\end{tabular}
\end{table}

\begin{table}[h]
  \centering
  \caption{Parameter Estimates for the Non centred parameterisation.}
  \label{tab:ds4fullresults:beta}
  \begin{tabular}{rrr}
    \hline\hline
    Parameter & Posterior mean & 95\% HPD Interval\\
    \hline
    $\beta_2$&    0.0083& (-0.0186  0.0361) \\
    $\beta_3$&   -0.0524& (-0.0748  -0.0302) \\
    $\beta_4$&   -0.1140& (-0.1583  -0.0702) \\
    $\beta_5$&   -0.0186& (-0.0397   0.0031) \\
    $\beta_6$&    0.1155& ( 0.0942   0.1366) \\
    $\beta_7$&    0.0256& (-0.0089   0.0610) \\
    $\beta_8$&   -0.0179& (-0.0503   0.0135) \\
    $\beta_9$&   -0.0610& (-0.0833  -0.0387) \\
    $\beta_{10}$ &-0.0367& (-0.0592  -0.0140) \\
    $\beta_{11}$ &-0.0522& (-0.0732  -0.0317) \\
    $\beta_{12}$ &-0.0419& (-0.0647  -0.0202) \\
        $\beta_{13}$ &   -0.0333& (-0.0559  -0.0110) \\
        $\beta_{14}$ &   -0.0208& (-0.0430   0.0005) \\
        $\beta_{15}$ &   0.0094& (-0.0128     0.0323) \\
        $\beta_{16}$ &   -0.0421& (-0.0645  -0.0202) \\
        $\beta_{17}$ &   -0.0518& (-0.0823  -0.0227) \\
        $\beta_{18}$ &   -0.0581& (-0.0956  -0.0205) \\
        $\beta_{19}$ &   -0.0471& (-0.0677  -0.0258) \\
        $\beta_{20}$ &   -0.0587& (-0.0813  -0.0356) \\
        $\beta_{21}$ &   -0.0420& (-0.0636  -0.0207) \\
        $\beta_{22}$ &   -0.0699& (-0.0931  -0.0469) \\
        $\beta_{23}$ &   -0.0371& (-0.0615  -0.0122) \\
        $\beta_{24}$ &   -0.0731& (-0.0935  -0.0525) \\
    \hline
  \end{tabular}
\end{table}


\begin{figure}
   \centering
   \includegraphics[width=0.8\textwidth]{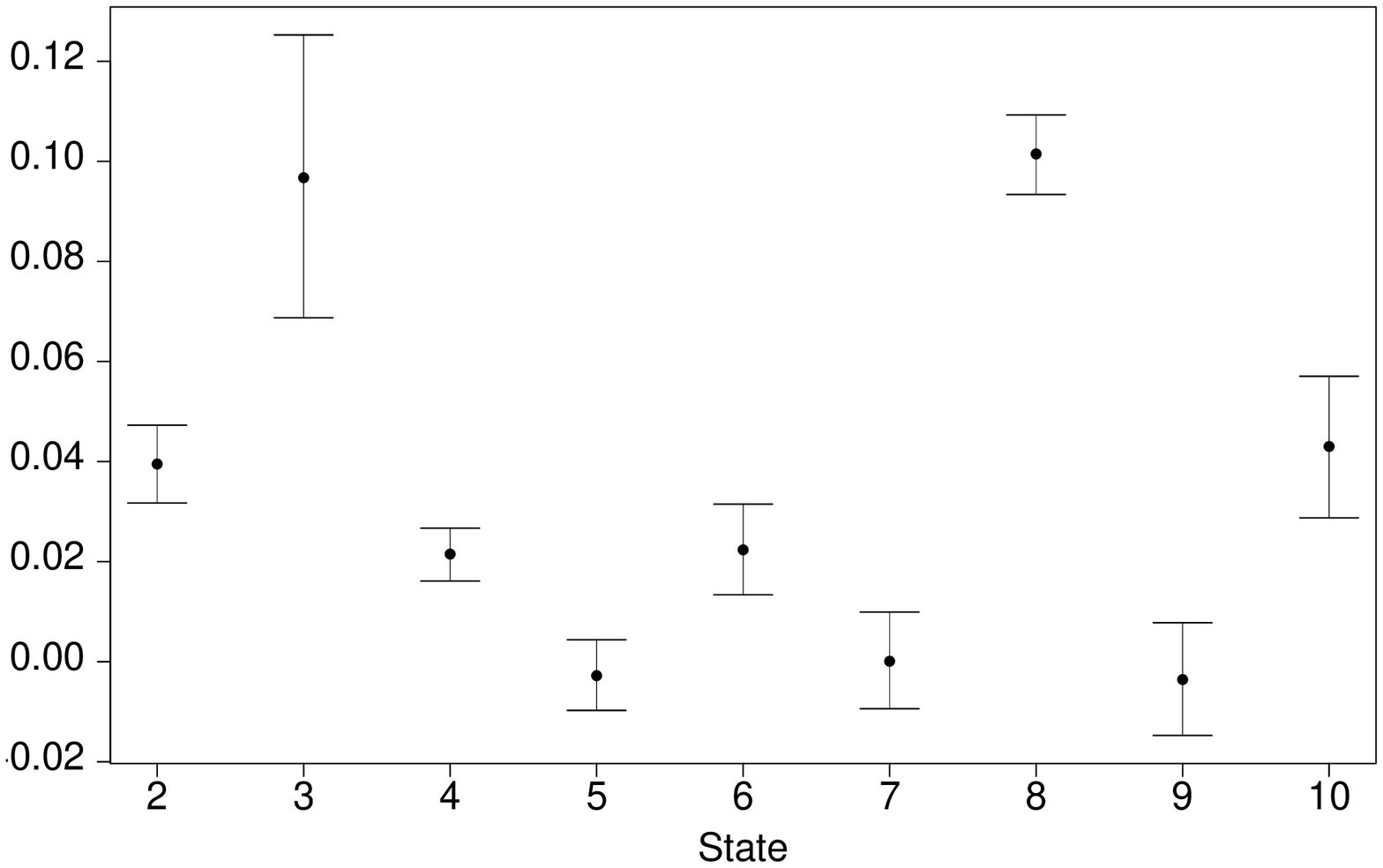}
   \caption{Boxplots on the State effects $\bfalpha$.}
   \label{fig:ds4-realphaboxp}
 \end{figure}

\begin{figure}
   \centering
   \includegraphics[width=0.8\textwidth]{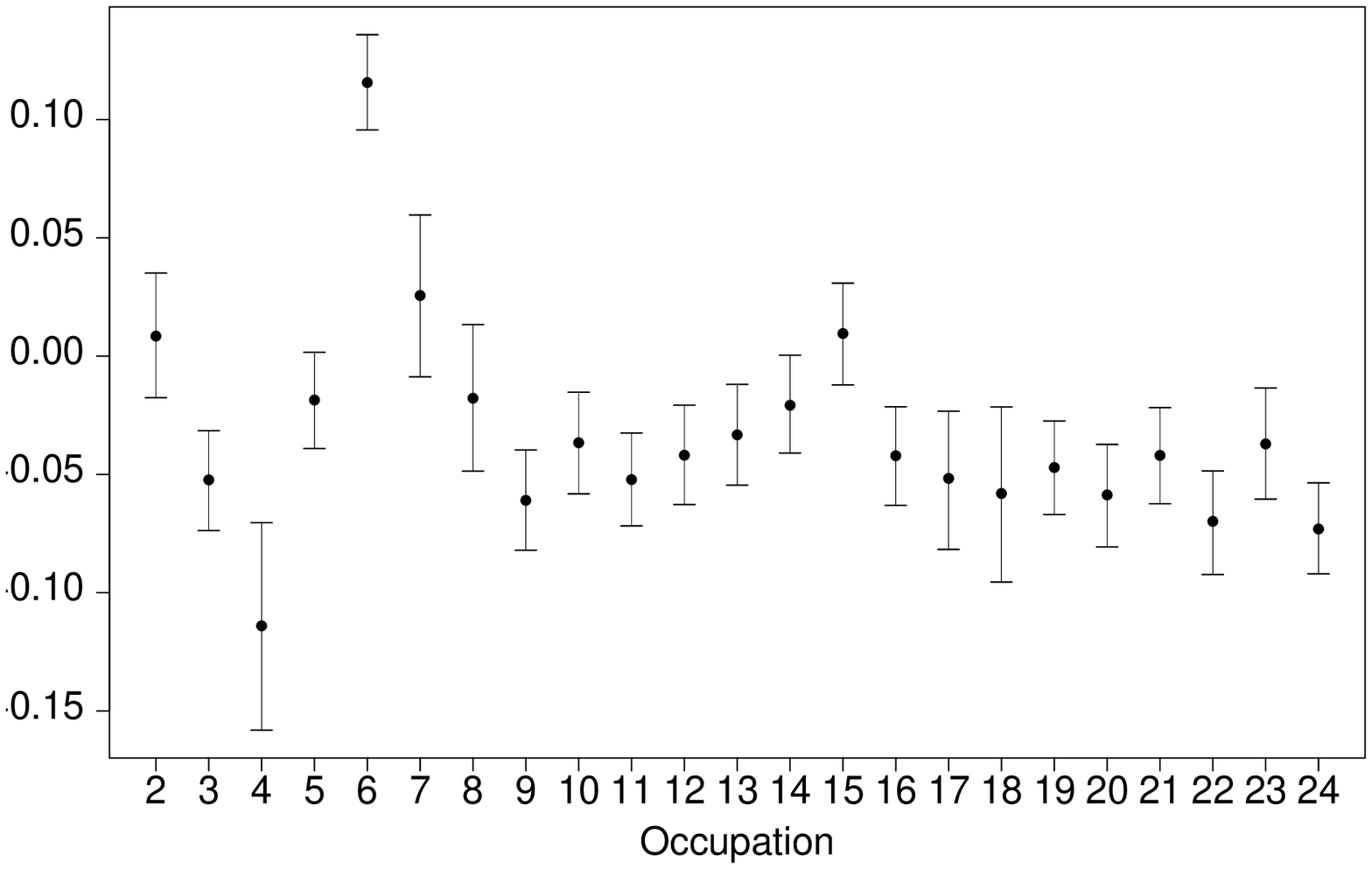}
   \caption{Boxplots on the Occupation effects $\bfbeta$.}
   \label{fig:ds4-rebetaboxp}
 \end{figure}

The lag--$k$ autocorrelation values are not directly comparable across the
models.  However, the smaller the absolute values of the correlation, the
better the chain is mixing. A comparative error plots of the state and
occupation effects are shown in Figures~\ref{fig:ds4-realphaboxp} and
~\ref{fig:ds4-rebetaboxp}.

\section{Model Discrimination}
This section introduces other models which are alternative models for
explaining the data. We then compare these one--way models with the two--way
model using the reversible jump method. The implementation of the reversible
jump algorithm will be based on the centering and scaling proposals of
~\shortciteN{brooks2003}, with some modifications.  The reversible jump
algorithm was introduced in Section~\ref{sec:reversible} as a method of model
selection and discrimination.
Model $M_1$ is the full model considered previously and has first level
\begin{equation*}
  R_{ijt} |\mu,\bfalpha,\bfbeta, \sigma , \bfE \sim
  \dnorm{\mu + \alpha_i +\beta_j}{(\sigma E_{ijt})^{-1}},
  \quad \alpha_1=0,\,\,\,\beta_1=0.
\end{equation*}
Model $M_2$ seeks to explain the data based on the State only.  It has first
level description of the data given by
\begin{equation*}
  R_{ijt} |\mu',\bfalpha', \sigma' ,\bfE \sim
  \dnorm{\mu' + \alpha'_i}{(\sigma' E_{ijt})^{-1}},
  \quad \alpha'_1=0.
\end{equation*}
The third model, denoted by $M_3$, has first level
\begin{equation*}
  R_{ijt} |\bfmu'',\bfbeta'', \sigma'', \bfE \sim
  \dnorm{\mu'' + \beta''_j}{(\sigma'' E_{ijt})^{-1}},
  \quad \beta''_1=0,
\end{equation*}
where the explanatory variables now depend on the occupations, indexed by
$j$. Model $M_1$ is designed to test the hypothesis that both occupation and
class effects are observed in the loss data, while models $M_1$ and $M_2$ are
designed to test the hypotheses that one of these factors is missing from the
observed data. We could plausibly add a fourth model to test the absence of
both class and occupation effects.  However, for our analysis, we assume that
there is at least one of these effects present.

\subsection{DIC Results}
We computed the $DIC$ values for each model. The results are tabulated in
Table~\ref{tab:dset4reparamdic}. 
\begin{table}[h]
  \centering
  \caption{$DIC$ Results.}\label{tab:dset4reparamdic}
  \begin{tabular}{r|rrrr}
    \hline\hline
    Model& $\overline{D(\bftheta_k)}$ & $D(\bar{\bftheta}_k)$ & $p_D$ & $DIC$\\
    \hline
    $M_1$ & -2687.25 & -2721.19 & 33.94 & -2653.31 \\
    $M_2$ & -1643.03 & -1653.91 & 10.88 & -1632.15 \\ 
    $M_3$ & -2114.05 & -2138.95 & 24.90 & -2089.15 \\
    \hline
  \end{tabular}
\end{table}
The results are consistent with the number of parameters and the hierarchical
levels in each model. They show that by using a hierarchical approach we have
essentially lost a fraction of a parameter in each model. Since
$|\bftheta_1|=34$, $|\bftheta_2| =11$, and $|\bftheta_3|=24$ the fraction
lost is seen to be very small though.

\subsection{Reversible Jump using Automatic Proposals}\label{sec:ds4-rj}
In this section, we use the reversible jump algorithm to explore the
possibility that the data are generated by some process which depends on the
state only, or even the occupation class only.  To implement this, we now
introduce two additional models. Both models are actually sub--models of
the more general model introduced in Section~\ref{sec:reparam}. Proposing
$\mu$, $\bfalpha$, $\bfbeta$ is desired, since we can take correlations into
consideration. In all observed cases, the off diagonal elements are negative
reflecting the fact that when $\mu$ increases, for example, $\alpha_i$
decreases.

For between--model moves, we propose to update all parameters when we change
model, so that the acceptance rates are of the form given in
Section~\ref{sec:efficientprop}.  This should help with between--model moves
as proposed parameters will be close to their modal values when we change
model.  The posterior distributions for $(\mu, \bfalpha, \bfbeta)$, $(\mu',
\bfalpha')$, and $(\mu'', \bfbeta'')$ in models $M_1$, $M_2$ and $M_3$,
respectively, are not standard.  Using the efficient proposal scheme of
~\shortciteN{brooks2003}, we can find a Gaussian density which approximates
the posterior conditional for these parameters.  

Notice that in all the given
acceptance probabilities, the Jacobian term is 1 since we simulate new values
for all the model parameters. Given that we are now at model $M_i$, we propose
a move to model $M_j$ according to the probability matrix $(r_{ij})$. For our
implementation, we take $r_{ij}=\tfrac1{2},\,i\neq j$. The inverse variance
parameter $\sigma$ differs across each of the three models. We therefore
propose to change $\sigma$ when we change models by simulating new values
form its marginal posterior distribution.

An initial attempt using weak non-identifiable centering to derive the
parameters for the proposal density, did not work well.  The algorithm when
started in a particular model, remains in that model and does not explore the
entire model space $\mathcal{M} = \{M_1, M_2, M_3 \}$.  Further analysis
showed that by using non--identifiable centering, the proposed parameters are
usually in an area of very small posterior probability, and since there are
lots of data, the posterior density of the proposed model is very small
compared with the current model and hence a small acceptance probability
results.  Weak non-identifiable centering does not work well for this
example, perhaps because of the large number of parameters which are added
or removed at each iteration. 

Consequently, we instead propose the conditional
maximisation approach.  In the conditional maximisation approach, the
centering point is chosen close to the posterior mean or mode so that the
joint distribution is maximised.  The remaining parameters are then derived
using the $k^{th}$ order methods described earlier.  For implementation of the conditional
maximisation scheme, we
\begin{itemize}
\item Run each model and compute posterior mean/modes
\item Use these posterior estimates as the centering point in the $k^{th}$
  order scheme
\item The method does not use weak non-identifiability centering since
  likelihood of both smaller and larger model are not identical at the
  centering point.
\end{itemize}
\subsection{Moves Between Models $M_1$ and $M_2$}\label{sec:modelsM1andM2}
For moves between models $M_1$ and $M_2$ we consider the ratio $A_{21}$
defined as
\begin{equation*}
  A_{21}=
  \frac{\pi(\mu,\bfalpha,\bfbeta,\sigma)}{\pi(\mu',\bfalpha',\sigma')}
  \frac{p(M_1)}{p(M_2)}   
  \frac{r_{12}}{r_{21}}
  \frac{q(\mu',\bfalpha',\sigma')}{q(\mu,\bfalpha,\bfbeta,\sigma)}
\end{equation*}
\begin{equation*}
  = 
  \frac{\pi(\mu,\bfalpha,\bfbeta,\sigma)}{\pi(\mu',\bfalpha',\sigma')}
  \frac{p(M_1)}{p(M_2)}
  \frac{q(\mu',\bfalpha',\sigma')}
  {q(\mu,\bfalpha,\bfbeta,\sigma)},
\end{equation*}
when $r_{ij}=\tfrac1{2}$.
The posterior marginal distributions for $\sigma$ and $\sigma'$ are different
under both models. Therefore, setting $\sigma=\sigma'$ will not work very well
for between--model moves. We therefore simulate $\sigma$ from its posterior
marginal distribution then simulate $(\mu,\bfalpha,\bfbeta)$ from
$q(\mu,\bfalpha,\bfbeta|\sigma)$.  Likewise, for the density
$q(\mu', \bfalpha',\sigma')$, we first simulate $\sigma'$ from its posterior
marginal distribution, then simulate $(\mu',\bfalpha') \sim
q(\mu',\bfalpha'|\sigma')$ so that the acceptance term becomes
\begin{equation*} A_{21} =
  \frac{\pi(\mu,\bfalpha,\bfbeta,\sigma)}{\pi(\mu',\bfalpha',\sigma')}
  \frac{p(M_1)}{p(M_2)}
  \frac{q(\mu',\bfalpha'|\sigma') q(\sigma')}
  {q(\mu,\bfalpha,\bfbeta|\sigma)q(\sigma)}.
\end{equation*}
Taking logs, we have $\log A_{21}= $
\begin{multline}\label{eq:logacceptM1M2}
  \log \pi(\mu,\bfalpha,\bfbeta,\sigma) 
  + \log q(\mu',\bfalpha'|\sigma') 
  - \log \pi(\mu',\bfalpha',\sigma')
  - \log q(\mu,\bfalpha,\bfbeta|\sigma) 
  + K_{12},
\end{multline}
where $K_{12}$ contains terms not involving $(\mu,\bfalpha,\bfbeta)$ or
$(\mu', \bfalpha')$.  We now recall the updating scheme proposed earlier.
We begin by simulating new values for the precision parameters $\sigma$ and $\sigma'$
depending on the proposed move. If the move is of type $M_1\Rightarrow M_2$,
then we simulate a new value for $\sigma'$; otherwise, if the move is of type
$M_2\Rightarrow M_1$, a new value for $\sigma$ is simulated from the posterior
marginal density. Having simulated new values of the precision parameters, we
can then simulate new values for the model parameters depending on the type
of move. Using this strategy, we can then see that the expression on the right
of Equation~\eqref{eq:logacceptM1M2} can be decomposed into four distinct
components when deriving the proposal densities.  Since
\begin{equation*}
  A_{21} =
  \frac{\pi(\mu,\bfalpha,\bfbeta,\sigma)}{\pi(\mu',\bfalpha',\sigma')}
  \frac{p(M_1)}{p(M_2)}
  \frac{q(\mu',\bfalpha'|\sigma')q(\sigma')}
  {q(\mu,\bfalpha,\bfbeta|\sigma)q(\sigma)},
\end{equation*}
the proposal density parameters for moves involving model $M_1$, $\mu$,
$\bfalpha$ and $\bfbeta$, can be obtained from
\begin{equation*}
  \nabla^k
  \log \pi(\mu, \bfalpha,\bfbeta,\sigma) -
  \log q(\mu,\bfalpha,\bfbeta|\sigma) 
  \Biggl |_{(\tilde\mu,\tilde\bfalpha,\tilde\bfbeta)} =0,
\end{equation*}
where $\nabla = (\partial / \partial\mu, \partial / \partial\bfalpha,
\partial / \partial\bfbeta)'$. 
Substituting
\begin{multline*}
  \pi(\mu,\bfalpha,\bfbeta,\sigma) \propto \\
  \exp\left\{ -\frac{1}{2} \left(
      \tau_{\mu}\mu^2+\tau_{\alpha}\sum \alpha_i^2+\tau_{\beta}\sum\beta_j^2+
      \sigma\sum_{ijt} E_{ijt}\left(R_{ijt}-\mu-\alpha_i-\beta_j \right)^2
    \right)\right\} \times \\
  \sigma^{\frac{mns}{2}+a-1} \exp\{ -b\sigma\}
\end{multline*}
and by also assuming that $q(\mu,\bfalpha,\bfbeta|\sigma)$ is a Gaussian density with
variance matrix $\varSigma_1$ and mean vector $\bfm_1$, we  then derive
the following proposal density parameters.
\begin{equation*} \varSigma_1^{-1} = 
  \begin{pmatrix}
    \nabla_{\mu}^2 f & \nabla_{\mu,\bfalpha}^2 f & \nabla_{\mu, \bfbeta}^2 f\\
    & \nabla_{\bfalpha}^2 f & \nabla_{\bfalpha, \bfbeta}^2 f\\
    & &  \nabla_{\bfbeta}^2 f
  \end{pmatrix} .
\end{equation*}
The mean vector satisfies
\begin{equation*}
  \varSigma_1^{-1} \left( \begin{pmatrix}
      \tilde\mu \\ \widetilde\bfalpha \\ \widetilde\bfbeta
    \end{pmatrix} - \bfm_1 \right) = 
  \begin{pmatrix}
  \nabla_{\mu} f \\
  \nabla_{\bfalpha} f \\
  \nabla_{\bfbeta} f
  \end{pmatrix},
\end{equation*}
which can be solved to give
\begin{equation*}
  \bfm_1 = \begin{pmatrix}  
    \tilde\mu \\ 
    \widetilde\bfalpha \\ 
    \widetilde\bfbeta \end{pmatrix} -
  \varSigma_1 \begin{pmatrix}
    \nabla_{\mu} f\\
    \nabla_{\bfalpha} f\\
    \nabla_{\bfbeta} f
    \end{pmatrix},
\end{equation*}
where $\nabla_{\mu} = (\partial / \partial\mu)$, $\nabla_{\bfalpha} =
(\partial/\partial\bfalpha)$, $\nabla_{\bfbeta}=(\partial/\partial\bfbeta)$,
and
\begin{equation*}
  f = 
    \tau_{\mu} \mu^2 + 
    \tau_{\alpha} \sum_{i=2}^m \alpha_i^2 +
    \tau_{\beta} \sum_{j=2}^n \beta_j^2 +
    \sigma   \sum_{ijt}E_{ijt}(R_{ijt} - \mu-\alpha_i-\beta_j) ^2
    .
\end{equation*}

Using Equation~\eqref{eq:logacceptM1M2},  we can also derive the parameters for
the proposal density for $(\mu', \bfalpha')$ for moves involving model $M_2$.
To derive the proposal parameters for jumps involving model $M_2$, we consider
only the terms involving $\mu'$ and $\bfalpha'$. 
Thus, we solve
\begin{equation*}
  \nabla^k
  \log \pi(\mu', \bfalpha',\sigma') -
  \log q(\mu',\bfalpha' |\sigma')
  \Bigl |_{(\tilde{\mu}',\tilde{\bfalpha}')}=0 ,
\end{equation*}
at our chosen centering, $(\tilde{\mu}',\tilde{\bfalpha}')$, point to get
proposal parameters for $q(\mu',\bfalpha'|\sigma')$, where $\nabla =
(\partial / \partial\mu', \partial / \partial\bfalpha')$.  Substituting
\begin{multline*}
  \pi(\mu',\bfalpha',\sigma') \propto \\
  \exp\left\{ -\frac{1}{2} \left(
      \tau_{\mu}\mu'^2 + \tau_{\alpha}\sum \alpha'^2_i + 
      \sigma'\sum_{ijt} E_{ijt}\left(R_{ijt}-\mu'-\alpha'_i \right)^2
    \right)\right\} \times \\
  \sigma'^{\frac{mns}{2}+a-1} \exp\{ -b\sigma'\}
\end{multline*}
and assuming that $q(\mu', \bfalpha'|\sigma')$ is a Gaussian density with variance
matrix $\varSigma_2$ and mean vector $\bfm_2$.
\begin{equation*} \varSigma_2^{-1}=
  \begin{pmatrix}
    \nabla_{\mu}^2 g & \nabla_{\mu, \bfalpha}^2 g \\
    & \nabla_{\bfalpha}^2 g
  \end{pmatrix} ,
\end{equation*}
and $\bfm_2$ satisfies
\begin{equation*}
  \varSigma_2^{-1}
  \left(\begin{pmatrix}\tilde\mu'\\\widetilde\bfalpha' \end{pmatrix} 
    - \bfm_2\right)=
  \begin{pmatrix}  \nabla_{\mu} g \\
    \nabla_{\bfalpha} g \end{pmatrix},
\end{equation*}
which gives the mean vector
\begin{equation*}
  \bfm_2 = \begin{pmatrix} \tilde\mu' \\ \widetilde\bfalpha' \end{pmatrix} - 
  \varSigma_2 \begin{pmatrix}
    \nabla_{\mu} g \\
   \nabla_{\bfalpha} g
  \end{pmatrix},
\end{equation*}
where $\nabla_{\mu} = (\partial / \partial\mu')$, and $\nabla_{\bfalpha} =
(\partial/\partial\bfalpha')$ and
\begin{equation*}
  g =
    \tau_{\mu}\mu'^2 +
    \tau_{\alpha} \sum_{i=2}^m {\alpha'}_i^2 +
    \sigma'\sum_{ijt} E_{ijt}(R_{ijt} - \mu' - {\alpha'}_i)^2
    .
\end{equation*}

\subsection{Moves Between Models $M_1$ and $M_3$}\label{sec:modelsM1andM3}
The acceptance probability for a proposed move between models $M_1$ and $M_3$
is given by $\min\{1, A_{31}\}$ where
\begin{equation*}
  A_{31}=
  \frac{\pi(\mu,\bfalpha,\bfbeta,\sigma)}{\pi(\mu'',\bfbeta'',\sigma'')}
  \frac{p(M_1)}{p(M_3)} 
  \frac{q(\mu'',\bfbeta'',\sigma'')}{q(\mu,\bfalpha,\bfbeta,\sigma)}.
\end{equation*}
Since the parameter values change between models, we first propose a new value
of $\sigma$ and then simulate new values for the other parameters given this
 value of $\sigma$. Thus, the term $A_{31}$ can further be written as
\begin{equation*}
  \frac{\pi(\mu,\bfalpha,\bfbeta,\sigma)}{\pi(\mu'',\bfbeta'',\sigma'')}
  \frac{p(M_1)}{p(M_3)}   
  \frac{q(\mu'',\bfbeta''|\sigma'')}{q(\mu,\bfalpha,\bfbeta|\sigma)}
  \frac{q(\sigma'')}{q(\sigma)} .
\end{equation*}
If we split the term $\log A_{31}$ into terms involving 
$(\mu,\bfalpha,\bfbeta)$ and $(\mu'', \bfbeta'')$,
we can see that $q(\mu, \bfalpha, \bfbeta|\sigma)$ will be identical to those
derived in Section~\ref{sec:modelsM1andM2}, since $ \log A_{31} = $
\begin{multline*}
  \log \pi(\mu,\bfalpha,\bfbeta, \sigma) 
  - q(\mu,\bfalpha, \bfbeta|\sigma) 
  - \log \pi(\mu'',\bfbeta'', \sigma'') 
  + q(\mu'',\bfbeta''|\sigma'') + K_{13},
\end{multline*}
We therefore refer to Section~\ref{sec:modelsM1andM2} for the proposal
parameters involving model $M_1$.  For moves from model $M_3$ to model $M_1$,
we are increasing the size of the parameter space by $23$ and such big
changes in the size of the parameter vector can have small acceptance rates.
For the reverse move, we decrease the parameter space from $34$ to $11$, a
decrease of $23$. Since the parameter values and interpretations change
between models, we propose to change all parameter values when we change
models. This means that even though the models are nested,  `down' moves
are not deterministic.

Assuming now that $q(\mu'',\bfbeta''|\sigma'')$ is Gaussian with variance
matrix $\varSigma_3$ and mean vector $\bfm_3$, then solving
\begin{equation*}
  \nabla^k \log \pi(\mu'', \bfbeta'', \sigma'') - 
  \log q(\mu'', \bfbeta''|\sigma'') 
  \Biggr|_{(\tilde{\mu}'', \tilde{\bfbeta}'')},
\end{equation*}
where $(\tilde\mu'', \tilde{\bfbeta}'')$ is the centering point, yields
\begin{equation*}\varSigma_3^{-1}=
  \begin{pmatrix}
    \nabla_{\mu}^2 h & \nabla_{\mu,\bfbeta}^2 h \\
    & \nabla_{\bfbeta}^2 h
  \end{pmatrix},
\end{equation*}
and
\begin{equation*}
  \varSigma_3^{-1}
  \left( \begin{pmatrix}\tilde\mu''\\ \widetilde\bfbeta''  \end{pmatrix}
    -\bfm_3 \right)=
  \begin{pmatrix} 
    \nabla_{\mu} h \\
    \nabla_{\bfbeta} h 
  \end{pmatrix},
\end{equation*}
so that
\begin{equation*}
  \bfm_3 =   
  \begin{pmatrix}\tilde\mu''\\ \widetilde\bfbeta''  \end{pmatrix}-
  \varSigma_3
    \begin{pmatrix} 
    \nabla_{\mu} h \\
    \nabla_{\bfbeta} h 
  \end{pmatrix},
\end{equation*}
where $\nabla_{\mu} = (\partial / \partial\mu'')$, and $\nabla_{\bfbeta} =
(\partial/\partial\bfbeta'')$, and
\begin{equation*}
  h =
    \tau_{\mu}\mu''^2 + 
    \tau_{\beta} \sum_{j=2}^n {\beta''}_j^2 +
    \sigma''  \sum_{ijt} E_{ijt}(R_{ijt} - \mu'' - {\beta''}_j)^2
    .
\end{equation*}

\subsection{Moves Between Models $M_2$ and $M_3$}\label{sec:modelsM2andM3}
The acceptance for moves between models $M_2$ and $M_3$ is given by
$\min\{1,A_{23}\}$ where
\begin{equation*}
  A_{23}=
  \frac{\pi(\mu'',\bfbeta'',\sigma'')}{\pi(\mu',\bfalpha',\sigma')}
  \frac{p(M_3)}{p(M_2)}  
  \frac{q(\mu',\bfalpha',\sigma')}{q(\mu'',\bfbeta'',\sigma'')}.
\end{equation*}
We further note that both models have different interpretation, and in
addition, there seems to be no natural diffeomorphism between the parameters
in both models. The precision parameter $\sigma$ is clearly different under
both models. We use again our strategy of simulating $\sigma$ then simulating
the remaining model parameters based on this new value of $\sigma$.
To reflect this the term $A_{23}$ can therefore be written as
\begin{equation*}
  \frac{\pi(\mu'',\bfbeta'',\sigma'')}{\pi(\mu',\bfalpha',\sigma')}
  \frac{p(M_3)}{p(M_2)}  
  \frac{q(\mu',\bfalpha'|\sigma')q(\sigma')}
  {q(\mu'',\bfbeta''|\sigma'')q(\sigma'')}.
\end{equation*}
For all the derived covariance matrices and mean vectors, the between--model
moves seem to be accepted with greater frequency, if we use a point close to
the posterior modes as the centering point. Using values dispersed with
respect to the posterior modes still result in the same stationary
distribution. However, the proposal parameters decline in quality if we do so.
This results in fewer between--model moves being accepted.  For example when
moving between models $M_2$ and $M_3$, centering at $\bfzero$ will result in
the same covariance matrix, since the covariance matrix depends only on the
data and $\sigma$. However, in computing the mean vector $\bfm_k$, the values
may not be close to the posterior modes which may also result in proposed
values being in a part of the space with very low probability. 

These values
will not affect the prior ratio or the proposal ratio. However, these values
will result in a small value for the likelihood and consequently a small
value for the acceptance probability. This seems to be dependent on the data,
as in this case, there are lots of data available which leads to a dominating
likelihood.  Thus small changes in the value of parameters can lead to
disproportionately large changes in the likelihood function. In fact, no such
between--model moves were observed in our simulations when the centering point
was not near the posterior modes.

\subsection{Simulation Study}\label{sec:ds4sim}
To investigate the possibility that model $M_1$ is superior simply because it
has more parameters, we simulate several datasets from models $M_2$ and
$M_3$, and apply the algorithm to see which model has greatest posterior
probability.  To test the accuracy of the reversible jump model selection we
simulate several datasets from models $M_2$ and $M_3$. We then apply the
algorithm to these datasets to observe the posterior model probabilities. If
the algorithm is working correctly then the model from which the data are
simulated should have the highest posterior probability.  For datasets that
are simulated from model $M_2$, both models $M_1$ and $M_2$ were able to
estimate these parameters; however model $M_3$ could not. For the reversible
jump algorithm however model $M_2$ had posterior probability equal to 1.

For data simulated from model $M_3$, both models $M_1$ and $M_3$ are able
to estimate accurately the parameter values. Model $M_3$ had posterior
probability equal to 1.
These results are consistent with what we expect. The model $M_1$ was
able to fit the data simulated from both models $M_2$ and $M_3$, since both
are sub--models of the bigger model. However model $M_2$ could not adequately
describe the data simulated from model $M_3$; likewise model $M_3$ could not
adequately describe the data simulated from model $M_2$. 
For data that are simulated from model $M_1$ both models $M_2$ and $M_3$
provided rather poor fits.
In each case the algorithm chose the model from which the data was simulated
with probability 1. This means that having additional parameters does not
provide a better fit to the data simulated from the smaller models.
\begin{table}[!b]
  \centering
  \caption{Simulation Studies.}\label{tab:simdataresults}
  \begin{tabular}{c|ccc}
    \hline\hline
    Data Origin  & \multicolumn{3}{c}{Fit}\\
            &   $M_1$ & $M_2$ & $M_3$ \\
            \hline
    $M_1$   & \checkmark & \Cross & \Cross \\
    $M_2$   & \checkmark & \checkmark &\Cross\\
    $M_3$   & \checkmark & \Cross & \checkmark\\
  \end{tabular}
\end{table}

\subsection{Sensitivity to Prior Parameters and Centering Point}
The results of our analysis are not sensitive to the prior distributions,
which is good. The convergence of our algorithm, however, depends on the
choice of a suitable centering point. A choice of centering point close to the posterior
mean of the model parameters results in an algorithm which converges much
faster.  Our method is firstly to run each model individually and record the
posteriors means once the algorithm has converged. These values are then used
as our centering point. 

Centering at the posterior modes is not strictly necessary for the algorithm
to work.  However, since moves of type $(M_2\Leftrightarrow M_3)$ are not
between nested models, centering at the posterior modes provides a useful
guide for determining the mean vector and covariance matrix of the proposal
densities $q(\mu', \bfalpha'|\sigma')$ and $q(\mu'' , \bfbeta''|\sigma'')$
since models $M_2$ and $M_3$ are not nested. Generally, centering at posterior
modes allow for non-nested model moves. If we considered only moves of type
$(M_1\Leftrightarrow M_2)$ or $(M_1\Leftrightarrow M_3)$ then the choice of a 
centering point would not matter since both models $M_2$ and $M_3$ are nested
sub--models of model $M_1$.

\subsection{Reversible Jump Results}
The full model with both state and occupation parameters is preferred to the
restricted models with either state only or occupation only parameters. The
posterior probability of the full model is approximately 1. It could be the
case that $M_1$ is preferred simply because it has more parameters and so is
better at explaining the data. However, as we shall explore in
Section~\ref{sec:ds4sim} with simulated data, if the data are from models
$M_2$ or $M_3$ both models would be selected with probability 1 and
would be preferred to the more complex model $M_1$. 


Even though the posterior model probabilities for $M_2$ and $M_3$ are very
small, practically 0, we can still get the reversible jump algorithm to
explore the model space by choosing appropriate prior model probabilities.
For convenience, i.e. for the algorithm to jump between models we take
$p(M_1) /p(M_2)$ = $e^{-334}$, $p(M_1)/ p(M_3)$ = $e^{-206}$ and hence
$p(M_3)/p(M_2)$ = $e^{-128}$.  These priors allow the algorithm to explore all
three models.  The resulting transition matrix, where the $(i,j)$ term gives
the probability of moving between model $M_i$ and $M_j$, is
\begin{equation*}
  \bordermatrix{& M_1 & M_2 & M_3 \cr
    M_1 & 0.554 &0.233 &0.213 \cr
    M_2 & 0.208 &0.565 &0.227 \cr
    M_3 & 0.198 &0.246 &0.556 \cr
  }
\end{equation*}
which has limiting probabilities $(0.308, 0.358, 0.334)$.  Taking into
consideration the prior model probabilities, the results clearly indicate that the
full model $M_1$ is more likely to describe the process which generated the
data. Models $M_2$ and $M_3$ are less likely to describe such a process. 
\begin{figure}
  \centering
  \includegraphics[width=0.8\textwidth]{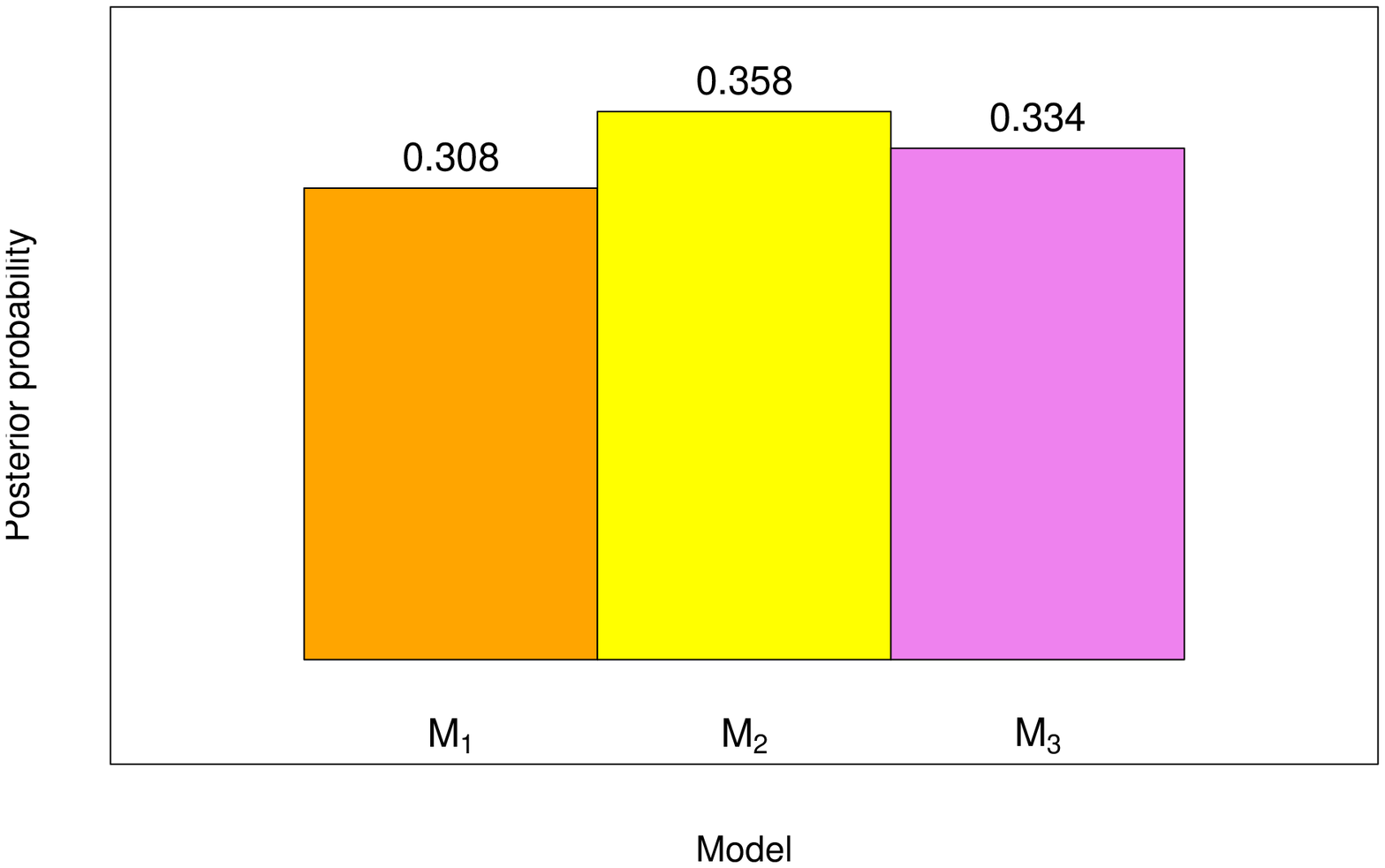}
  \caption{Posterior model probabilities. The prior probabilities
    have been chosen so that the models have approximately equal posterior
    probabilities.}
\end{figure}
\begin{figure}
  \centering
  \includegraphics[width=0.8\textwidth]{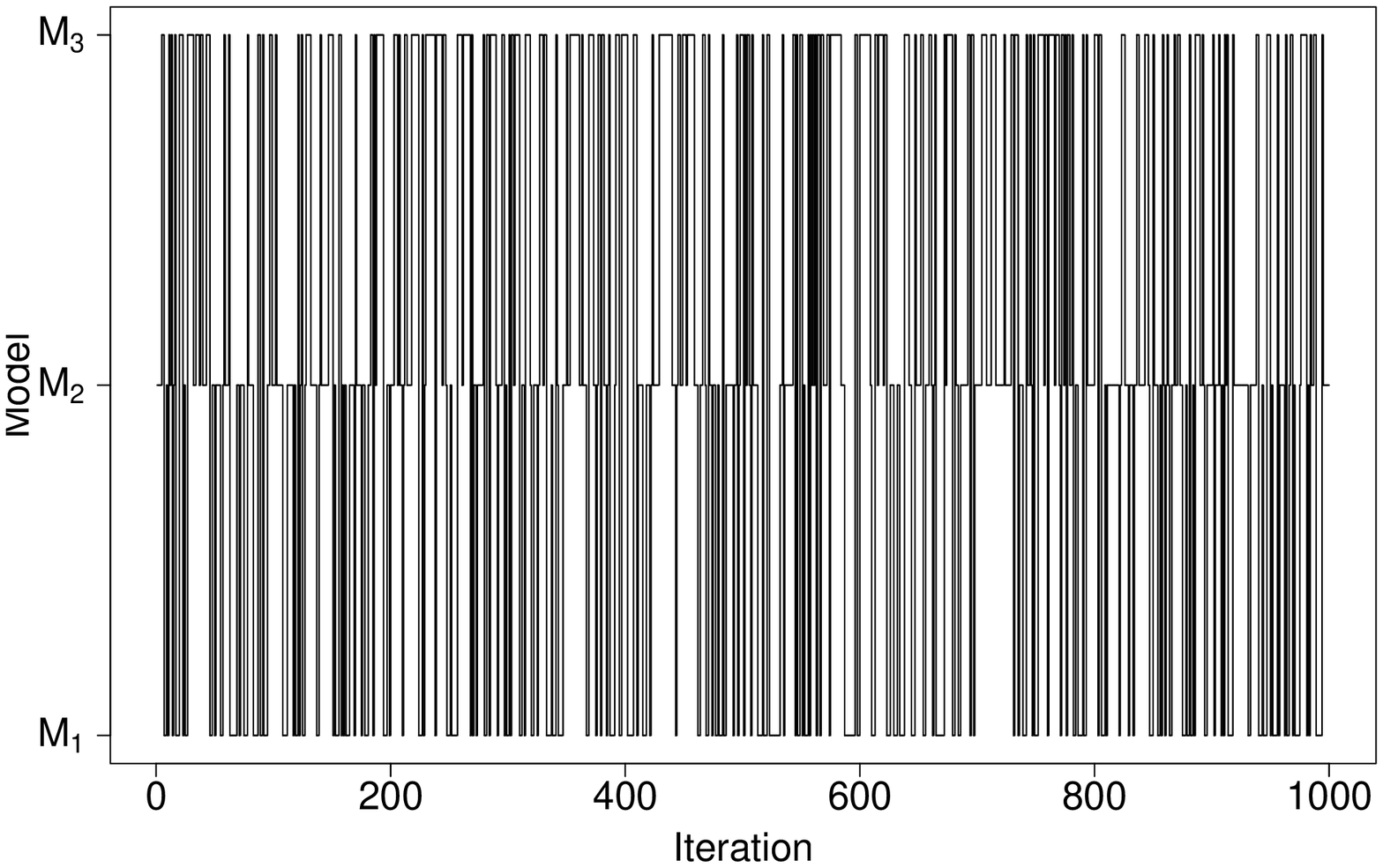}
  \caption{Model trace indicator for the reversible jump algorithm.}
\end{figure}
\begin{figure}
    \centering
    \includegraphics[width=0.8\textwidth]{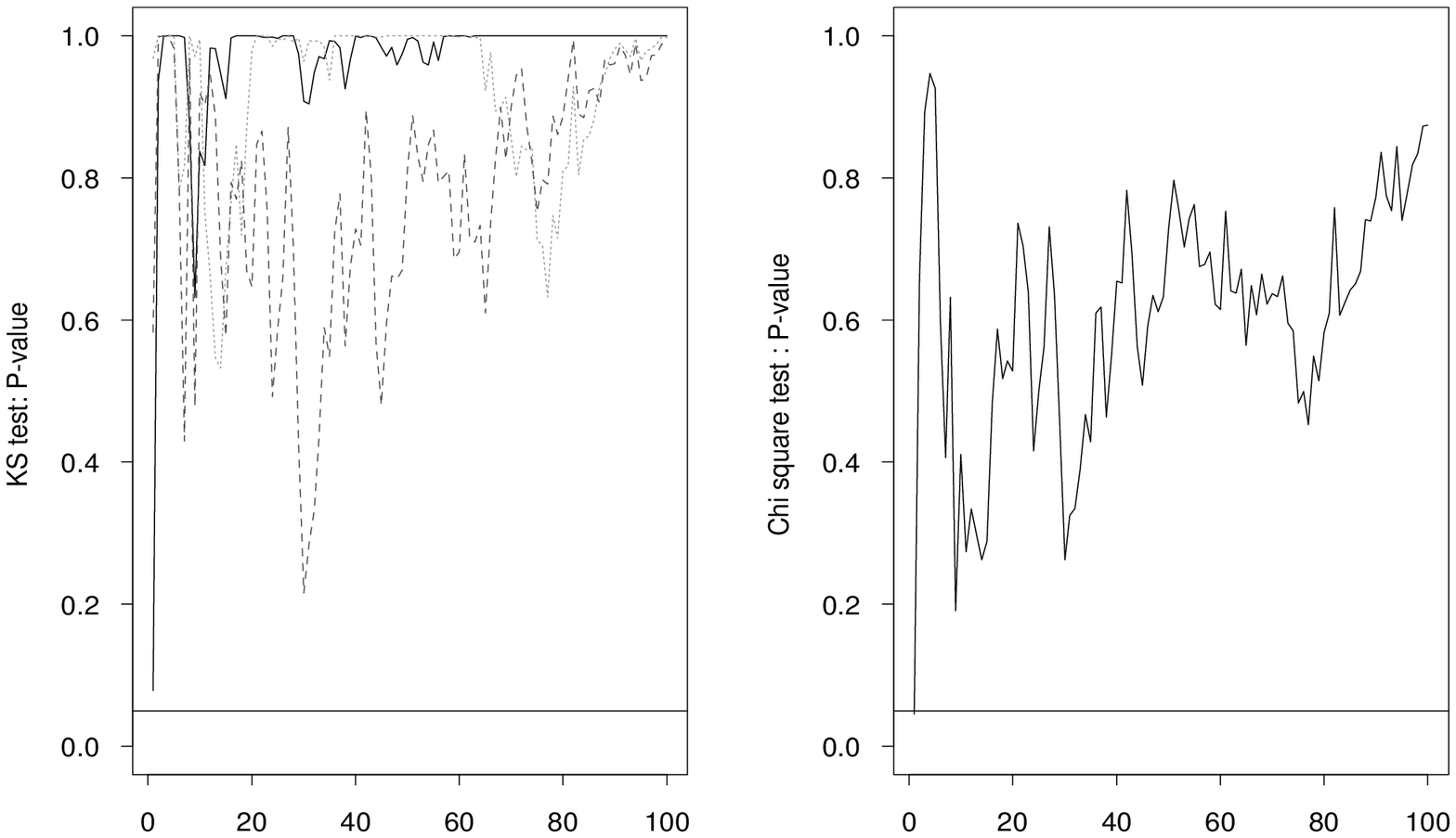}
    \caption{Convergence diagnostics.}\label{fig:ds4:conv}
\end{figure}
The centering points used are the posterior means of the model
parameters.  For the scale parameters, we simulate those from their marginal
posterior distributions, and then simulate the other proposal parameters
conditional on this value of the scale parameter. The results clearly show
the $\pi(M_1|\bfR) \approx 1$, $\pi(M_2 |\bfR) \approx 0$, and $\pi(M_3|\bfR)
\approx 0$.  To assess convergence of the algorithm, we simulate 3 chains
using different starting values and different random number seeds for a total
of 100000 iterations. Both the $\chi$-square and Kolmogorov--Smirnov
diagnostics are computed. These diagnostics are plotted in
Figure~\ref{fig:ds4:conv}. In both cases, the diagnostic is well above the
critical 5\% value. The methods employed in this paper are not exactly weak
non-identifiable centering methods.  The approach that we use is general
enough so that down moves, say from model $M_1$ to models $M_2$ or $M_3$, are
not deterministic. This approach also allow the use of the posterior mean
as the mean of or proposal density, from which the posterior variance
matrix can be derived using the centering methods described.
\section{Summary}
Anova models arise in many areas of insurance credibility theory. The
B{\"u}hlmann--Straub model can be reparameterised as a one--way model. In this
paper, we use posterior model probabilities to compare one--way and 
two--way models. The results can be extended to cover yet more general models
such as the \citeN{jewell1975} and \citeN{taylor1974} models. Using the
conditional maximisation scheme of \shortciteN{brooks2003}, we 
constructed an algorithm which can be used to compute posterior model
probabilities for model discrimination. When applied to loss ratios extracted 
from datasets of worker's compensation insurance in the United States,
 there is
overwhelming posterior odds in favour of the full model. This seems quite
plausible given the structure and size of the data used. Model discrimination
measures computed using the deviance information criterion, $DIC$,  give results which are
consistent with the reversible jump model probabilities obtained.

\bibliographystyle{chicago}\bibliography{paper3}

\end{document}